\documentclass[paper,notoc]{JHEP3}

  \usepackage{graphicx}
  \usepackage{epsfig}
  \usepackage{axodraw}
  \usepackage{amsmath}
  \usepackage{amssymb}
  \usepackage{subfigure}
  \usepackage{placeins}

  \usepackage{soul}
  
  \newcommand{\eps}{\epsilon}

\newcommand{\be}{\begin{equation}}
\newcommand{\ee}{\end{equation}}
\newcommand{\nn}{\nonumber}
\newcommand{\bea}{\begin{eqnarray}}
\newcommand{\eea}{\end{eqnarray}}
\newcommand{\bfig}{\begin{figure}}
\newcommand{\efig}{\end{figure}}
\newcommand{\bc}{\begin{center}}
\newcommand{\ec}{\end{center}}
\newcommand{\ep}{\epsilon}
\newcommand{\varep}{\varepsilon}

\graphicspath{ {fig/} }

\allowdisplaybreaks[4]

\title{Next-to-Leading Order QCD Corrections to the Decay Width $H \to Z \gamma$}

\author{Roberto Bonciani\\
Dipartimento di Fisica, Universit\`a di Roma ``La Sapienza'' and 
INFN Sezione di Roma, 00185 Roma, Italy\\
       E-mail: \email{roberto.bonciani@roma1.infn.it}}

\author{Vittorio Del Duca\\
Institute for Theoretical Physics, ETH Zurich, 8093 Zurich, Switzerland\\
and INFN Laboratori Nazionali di Frascati, 00044 Frascati (Roma), Italy\\
       E-mail: \email{delducav@itp.phys.ethz.ch},\email{delduca@lnf.infn.it}}

\author{Hjalte Frellesvig\\
Institute of Nuclear and Particle Physics, NCSR ``Demokritos'',\\
Agia Paraskevi, 15310, Greece\\
       E-mail: \email{frellesvig@inp.demokritos.gr}}

\author{Johannes M. Henn\\
Institute for Advanced Study, Princeton, NJ 08540, USA\\
       E-mail: \email{jmhenn@ias.edu}}
       
\author{Francesco Moriello\\
Dipartimento di Fisica, Universit\`a di Roma ``La Sapienza'' and 
INFN Sezione di Roma, 00185 Roma, Italy\\
       E-mail: \email{francesco.moriello@roma1.infn.it}}

\author{Vladimir A. Smirnov\\
Skobeltsyn Inst. of Nuclear Physics of Moscow State University, 119991 Moscow, Russia\\
       E-mail: \email{smirnov@theory.sinp.msu.ru}}

\abstract{We present the analytic calculation of the two-loop QCD corrections to the decay
width of a Higgs boson into a photon and a $Z$ boson. 
The calculation is carried out using integration-by-parts identities for the reduction to master integrals of the scalar integrals, 
in terms of which we express the amplitude. The calculation of the master integrals is performed using
differential equations applied to a set of functions suitably chosen to be of uniform weight.
The final result is expressed in terms of logarithms and polylogarithmic functions $\text{Li}_2$, $\text{Li}_3$, $\text{Li}_4$
and $\text{Li}_{2,2}$.}

\preprint{~}

\begin{document}

\section{Introduction}
\label{sec:intro}

The discovery at the Large Hadron Collider (LHC) of the Higgs boson of the 
Standard Model (SM)~\cite{Aad:2012tfa,Chatrchyan:2012ufa} calls for the investigation of the its properties, 
with the degree of agreement between the observed and the predicted behaviour being still an open question.
The new boson decays into two photons or two electroweak $W/Z$ bosons. 
It should also decay into a photon and a $Z$ boson. If it is a SM Higgs boson with a mass of
125.1~GeV, the branching ratio is $B(H\to Z \gamma) = 1.55 \cdot 10^{-3}$, with an uncertainty
of about 9\%~\cite{Heinemeyer:2013tqa}. The $H\to Z \gamma$ decay is being searched by the
CMS~\cite{Chatrchyan:2013vaa} and ATLAS~\cite{Aad:2014fia} Collaborations at the LHC.

The leading order evaluation of the SM $H\to Z \gamma$ decay was performed long ago~\cite{Cahn:1978nz,Bergstrom:1985hp}.
As the Higgs boson has no electric charge, it does not couple directly to photons. 
Then the decay of a Higgs boson into a photon and a $Z$ boson must be mediated at loop level
by charged particles. At one loop, it
is mediated by a heavy-quark loop~\cite{Cahn:1978nz} or a $W$-boson loop~\cite{Bergstrom:1985hp}, 
just as for the $H\to \gamma \gamma$ decay.
$H\to Z \gamma$ may provide information on new physics,
as a different $H\to Z \gamma$ decay rate is expected if $H$ is a non-SM scalar 
boson~\cite{Low:2011gn,Low:2012rj}, or a composite state~\cite{Azatov:2013ura},
or if different particles circulate in the loop~\cite{Carena:2012xa,Chiang:2012qz,Chen:2013vi}.

The two--loop QCD corrections to the $H\to Z \gamma$ decay were computed numerically in Ref.~\cite{Spira:1991tj}.
They correspond to two-loop corrections generated by a gluonic exchange within the heavy-quark loop.
In this paper, we present an analytic calculation of the QCD corrections 
to $H \to Z \gamma$. 

The paper is organised as follows: in Sect.~\ref{sec:HZgammaLO}, we recall the definition and the explicit expressions
of the amplitude for the decay $H\to Z\gamma$ at leading order; in Sect.~\ref{sec:HZgammaNLOQCD}, we describe 
how we perform the analytic computation of the NLO QCD corrections to the decay width; in Sect.~\ref{sec:numerics},
we provide numerical results as a function of the Higgs mass; in Sect.~\ref{sec:conclusions}, we draw our conclusions. 
The appendices contain details on the projection operators used to define the decay amplitude (App.~\ref{sec:appa}),
the master integrals (App.~\ref{sec:appb}), the matrices of the system of differential equations for the master integrals (App.~\ref{sec:appc})
and the analytic properties of the functions occurring in the master integrals (App.~\ref{sec:appd}).


\section{$H \to Z \gamma$ at Leading Order}
\label{sec:HZgammaLO}

\subsection{The amplitude for the decay $H\to Z\gamma$}
\label{sec:vertHZgamma}

At leading order the SM Higgs boson decays into a photon and a $Z$ boson via either a heavy-fermion loop or a $W$-boson loop. 
The corresponding Feynman diagrams are shown in Fig.~\ref{fig1}.
%
\begin{figure}
\bc
\[ \vcenter{
\hbox{
  \begin{picture}(0,0)(0,0)
\SetScale{1.}
  \SetWidth{.4}
\DashLine(-50,0)(-35,0){2}
\Photon(10,-30)(45,-30){3}{6}
\Photon(10,30)(45,30){3}{6}
%
%
  \SetWidth{1.4}
\ArrowLine(-35,0)(10,30)
\ArrowLine(10,30)(10,-30)
\ArrowLine(10,-30)(-35,0)
\Text(0,-60)[c]{(a)}
\Text(55,35)[c]{$Z$}
\Text(55,-35)[c]{$\gamma$}
\Text(-20,-25)[c]{$Q$}
\Text(-55,10)[c]{$H$}
\end{picture}}
}
\hspace{5.cm}
\vcenter{
\hbox{
  \begin{picture}(0,0)(0,0)
\SetScale{1.}
  \SetWidth{.4}
\DashLine(-50,0)(-35,0){2}
\Photon(10,-30)(45,-30){3}{6}
\Photon(10,30)(45,30){3}{6}
%
%
\Photon(-35,0)(10,30){3}{6}
\Photon(10,30)(10,-30){3}{6}
\Photon(10,-30)(-35,0){3}{6}
\Text(0,-60)[c]{(b)}
\Text(55,35)[c]{$Z$}
\Text(55,-35)[c]{$\gamma$}
\Text(-20,-25)[c]{$W$}
\Text(-20,25)[c]{$W$}
\Text(25,2)[c]{$W$}
\Text(-55,10)[c]{$H$}
\end{picture}}
}
\hspace{5.cm}
%
\vcenter{
\hbox{
  \begin{picture}(0,0)(0,0)
\SetScale{1.}
  \SetWidth{.4}
\DashLine(-50,0)(-35,0){2}
\Photon(10,0)(45,-30){3}{6}
\Photon(10,0)(45,26){3}{6}
%
%
%
%
\PhotonArc(-12.5,0)(20,0,180){3}{6}
\PhotonArc(-12.5,0)(20,180,360){3}{6}
\Text(0,-60)[c]{(c)}
\Text(55,35)[c]{$Z$}
\Text(55,-35)[c]{$\gamma$}
\Text(-18,-33)[c]{$W$}
\Text(-18,33)[c]{$W$}
\Text(-55,10)[c]{$H$}
\end{picture}}
}
\]
\vspace*{15mm}
\caption{\label{fig1} Feynman diagrams for the decay process 
$H \to Z \gamma$ at leading order.}
\ec
\efig
%
Let us label the $Z$-boson momentum as $p_1$ and the photon momentum as $p_2$. The general Lorentz structure $T^{\mu \nu}$ of the amplitude,
\be
{\cal M} = T^{\mu \nu} \varep_\mu(p_1) \varep_\nu(p_2) \, ,
\ee
for the decay of a Higgs boson into a $Z$ boson and a photon with polarization vectors $\varep_\mu(p_1)$ and $\varep_\nu(p_2)$ respectively, is
\be
T^{\mu \nu} =   p_1^{\mu} p_1^{\nu} \, T_1 
              + p_2^{\mu} p_2^{\nu} \, T_2
	      + p_1^{\mu} p_2^{\nu} \, T_3 
              + p_2^{\mu} p_1^{\nu} \, T_4
	      + \delta^{\mu \nu} \, T_5
	      + \ep^{\mu \nu \rho \sigma} p_{1 \, \rho} p_{2 \, \sigma} \, T_6 \, ,
\label{ffactgen}
\ee
where $p^2 = \sum_{i=1}^{3} p_i^2 - p_0^2$ and $p_1^2 = - m_Z^2$,  $p_2^2 = 0$,  $(p_1+p_2)^2 = - m_H^2$. 
$m_H$ is the Higgs mass and $m_Z$ is the $Z$-boson mass. 
%
The coefficients $T_i$ are functions of the Mandelstam invariants of the problem under consideration and can be extracted using projector operators $P_i^{\mu \nu}$ such that $T_i = P_{i \, \mu \nu} T^{\mu \nu}$. The projector operators are collected in Appendix~\ref{sec:appa}.

Requiring the photon gauge invariance, $T^{\mu \nu} p_{2\nu} = 0$, in Eq.~(\ref{ffactgen}), we obtain
\bea
T_1 & = & 0 \, , \nonumber\\
T_5 & = & - p_1 \cdot p_2\, T_4 = \frac{m_H^2-m_Z^2}{2} \, T_4 \, .
\eea
Furthermore, $T_2$, $T_3$ and $T_6$\footnote{Note that $T_6$ receives contribution only from the 
axial-vector part of the $ZQ\bar{Q}$ vertex. 
However, since the Higgs particle is a C-even state, and the photon is C-odd, only the C-odd
coupling of the $Z$ ({\it i.e.} the vector coupling and not the axial-vector one) contributes to the decay 
width. This implies that $T_6$ does not contribute to the general form factor (\ref{ffactgen}).
In fact, when we add together the diagrams with an opposite flow of the fermionic
arrow, the contribution to $T_6$ changes in sign, in such a way that in
the sum $T_6=0$.} 
do not contribute to the squared amplitude $|{\cal M}|^2$.
Thus, up to contributions which vanish in $|{\cal M}|^2$, the form factor $T^{\mu \nu}$ can be written as,
\be
T^{\mu \nu} = \left( p_2^{\mu} p_1^{\nu} - p_1 \cdot p_2\, \delta^{\mu \nu} \right) \, T_4 \, .
\ee
%
%

\subsection{Leading Order Contribution}

The width for the decay of a Higgs boson into a photon and a $Z$ boson can be 
cast in the following form,
\be
\Gamma_{H \to Z \gamma} = \frac{G_F \alpha^2}{64 \sqrt{2} \pi^3 m_H} \, \frac{(m_H^2 - m_Z^2)^3}{m_H^2} \, | {\mathcal F} | ^2 \, ,
\label{decaywidthfin}
\ee
where $\alpha$ is the fine structure constant, $G_F$ is the Fermi constant and
where we introduced the function ${\mathcal F}$ related to the form factor $T_4$ by the following equation,
\be
{\mathcal F} = \frac{16 \pi^2 m_W}{g^3 s_W^2} \, T_4 \, .
\label{FinT4}
\ee
In Eq.~(\ref{FinT4}), $g$ is the weak coupling constant, $s_W = \sin{\theta_W}$ is the sine of the weak mixing angle, and $m_W$ is the $W$-boson mass.

The function ${\mathcal F} $ can be expanded in powers of the coupling constants, starting with the one-loop contribution. 
At leading order we have,
\be
{\mathcal F}^{(1l)} = \frac{c_W}{s_W} \, {\mathcal F}^{(1l)}_{W} 
+ N_c \, Q_q \, \frac{\left( \frac{T_q^3}{2}-Q_q s_W^2 \right)}{s_W c_W} \, {\mathcal F}^{(1l)}_{q} 
\, ,
\ee
where $N_c$ is the number of colors, $q = t, b$ labels the type of heavy quark, either a top or a bottom, circulating in the loop, 
$Q_q$ is the heavy--quark electric charge in units of $e$, $T_q^3$ is the 
third component of the heavy--quark isospin, and $c_W = \cos{\theta_W}$ is the cosine of the weak mixing angle. 
The actual expressions for ${\mathcal F}^{(1l)}_{W}$ and ${\mathcal F}^{(1l)}_{q}$ are,
\bea
{\mathcal F}^{(1l)}_{W} & = & 
\frac{(m_H^2m_Z^2 + 2m_Z^2m_W^2 - 2m_H^2m_W^2 - 12m_W^4)}{m_W^2(m_H^2-m_Z^2)} \nn\\
& & + \frac{(2m_H^2m_Z^2m_W^2 - 2m_Z^4m_W^2 - m_Z^4m_H^2 - 12m_Z^2m_W^4)}{m_H^2(m_H^2-m_Z^2)^2} \, 
I(x_W,y_W) \nn\\
& & + \frac{(4m_Z^4-2m_Z^2m_H^2+12m_W^2m_H^2-12m_W^2m_Z^2-24m_W^4)}{(m_H^2-m_Z^2)^2} \, 
J(x_W,y_W) \, , \\
{\mathcal F}^{(1l)}_{q} & = & \frac{8m_q^2}{(m_H^2-m_Z^2)} 
+ \frac{8m_q^2 \, m_Z^2}{(m_H^2-m_Z^2)^2} \, I(x_q,y_q)  + \frac{4 m_q^2(4 m_q^2-m_H^2+m_Z^2)}{(m_H^2-m_Z^2)^2} \, J(x_q,y_q) , 
\eea
with
\bea
I(x_f,y_f) & = & \sqrt{1-\frac{4 m_f^2}{m_H^2}} \, \log(x_f)- \sqrt{1-\frac{4 m_f^2}{m_H^2}} \, \log(y_f)  \, , \\
J(x_f,y_f) & = & \frac{1}{2} \, \log^2(x_f) - \frac{1}{2} \, \log^2(y_f) \, ,
\eea
where the variables $x_f$ and $y_f$ are defined through,
\be
m_H^2  =  - m_f^2 \frac{(1-x_f)^2}{x_f} \, , \quad
m_Z^2  =  - m_f^2 \frac{(1-y_f)^2}{y_f} \, ,
\label{xandydef}
\ee
with $f=W, q$ and $q = t, b$.


Let us consider real values of $m_t$, $m_W$, $m_Z$
and $m_H$. In the 
region $0<m_H<2 m_f$, $x_f$ has an imaginary part, with unit modulus and phase between 0 and $\pi$,
\be
x_f = \exp \left\{ i \arctan{\frac{\sqrt{m_H^2(4m_f^2-m_H^2)}}{2 m_f^2-m_H^2}} \right\} \, .
\ee
In the region $m_H>2 m_f$, we have $-1<x_f<0$; we define $x_f = - x_f'+i 0$ with
\be
x_f' = \frac{\sqrt{m_H^2}-\sqrt{m_H^2-4m_f^2}}{\sqrt{m_H^2}+\sqrt{m_H^2-4m_f^2}} \, .
\ee
Analogously, in the case $0<m_Z<2 m_f$ the variables $y_f$ are on the unit circle
\be
y_f = \exp \left\{ i \arctan{\frac{\sqrt{m_Z^2(4m_f^2-m_Z^2)}}{2 m_f^2-m_Z^2}} \right\} \, ,
\ee
while for $m_Z>2 m_f$ we define $y_f = - y_f'+i 0$ with
\be
y_f' = \frac{\sqrt{m_Z^2}-\sqrt{m_Z^2-4m_f^2}}{\sqrt{m_Z^2}+\sqrt{m_Z^2-4m_f^2}} \,.
\ee


\section{NLO QCD Corrections}
\label{sec:HZgammaNLOQCD}

The diagrams involved in the calculation of the NLO QCD corrections to the decay width of a Higgs boson 
into a photon and a $Z$ boson are shown in Fig.~\ref{fig2}. 
%
\begin{figure}
\bc
\[ \vcenter{
\hbox{
  \begin{picture}(0,0)(0,0)
\SetScale{1.}
  \SetWidth{.4}
\DashLine(-50,0)(-35,0){2}
\Photon(10,-30)(45,-30){3}{6}
\Photon(10,30)(45,30){3}{6}
%
%
\Gluon(-5,20)(-5,-20){3}{5}
  \SetWidth{1.4}
\ArrowLine(-35,0)(10,30)
\ArrowLine(10,30)(10,-30)
\ArrowLine(10,-30)(-35,0)
\Text(0,-60)[c]{(a)}
\Text(55,35)[c]{$Z$}
\Text(55,-35)[c]{$\gamma$}
\Text(-20,-25)[c]{$t$}
\Text(-55,10)[c]{$H$}
\end{picture}}
}
\hspace{5.cm}
\vcenter{
\hbox{
  \begin{picture}(0,0)(0,0)
\SetScale{1.}
  \SetWidth{.4}
\DashLine(-50,0)(-35,0){2}
\Photon(10,-30)(45,-30){3}{6}
\Photon(10,30)(45,30){3}{6}
%
%
\Gluon(-10,16)(10,-5){3}{5}
  \SetWidth{1.4}
\ArrowLine(-35,0)(10,30)
\ArrowLine(10,30)(10,-30)
\ArrowLine(10,-30)(-35,0)
\Text(0,-60)[c]{(b)}
\Text(55,35)[c]{$Z$}
\Text(55,-35)[c]{$\gamma$}
\Text(-20,-25)[c]{$t$}
\Text(-55,10)[c]{$H$}
\end{picture}}
}
\hspace{5.cm}
%
\vcenter{
\hbox{
  \begin{picture}(0,0)(0,0)
\SetScale{1.}
  \SetWidth{.4}
\DashLine(-50,0)(-35,0){2}
\Photon(10,-30)(45,-30){3}{6}
\Photon(10,30)(45,30){3}{6}
%
%
\Gluon(-10,-16)(10,5){3}{5}
  \SetWidth{1.4}
\ArrowLine(-35,0)(10,30)
\ArrowLine(10,30)(10,-30)
\ArrowLine(10,-30)(-35,0)
\Text(0,-60)[c]{(c)}
\Text(55,35)[c]{$Z$}
\Text(55,-35)[c]{$\gamma$}
\Text(-20,-25)[c]{$t$}
\Text(-55,10)[c]{$H$}
\end{picture}}
}
\]
\vspace*{3.5cm}
\[ \vcenter{
\hbox{
  \begin{picture}(0,0)(0,0)
\SetScale{1.}
  \SetWidth{.4}
\DashLine(-50,0)(-35,0){2}
\Photon(10,-30)(45,-30){3}{6}
\Photon(10,30)(45,30){3}{6}
%
%
\GlueArc(-12,14)(13,40,210){3}{5}
  \SetWidth{1.4}
\ArrowLine(-35,0)(10,30)
\ArrowLine(10,30)(10,-30)
\ArrowLine(10,-30)(-35,0)
\Text(0,-60)[c]{(d)}
\Text(55,35)[c]{$Z$}
\Text(55,-35)[c]{$\gamma$}
\Text(-20,-25)[c]{$t$}
\Text(-55,10)[c]{$H$}
\end{picture}}
}
\hspace{5.cm}
\vcenter{
\hbox{
  \begin{picture}(0,0)(0,0)
\SetScale{1.}
  \SetWidth{.4}
\DashLine(-50,0)(-35,0){2}
\Photon(10,-30)(45,-30){3}{6}
\Photon(10,30)(45,30){3}{6}
%
%
\GlueArc(-14,-16)(13,140,330){3}{5}
  \SetWidth{1.4}
\ArrowLine(-35,0)(10,30)
\ArrowLine(10,30)(10,-30)
\ArrowLine(10,-30)(-35,0)
\Text(0,-60)[c]{(e)}
\Text(55,35)[c]{$Z$}
\Text(55,-35)[c]{$\gamma$}
\Text(-20,25)[c]{$t$}
\Text(-55,10)[c]{$H$}
\end{picture}}
}
\hspace{5.cm}
%
\vcenter{
\hbox{
  \begin{picture}(0,0)(0,0)
\SetScale{1.}
  \SetWidth{.4}
\DashLine(-50,0)(-35,0){2}
\Photon(10,-30)(45,-30){3}{6}
\Photon(10,30)(45,30){3}{6}
%
%
\GlueArc(10,-0)(13,-90,90){3}{5}
  \SetWidth{1.4}
\ArrowLine(-35,0)(10,30)
\ArrowLine(10,30)(10,-30)
\ArrowLine(10,-30)(-35,0)
\Text(0,-60)[c]{(f)}
\Text(55,35)[c]{$Z$}
\Text(55,-35)[c]{$\gamma$}
\Text(-20,-25)[c]{$t$}
\Text(-55,10)[c]{$H$}
\end{picture}}
}
\]
\vspace*{15mm}
\caption{\label{fig2} Feynman diagrams for the NLO QCD corrections to the decay process 
$H \to Z \gamma$. Diagrams with the reversed direction of the fermionic arrow are not shown.
We can easily consider the $N_h^2$ contribution (one top-quark loop coupled to the Higgs and the other correcting the $Z$ or $\gamma$ propagator).}
\ec
\efig
%
Their contribution to the form factors can be extracted using the projectors defined in 
Appendix \ref{sec:appa}. Expanding in the strong coupling constant we have,
\be
{\mathcal F} = {\mathcal F}^{(1l)} + \frac{\alpha_S}{\pi} {\mathcal F}^{(2l)}_{0} + ... \, ,
\ee
At this order in $\alpha_S$, the bare form factor, ${\mathcal F}^{(2l)}_{0}$, is UV divergent and needs to 
be renormalized. 
The only renormalization required is the  heavy-quark mass renormalization, in the fermionic propagators and 
in the coupling of the Higgs boson to the heavy-quark pair. 
We choose to perform the mass renormalization in the 
on-shell (OS) scheme \cite{Bardeen:1978yd},
\be
\delta m_{{\mathrm OS}}^{(1l)} \Bigl( \epsilon,m_q,\frac{\mu^2}{m_q^2} \Bigr) 
=  - \, m_q \ \frac{\alpha_S}{\pi}  \, C(\epsilon) \, 
\left( \frac{\mu^{2}}{m_q^2} \right)^{\epsilon} \
\frac{C_{F}}{4} \frac{(3-2 \epsilon)}{\epsilon \, (1-2 \epsilon)} \ ,
\ee
where $\mu$ is the renormalization scale, $C_F = (N_c^2-1)/(2 N_c)$ with $N_c$ the number of colors,
$C(\epsilon) = (4 \pi)^{\epsilon} \Gamma(1+\epsilon)$ and $\epsilon = (4-d)/2$, with $d$ the space-time dimension.
Indicating with ${\mathcal F}^{(2l)}$ the renormalized form factor, we have
\be
{\mathcal F}^{(2l)} = {\mathcal F}^{(2l)}_{0} + \delta m_{{\mathrm OS}}^{(1l)} \, CT \, .
\ee
The contributions to the counterterm come from the diagrams shown in Fig.~\ref{figCT}.
\begin{figure}
\bc
\[ \vcenter{
\hbox{
  \begin{picture}(0,0)(0,0)
\SetScale{1.}
  \SetWidth{.4}
\DashLine(-50,0)(-35,0){2}
\Photon(10,-30)(30,-30){3}{3}
\Photon(10,30)(30,30){3}{3}
%
%
  \SetWidth{1.4}
\ArrowLine(-35,0)(10,30)
\ArrowLine(10,30)(10,-30)
\ArrowLine(10,-30)(-35,0)
\CBoxc(-35,0)(7,7){0.1}{0.9}
\Text(0,-60)[c]{(a)}
\Text(-35,15)[c]{$\delta m$}
\Text(30,43)[c]{$Z$}
\Text(30,-40)[c]{$\gamma$}
\Text(-55,10)[c]{$H$}
\end{picture}}
}
\hspace{3.5cm}
\vcenter{
\hbox{
  \begin{picture}(0,0)(0,0)
\SetScale{1.}
  \SetWidth{.4}
\DashLine(-50,0)(-35,0){2}
\Photon(10,-30)(30,-30){3}{3}
\Photon(10,30)(30,30){3}{3}
%
%
  \SetWidth{1.4}
\ArrowLine(-35,0)(10,30)
\ArrowLine(10,30)(10,-30)
\ArrowLine(10,-30)(-35,0)
\CBoxc(-13,15)(7,7){0.1}{0.9}
\Text(0,-60)[c]{(b)}
\Text(-14,30)[c]{$\delta m$}
\Text(30,43)[c]{$Z$}
\Text(30,-40)[c]{$\gamma$}
\Text(-55,10)[c]{$H$}
\end{picture}}
}
\hspace{3.5cm}
\vcenter{
\hbox{
  \begin{picture}(0,0)(0,0)
\SetScale{1.}
  \SetWidth{.4}
\DashLine(-50,0)(-35,0){2}
\Photon(10,-30)(30,-30){3}{3}
\Photon(10,30)(30,30){3}{3}
%
%
  \SetWidth{1.4}
\ArrowLine(-35,0)(10,30)
\ArrowLine(10,30)(10,-30)
\ArrowLine(10,-30)(-35,0)
\CBoxc(-12,-15)(7,7){0.1}{0.9}
\Text(0,-60)[c]{(c)}
\Text(-14,-30)[c]{$\delta m$}
\Text(30,43)[c]{$Z$}
\Text(30,-40)[c]{$\gamma$}
\Text(-55,10)[c]{$H$}
\end{picture}}
}
\hspace{3.5cm}
\vcenter{
\hbox{
  \begin{picture}(0,0)(0,0)
\SetScale{1.}
  \SetWidth{.4}
\DashLine(-50,0)(-35,0){2}
\Photon(10,-30)(30,-30){3}{3}
\Photon(10,30)(30,30){3}{3}
%
%
  \SetWidth{1.4}
\ArrowLine(-35,0)(10,30)
\ArrowLine(10,30)(10,-30)
\ArrowLine(10,-30)(-35,0)
\CBoxc(10,0)(7,7){0.1}{0.9}
\Text(0,-60)[c]{(d)}
\Text(30,0)[c]{$\delta m$}
\Text(30,43)[c]{$Z$}
\Text(30,-40)[c]{$\gamma$}
\Text(-55,10)[c]{$H$}
\end{picture}}
}
\]
\vspace*{15mm}
\caption{\label{figCT} Counterterm diagrams involved in the heavy-quark mass renormalization.}
\ec
\efig

Retaining only terms of ${\mathcal O}(\alpha_S)$, Eq.~(\ref{decaywidthfin}) can be written as,
\bea
\Gamma_{H \to Z \gamma} 
& = & \frac{G_{\mu} \alpha^2}{64 \sqrt{2} \pi^3 m_H} \, \frac{(m_H^2 - m_Z^2)^3}{m_H^2} \, 
\left\{ \Re{({\mathcal F}^{(1l)})}^2 + \Im{({\mathcal F}^{(1l)})}^2 \right. \nn\\
& & \hspace{25mm} \left. + 2 \, \frac{\alpha_S}{\pi} \left[ \Re{({\mathcal F}^{(1l)})}\Re{({\mathcal F}^{(2l)})} + \Im{({\mathcal F}^{(1l)})}\Im{({\mathcal F}^{(2l)})}    \right]  \right\}  \\
& = & \Gamma_{H \to Z \gamma}^{(1l)} \bigl( 1 + \delta_{QCD} ) \,, \nn
\eea
where we defined
\be
\delta_{QCD} = 2 \, \frac{\alpha_S}{\pi} \frac{\Re{({\mathcal F}^{(1l)})}\Re{({\mathcal F}^{(2l)})} 
+ \Im{({\mathcal F}^{(1l)})}\Im{({\mathcal F}^{(2l)})} }{\Re{({\mathcal F}^{(1l)})}^2 
+ \Im{({\mathcal F}^{(1l)})}^2} 
\label{deltaQCD}
\ee
as  of the NLO QCD corrections with respect to the LO contribution.

\subsection{Calculation of the master integrals}

%
\begin{figure}[!htp]
\bc
\[ \vcenter{
\hbox{
  \begin{picture}(0,0)(0,0)
\SetScale{0.9}
  \SetWidth{2}
%
\CArc(-15,0)(15,0,180)
\CArc(-15,0)(15,180,360)
\CArc(15,0)(15,0,180)
\CArc(15,0)(15,180,360)
\CCirc(-15,15){4}{0.9}{0.9}
\CCirc(15,15){4}{0.9}{0.9}
%
%
%
%
\Text(0,-37)[c]{($M_1$)}
\end{picture}}
}
\hspace{3.3cm}
\vcenter{
\hbox{
  \begin{picture}(0,0)(0,0)
\SetScale{0.9}
  \SetWidth{.4}
\DashLine(-35,0)(-20,0){2}
\DashLine(20,0)(35,0){2}
  \SetWidth{2}
\CArc(0,0)(20,0,180)
\CArc(0,0)(20,180,360)
\CArc(60,0)(40,150,180)
\CArc(0,34.6)(40,300,330)
\CArc(30.20,17.60)(5.28,-34,153)
\CCirc(0,20){4}{0.9}{0.9}
\CCirc(32,21){3}{0.9}{0.9}
  \SetWidth{.4}
\Text(0,-37)[c]{($M_2$-$M_3$)}
\Text(-40,10)[c]{$p^2$}
\end{picture}}
}
\hspace{3.3cm}
\vcenter{
\hbox{
  \begin{picture}(0,0)(0,0)
\SetScale{0.9}
  \SetWidth{.4}
\DashLine(-35,0)(-20,0){2}
\DashLine(20,0)(35,0){2}
  \SetWidth{2}
\CArc(0,0)(20,0,180)
\CArc(0,0)(20,180,360)
%
%
\CCirc(0,20){4}{0.9}{0.9}
\CCirc(0,-20){4}{0.9}{0.9}
%
  \SetWidth{.4}
\Line(-20,0)(20,0)
\Text(0,-37)[c]{($M_4$-$M_5$)}
\Text(-40,10)[c]{$p^2$}
\end{picture}}
}
\hspace{3.3cm}
\vcenter{
\hbox{
  \begin{picture}(0,0)(0,0)
\SetScale{0.9}
  \SetWidth{.4}
\DashLine(-35,0)(-20,0){2}
\DashLine(20,0)(35,0){2}
  \SetWidth{2}
\CArc(0,0)(20,0,180)
\CArc(0,0)(20,180,360)
\CCirc(0,20){4}{0.9}{0.9}
\CCirc(0,0){4}{0.9}{0.9}
  \SetWidth{.4}
\Line(-20,0)(20,0)
\Text(0,-37)[c]{($M_6$-$M_7$)}
\Text(-40,10)[c]{$p^2$}
\end{picture}}
}
\]
\vspace*{1.6cm}
\[ 
\vcenter{
\hbox{
  \begin{picture}(0,0)(0,0)
\SetScale{0.9}
  \SetWidth{.4}
\DashLine(-45,0)(-30,0){2}
\Line(15,30)(30,30)
\Line(15,-30)(30,-30)
  \SetWidth{2}
\Line(-30,0)(15,30)
\Line(15,30)(15,-30)
\Line(15,-30)(-30,0)
\CArc(-3.2,29)(40,185,227)
\CArc(-57.2,29)(40,313,355)
\CArc(-30,28)(13,-10,190)
\CCirc(-30,41){4}{0.9}{0.9}
  \SetWidth{.4}
\Text(0,-45)[c]{($M_8$)}
\Text(-40,-10)[c]{$m_H^2$}
\Text(30,35)[c]{$m_Z^2$}
\end{picture}}
}
\hspace{3.0cm}
\vcenter{
\hbox{
  \begin{picture}(0,0)(0,0)
\SetScale{0.9}
  \SetWidth{.4}
\DashLine(-45,0)(-30,0){2}
\DashLine(30,0)(45,0){2}
  \SetWidth{2}
\CArc(-15,0)(15,0,180)
\CArc(-15,0)(15,180,360)
\CArc(15,0)(15,0,180)
\CArc(15,0)(15,180,360)
\CCirc(-15,15){4}{0.9}{0.9}
\CCirc(15,15){4}{0.9}{0.9}
  \SetWidth{.4}
\Text(0,-45)[c]{($M_9$-$M_{10}$)}
\Text(-40,10)[c]{$p^2$}
\end{picture}}
}
\hspace{3.3cm}
\vcenter{
\hbox{
  \begin{picture}(0,0)(0,0)
\SetScale{0.9}
  \SetWidth{.4}
\DashLine(-45,0)(-30,0){2}
\Line(30,0)(45,0)
\Line(0,0)(0,-30)
  \SetWidth{2}
\CArc(-15,0)(15,0,180)
\CArc(-15,0)(15,180,360)
\CArc(15,0)(15,0,180)
\CArc(15,0)(15,180,360)
\CCirc(-15,15){4}{0.9}{0.9}
\CCirc(15,15){4}{0.9}{0.9}
  \SetWidth{.4}
\Text(0,-45)[c]{($M_{11}$)}
\Text(-40,10)[c]{$m_H^2$}
\Text(40,10)[c]{$m_Z^2$}
\end{picture}}
}
\hspace{4.0cm}
\vcenter{
\hbox{
  \begin{picture}(0,0)(0,0)
\SetScale{0.9}
  \SetWidth{.4}
\DashLine(-50,0)(-35,0){2}
\Line(10,30)(25,30)
\Line(10,-30)(25,-30)
\CArc(-25,34)(35,255,350)
  \SetWidth{2}
\Line(-35,0)(10,30)
\Line(10,30)(10,-30)
\Line(10,-30)(-35,0)
\CCirc(-13,15){4}{0.9}{0.9}
  \SetWidth{.4}
\Text(0,-45)[c]{($M_{12}$)}
\Text(-40,-10)[c]{$m_H^2$}
\Text(26,35)[c]{$m_Z^2$}
\end{picture}}
}
\]
\vspace*{2.1cm}
\[ 
\vcenter{
\hbox{
  \begin{picture}(0,0)(0,0)
\SetScale{0.9}
  \SetWidth{.4}
\DashLine(-50,0)(-35,0){2}
\Line(10,30)(25,30)
\Line(10,-30)(25,-30)
\CArc(30,0)(35,125,235)
  \SetWidth{2}
\Line(-35,0)(10,30)
\Line(10,30)(10,-30)
\Line(10,-30)(-35,0)
\CCirc(10,0){4}{0.9}{0.9}
  \SetWidth{.4}
\Text(0,-45)[c]{($M_{13}$)}
\Text(-45,-10)[c]{$m_H^2$}
\Text(25,35)[c]{$m_Z^2$}
\end{picture}}
}
\hspace{3.3cm}
\vcenter{
\hbox{
  \begin{picture}(0,0)(0,0)
\SetScale{0.9}
  \SetWidth{.4}
\DashLine(-50,0)(-35,0){2}
\Line(10,30)(25,30)
\Line(10,-30)(25,-30)
\CArc(30,0)(35,125,235)
  \SetWidth{2}
\Line(-35,0)(10,30)
\Line(10,30)(10,-30)
\Line(10,-30)(-35,0)
\CCirc(10,7){4}{0.9}{0.9}
\CCirc(10,-7){4}{0.9}{0.9}
  \SetWidth{.4}
\Text(0,-45)[c]{($M_{14}$)}
\Text(-45,-10)[c]{$m_H^2$}
\Text(30,35)[c]{$m_Z^2$}
\end{picture}}
}
\hspace{3.3cm}
\vcenter{
\hbox{
  \begin{picture}(0,0)(0,0)
\SetScale{0.9}
  \SetWidth{.4}
\DashLine(-50,0)(-35,0){2}
\Line(10,30)(25,30)
\Line(10,-30)(25,-30)
\CArc(30,0)(35,125,235)
  \SetWidth{2}
\Line(-35,0)(10,30)
\Line(10,30)(10,-30)
\Line(10,-30)(-35,0)
\CCirc(10,0){4}{0.9}{0.9}
\CCirc(-13,15){4}{0.9}{0.9}
  \SetWidth{.4}
\Text(0,-45)[c]{($M_{15}$)}
\Text(-45,-10)[c]{$m_H^2$}
\Text(25,35)[c]{$m_Z^2$}
\Text(-25,30)[c]{$[k_{12}^2\!+\!m^2]$}
\end{picture}}
}
\hspace{3.3cm}
\vcenter{
\hbox{
  \begin{picture}(0,0)(0,0)
\SetScale{0.9}
  \SetWidth{.4}
\DashLine(-50,0)(-35,0){2}
\Line(10,30)(25,30)
\Line(10,-30)(25,-30)
\CArc(-25,-34)(35,10,105)
  \SetWidth{2}
\Line(-35,0)(10,30)
\Line(10,30)(10,-30)
\Line(10,-30)(-35,0)
\CCirc(-13,-15){4}{0.9}{0.9}
  \SetWidth{.4}
\Text(0,-45)[c]{($M_{16}$)}
\Text(-45,-10)[c]{$m_H^2$}
\Text(25,35)[c]{$m_Z^2$}
\end{picture}}
}
\]
\vspace*{2.1cm}
\[ 
\hspace{-0.3cm}
\vcenter{
\hbox{
  \begin{picture}(0,0)(0,0)
\SetScale{0.9}
  \SetWidth{.4}
\DashLine(-50,0)(-35,0){2}
\Line(10,30)(25,30)
\Line(10,-30)(25,-30)
\CArc(-25,-34)(35,10,105)
  \SetWidth{2}
\Line(-35,0)(10,30)
\Line(10,30)(10,-30)
\Line(10,-30)(-35,0)
\CCirc(-18,-12){4}{0.9}{0.9}
\CCirc(-8,-18){4}{0.9}{0.9}
  \SetWidth{.4}
\Text(0,-45)[c]{($M_{17}$)}
\Text(-45,-10)[c]{$m_H^2$}
\Text(25,35)[c]{$m_Z^2$}
\end{picture}}
}
\hspace{3.3cm}
\vcenter{
\hbox{
  \begin{picture}(0,0)(0,0)
\SetScale{0.9}
  \SetWidth{.4}
\DashLine(-50,0)(-35,0){2}
\Line(10,30)(25,30)
\Line(10,-30)(25,-30)
\CArc(-25,-34)(35,10,105)
  \SetWidth{2}
\Line(-35,0)(10,30)
\Line(10,30)(10,-30)
\Line(10,-30)(-35,0)
\CCirc(-13,-15){4}{0.9}{0.9}
\CCirc(10,0){4}{0.9}{0.9}
  \SetWidth{.4}
\Text(0,-45)[c]{($M_{18}$)}
\Text(-45,-10)[c]{$m_H^2$}
\Text(30,35)[c]{$m_Z^2$}
\Text(-22,30)[c]{$(p_2\!+\! k_{12})^2$}
\end{picture}}
}
\hspace{3.0cm}
\vcenter{
\hbox{
  \begin{picture}(0,0)(0,0)
\SetScale{0.9}
  \SetWidth{.4}
\DashLine(-45,0)(-30,0){2}
\Line(25,30)(40,30)
\Line(25,-30)(40,-30)
  \SetWidth{2}
\CArc(-15,0)(15,0,180)
\CArc(-15,0)(15,180,360)
\Line(0,0)(25,30)
\Line(25,30)(25,-30)
\Line(25,-30)(0,0)
\CCirc(-15,15){4}{0.9}{0.9}
  \SetWidth{.4}
\Text(0,-45)[c]{($M_{19}$)}
\Text(-40,-10)[c]{$m_H^2$}
\Text(30,35)[c]{$m_Z^2$}
\end{picture}}
}
\hspace{3.3cm}
\vcenter{
\hbox{
  \begin{picture}(0,0)(0,0)
\SetScale{0.9}
  \SetWidth{.4}
\Line(-45,0)(-30,0)
\Line(25,30)(40,30)
\Line(25,-30)(40,-30)
\Line(0,0)(0,-30)
  \SetWidth{2}
\CArc(-15,0)(15,0,180)
\CArc(-15,0)(15,180,360)
\Line(0,0)(25,30)
\Line(25,30)(25,-30)
\Line(25,-30)(0,0)
\CCirc(-15,15){4}{0.9}{0.9}
  \SetWidth{.4}
\Text(0,-45)[c]{($M_{20}$)}
\Text(-40,-10)[c]{$m_Z^2$}
\Text(35,35)[c]{$m_Z^2$}
\end{picture}}
}
\]
\vspace*{2.1cm}
\[ 
\vcenter{
\hbox{
  \begin{picture}(0,0)(0,0)
\SetScale{0.9}
  \SetWidth{.4}
\DashLine(-50,0)(-35,0){2}
\Line(10,30)(25,30)
\Line(10,-30)(25,-30)
\Line(-35,0)(10,0)
%
  \SetWidth{2}
\Line(-35,0)(10,30)
\Line(10,30)(10,-30)
\Line(10,-30)(-35,0)
%
%
  \SetWidth{.4}
\Text(0,-45)[c]{($M_{21}$)}
\Text(-45,-10)[c]{$m_H^2$}
\Text(25,35)[c]{$m_Z^2$}
\end{picture}}
}
\hspace{3.3cm}
\vcenter{
\hbox{
  \begin{picture}(0,0)(0,0)
\SetScale{0.9}
  \SetWidth{.4}
\DashLine(-50,0)(-35,0){2}
\Line(10,30)(25,30)
\Line(10,-30)(25,-30)
\Line(-35,0)(10,0)
%
  \SetWidth{2}
\Line(-35,0)(10,30)
\Line(10,30)(10,-30)
\Line(10,-30)(-35,0)
\CCirc(-13,15){4}{0.9}{0.9}
  \SetWidth{.4}
\Text(0,-45)[c]{($M_{22}$)}
\Text(-45,-10)[c]{$m_H^2$}
\Text(25,35)[c]{$m_Z^2$}
\end{picture}}
}
\hspace{3.3cm}
\vcenter{
\hbox{
  \begin{picture}(0,0)(0,0)
\SetScale{0.9}
  \SetWidth{.4}
\DashLine(-50,0)(-35,0){2}
\Line(10,30)(25,30)
\Line(10,-30)(25,-30)
\Line(-13,-15)(10,30)
%
  \SetWidth{2}
\Line(-35,0)(10,30)
\Line(10,30)(10,-30)
\Line(10,-30)(-35,0)
%
%
  \SetWidth{.4}
\Text(0,-45)[c]{($M_{23}$)}
\Text(-45,-10)[c]{$m_H^2$}
\Text(25,35)[c]{$m_Z^2$}
\end{picture}}
}
\hspace{3.3cm}
\vcenter{
\hbox{
  \begin{picture}(0,0)(0,0)
\SetScale{0.9}
  \SetWidth{.4}
\DashLine(-50,0)(-35,0){2}
\Line(10,30)(25,30)
\Line(10,-30)(25,-30)
\Line(-13,-15)(10,30)
%
  \SetWidth{2}
\Line(-35,0)(10,30)
\Line(10,30)(10,-30)
\Line(10,-30)(-35,0)
\CCirc(-13,15){4}{0.9}{0.9}
  \SetWidth{.4}
\Text(0,-45)[c]{($M_{24}$)}
\Text(-45,-10)[c]{$m_H^2$}
\Text(25,35)[c]{$m_Z^2$}
\end{picture}}
}
\]
\vspace*{2.1cm}
\[ 
\vcenter{
\hbox{
  \begin{picture}(0,0)(0,0)
\SetScale{0.9}
  \SetWidth{.4}
\DashLine(-50,0)(-35,0){2}
\Line(10,30)(25,30)
\Line(10,-30)(25,-30)
\Line(-13,15)(10,-30)
%
  \SetWidth{2}
\Line(-35,0)(10,30)
\Line(10,30)(10,-30)
\Line(10,-30)(-35,0)
%
%
  \SetWidth{.4}
\Text(0,-45)[c]{($M_{25}$)}
\Text(-45,-10)[c]{$m_H^2$}
\Text(25,35)[c]{$m_Z^2$}
\end{picture}}
}
\hspace{3.3cm}
\vcenter{
\hbox{
  \begin{picture}(0,0)(0,0)
\SetScale{0.9}
  \SetWidth{.4}
\DashLine(-50,0)(-35,0){2}
\Line(10,30)(25,30)
\Line(10,-30)(25,-30)
\Line(-13,15)(10,-30)
%
  \SetWidth{2}
\Line(-35,0)(10,30)
\Line(10,30)(10,-30)
\Line(10,-30)(-35,0)
\CCirc(10,0){4}{0.9}{0.9}
  \SetWidth{.4}
\Text(0,-45)[c]{($M_{26}$)}
\Text(-45,-10)[c]{$m_H^2$}
\Text(25,35)[c]{$m_Z^2$}
\end{picture}}
}
\hspace{3.3cm}
\vcenter{
\hbox{
  \begin{picture}(0,0)(0,0)
\SetScale{0.9}
  \SetWidth{.4}
\DashLine(-50,0)(-35,0){2}
\Line(10,30)(25,30)
\Line(10,-30)(25,-30)
\Line(-13,15)(10,-30)
%
  \SetWidth{2}
\Line(-35,0)(10,30)
\Line(10,30)(10,-30)
\Line(10,-30)(-35,0)
\CCirc(-13,-15){4}{0.9}{0.9}
  \SetWidth{.4}
\Text(0,-45)[c]{($M_{27}$)}
\Text(-45,-10)[c]{$m_H^2$}
\Text(25,35)[c]{$m_Z^2$}
\end{picture}}
}
\hspace{3.3cm}
\vcenter{
\hbox{
  \begin{picture}(0,0)(0,0)
\SetScale{0.9}
  \SetWidth{.4}
\DashLine(-50,0)(-35,0){2}
\Line(10,30)(25,30)
\Line(10,-30)(25,-30)
\Line(-13,15)(10,-30)
%
  \SetWidth{2}
\Line(-35,0)(10,30)
\Line(10,30)(10,-30)
\Line(10,-30)(-35,0)
\CCirc(10,0){4}{0.9}{0.9}
\CCirc(-13,-15){4}{0.9}{0.9}
  \SetWidth{.4}
\Text(0,-45)[c]{($M_{28}$)}
\Text(-45,-10)[c]{$m_H^2$}
\Text(25,35)[c]{$m_Z^2$}
\end{picture}}
}
\]
\vspace*{13mm}
\caption{\label{misqcd} Set of master integrals for the NLO QCD corrections to the decay $H\to Z \gamma$. 
The master integrals $M_2$, $M_4$, $M_6$, $M_9$ are functions of $p^2=-m_H^2$, while the master integrals 
$M_3$, $M_5$, $M_7$, $M_{10}$ are functions of $p^2=-m_Z^2$. $M_{15}$ and $M_{18}$ have the numerator 
explicitly written. A dot on a propagator indicates that the propagator is raised to power 2. Two dots 
means that the propagator is raised to power 3. See Appendix B.}
\ec
\efig
%

${\mathcal F}^{(2l)}$ is expressed in terms of a large number of scalar integrals that are individually
ultraviolet divergent. To deal with these divergences, we perform the integrals in dimensional 
regularization~\cite{'tHooft:1972fi,Bollini:1972bi,Bollini:1972ui,Ashmore:1972uj,Cicuta:1972jf,Gastmans:1973uv}. 
The dimensionally regularized scalar integrals are not all independent. The reduction to a set
of independent integrals, called master integrals, is carried out using two different computer programs, 
FIRE~\cite{Smirnov:2008iw,Smirnov:2013dia,Smirnov:2014hma} and 
Reduze~\cite{Studerus:2009ye,vonManteuffel:2012np}, that implement in an automatic way
the solution of the linear system of integration-by-parts identities~\cite{Tkachov:1981wb,Chetyrkin:1981qh}
which relate the Feynman integrals.

\FloatBarrier

We calculate the master integrals using the Differential Equations Method 
\cite{Kotikov:1990kg,Kotikov:1991hm,Kotikov:1991pm,Remiddi:1997ny,Caffo:1998du,Gehrmann:1999as,Argeri:2007up}, 
following recent developments 
\cite{Henn:2013pwa,Henn:2013woa,Henn:2013nsa,Henn:2014lfa,Caola:2014lpa} (see also 
\cite{Caron-Huot:2014lda,Gehrmann:2014bfa,Argeri:2014qva,Hoschele:2014qsa,Dulat:2014mda,Bell:2014zya} and \cite{Henn:2014qga,Lee:2014ioa} for further studies of the method).


The system of linear differential equations is cast, via a suitable basis choice for the master integrals, 
in the canonical form \cite{Henn:2013pwa},
\be
d{\bf f} = \epsilon \, d\tilde{A} \, {\bf f} \, .
\label{sysdeq}
\ee
The $\epsilon$ dependence is completely factorized from the 
matrix $d \tilde{A}$. The latter depends only on the dimensionless variables $x_q$ and $y_q$, defined in Eq.~(\ref{xandydef}). 

${\bf f}$ is a vector of 28 functions, $f_i(x_q,y_q,\epsilon)$, defined in terms of the integrals drawn in Fig.~\ref{misqcd},
\bea
f_1 & = & 16 \, \epsilon^2 \, M_1 \, , \label{af1}\\
f_2 & = & 16 \, \epsilon^2 \, m_H^2 \, M_{14} \, , \label{af2}\\
f_3 & = & 8 \, \epsilon^2 \, \sqrt{-m_H^2( 4m_q^2 - m_H^2)} \left( M_4+2 M_6 \right) 
\, ,\label{af3}\\
f_4 & = & 16 \, \epsilon^2 \, m_Z^2 \, M_5 \, , \label{af4}\\
f_5 & = & 8 \, \epsilon^2 \, \sqrt{-m_Z^2( 4m_q^2 - m_Z^2)} \left( M_5+2 M_7 \right) 
\, ,\label{af5}\\
f_6 & = & 16 \, \epsilon^2 \, \sqrt{-m_Z^2( 4m_q^2 - m_Z^2)} M_3 \, , \label{af6}\\
f_7 & = & 16 \, \epsilon^2 \, \sqrt{-m_H^2( 4m_q^2 - m_H^2)} M_2 \, , \label{af7}\\
f_8 & = & 16 \, \epsilon^3 \, (m_H^2- m_Z^2) \, M_{16} \, , \label{af8}\\
f_9 & = & 16 \, \epsilon^2 \, m_q^2 \, (m_H^2- m_Z^2) \, M_{17} \, , \label{af9}\\
f_{10} & = & 4 \epsilon^2 \sqrt{1- \frac{4m_q^2}{m_Z^2}} \left[ m_H^2 
\bigl( 4 m_q^2 M_{17} - M_4 \bigr) + 4 m_Z^2 \bigl( m_q^2 M_{17} - M_{18} \bigr) \right.
\nn\\
& & \left.  - 2 \, \epsilon \, \left( m_H^2+m_Z^2 \right) M_{16} \right] \, , \label{af10}\\
f_{11} & = & 16 \, \epsilon^3 \, \left( m_H^2-m_Z^2 \right) M_{13} \, , \label{af11}\\
f_{12} & = & 16 \, \epsilon^2 \, m_q^2 \, \left( m_H^2-m_Z^2 \right) M_{14} \, , \label{af12}\\
f_{13} & = & 4 \, \epsilon^2 \, \frac{1}{2 m_q^2 - m_H^2} \, \sqrt{1-\frac{4 m_q^2}{m_H^2}}
\biggl\{ m_Z^2 \frac{4 m_q^2 m_H^2 - m_H^2 m_Z^2 + m_Z^4}{m_H^2-m_Z^2} \, M_5 \nn\\
& & -  4 \, m_H^2 \frac{m_q^2 m_H^2 - m_H^2 m_Z^2 + m_Z^4}{m_H^2-m_Z^2} \, M_{15} 
- 4 m_q^2 (m_H^2-m_Z^2) m_Z^2 \, M_{14}  \nn\\
& & + 2 \epsilon 
      (2 m_q^2 m_H^2 + m_H^2 m_Z^2 - m_Z^4) M_{13} \biggr\} \, , \label{af13}\\
f_{14} & = & 16 \, \epsilon^3 \, (m_H^2-m_Z^2) \, M_{12} \, , \label{af14}\\
f_{15} & = & 16 \, \epsilon^2 \, m_Z^2 (4 m_q^2-m_Z^2) \, M_{10} \, , \label{af15}\\
f_{16} & = & 16 \, \epsilon^2 \, \sqrt{-m_H^2( 4m_q^2 - m_H^2)} \,
\sqrt{-m_Z^2( 4m_q^2 - m_Z^2)} \, M_{11} \, , \label{af16}\\
f_{17} & = & 16 \, \epsilon^3 \, (m_H^2-m_Z^2) \, \sqrt{-m_Z^2( 4m_q^2 - m_Z^2)} \, M_{20} 
\, , \label{af17}\\
f_{18} & = & 16 \, \epsilon^4 \, (m_H^2-m_Z^2) \, M_{21} \, , \label{af18}\\
f_{19} & = & 16 \, \epsilon^3 \, (m_H^2-m_Z^2) \, \sqrt{-m_Z^2( 4m_q^2 - m_Z^2)} \, M_{22} 
\, , \label{af19}\\
f_{20} & = & 16 \, \epsilon^4 \, (m_H^2-m_Z^2) \, M_{25} \, , \label{af20}\\
f_{21} & = & - 16 \, \epsilon^3 \, m_Z^2 \, (m_H^2-m_Z^2) \, \sqrt{-m_Z^2( 4m_q^2 - m_Z^2)} 
\, M_{26} \, , \label{af21}\\
f_{22} & = & 16 \, \epsilon^3 \,(m_H^2-m_Z^2) \, \sqrt{-m_H^2( 4m_q^2 - m_H^2)} \, M_{27} 
\, , \label{af22}\\
f_{23} & = & 16 \, \epsilon^2 \, \left[ 2 \bigl( m_H^2 m_Z^2 - 2 m_q^2 m_H^2 - 2 m_q^2 m_Z^2 \bigr)
M_{11} + m_q^2 \, (m_H^2-m_Z^2)^2 M_{28} \right. \nn\\
& & + \left. \epsilon (m_H^2-m_Z^2) ( m_Z^2 \, M_{26} - m_H^2 \, M_{27} ) \right] 
\, , \label{af23}\\
f_{24} & = & 16 \, \epsilon^4 \, (m_H^2-m_Z^2) \, M_{23} \, , \label{af24}\\
f_{25} & = & 16 \, \epsilon^3 \, (m_H^2-m_Z^2) \, \sqrt{-m_H^2( 4m_q^2 - m_H^2)} \, M_{24} 
\, , \label{af25}\\
f_{26} & = & 16 \, \epsilon^3 \,(m_H^2-m_Z^2) \, M_8 \, , \label{af26}\\
f_{27} & = & 16 \, \epsilon^2 \, m_H^2 (4 m_q^2-m_H^2) \, M_9 \, , \label{af27}\\
f_{28} & = & 16 \, \epsilon^3 \,(m_H^2-m_Z^2) \,  \sqrt{-m_H^2( 4m_q^2 - m_H^2)} \, M_{19} 
\, .\label{af28}
\label{ffunc}
\eea
The functions $f_i(x_q,y_q,\epsilon)$ are chosen in such a way to be pure and of uniform weight,
in the sense of \cite{Henn:2013pwa}.

%
The explicit definition of the integrals $M_1, \cdots, M_{28}$ is given in Appendix~\ref{sec:appb}\footnote{The 
calculation of some of the master integrals in Fig.~\ref{misqcd} was already performed in Ref.~\cite{Christian}.
}.
To set the normalization, we define $M_1$ as the following integral,
\be
M_1 = \int {\mathcal D}^Dk_1 \, {\mathcal D}^Dk_2 \, \frac{1}{(k_1^2+m_q^2)^2} \, \frac{1}{(k_2^2+m_q^2)^2} = \frac{1}{16 \epsilon^2} \, ,
\ee
\noindent where the integration measure is such that,
\be
d^Dk_1 = 4 \pi^{2-\epsilon} \Gamma(1+\epsilon) \left( \frac{\mu^2}{m_q^2} \right)^{\epsilon} \, {\mathcal D}^Dk_1 \, .
\ee
With this normalization, 
\be
f_1 = 1 \, .
\ee

The matrix $d\tilde{A}$ is a differential depending on $x_q$ and $y_q$,
\bea\label{defdatilde}
d\tilde{A} & = & \mathbb{S}_1 \, d\log{x_q} + \mathbb{S}_2 \, d\log{(1-x_q)} + \mathbb{S}_3 \, d\log{(1+x_q)} 
+ \mathbb{S}_{4} \, d\log{y_q} + \mathbb{S}_{5} \, d\log{(1-y_q)} \nn\\
& & + \mathbb{S}_{6} \, d\log{(1+y_q)}
+ \mathbb{S}_{7} \, d\log{(x_q-y_q)} + \mathbb{S}_8 \, d\log{(1-x_q y_q)} \nn\\
& & + \mathbb{S}_9 \, d\log{(1 - x_q - x_q y_q + x_q^2)} + \mathbb{S}_{10} \, d\log{(1 - y_q - x_q y_q + y_q^2)} \nn\\
& & + \mathbb{S}_{11} \, d\log{(x_q - y_q + x_q y_q - x_q^2 y_q)} + \mathbb{S}_{12} \, d\log{(x_q - y_q - x_q y_q + x_q y_q^2)}  \, .
\label{differential}
\eea
The sparse matrices $\mathbb{S}_i$ are purely numerical and they are collected in Appendix~\ref{sec:appc}. 

To fully describe the $\bf{f}$, we need to complement the differential equations with boundary conditions.
We can, in principle, choose any kinematic point $(x_q, y_q)$. However, it is well known that often, 
(spurious) singularities of the differential equations allow one to fix the boundary condition without
calculation, using the physical insight that certain limits must be non-singular. This is also the case here.
We notice that the
integrals defining the primary basis (Fig.~\ref{misqcd}) are regular in the point $s=m_Z^2=0$.
Moreover the rational prefactors of the combinations given in Eqs.~(\ref{af1}--\ref{af28}) vanish in
the same limit. Therefore, all the integrals $f_i(x_q,y_q)$, $i=2,...,28$, vanish in this limit. The
only exception is represented by $f_1$, which is identically equal to $1$. We can write the full set
of boundary conditions in the compact form,
\be
f_i(1,1)=\delta_{1,i} \, .
\label{boundary}
\ee
The system (\ref{sysdeq}), together with the boundary condition (\ref{boundary}), makes it obvious that the solution, 
i.e. the functions $f_i(x_q,y_q)$, have a number of desirable properties.
At any order in the expansion in $\eps$, they are given by iterated integrals \cite{Chen} over the one-form $d \tilde{A}$.
Defining the weight of an iterated integral as the number of integrations, we see that at order $\eps^k$, $\bf{f}$
is given by a $\mathbb{Q}$-linear combination of iterated integrals of weight $k$. 
Such functions are referred to as pure functions of uniform weight.

The basis choice for ${\bf f}$ leading to this form was achieved using the ideas outlined in Ref.~\cite{Henn:2013pwa}.
Specifically, generalized unitarity cuts in four dimensions were used to project onto subsets of the differential equations.
This typically leads to an answer close to the canonical form, where unwanted terms e.g. due to integrals
vanishing on the cuts can be easily removed algorithmically, see e.g. Refs.~\cite{Caron-Huot:2014lda,Gehrmann:2014bfa}.
A canonical basis for very similar integrals was found in Ref.~\cite{Caron-Huot:2014lda}, and using the results from that
paper the present basis choice was rather straightforward.

The one-form $d\tilde{A}$ characterizes the type of iterated integrals that are needed.
In particular, given the definition $d\tilde{A} = \sum_{i=1}^{12} \mathbb{S}_{i} d \log(\alpha_{i})$,
the individual $\alpha_i$ are called the {letters},  and iterated integrals correspond to words in
those letters. The set $\{ \alpha_{i} \}$ is called alphabet.

For the specific one-form $d\tilde{A}$, the system in Eq.~(\ref{sysdeq}) can be solved in terms of a subset of generalized (or Goncharov) polylogarithms (GPL) 
\cite{Goncharov:1998kja,Broadhurst:1998rz,Remiddi:1999ew}, 
defined as iterated integrations over a set of basic polynomials $\{ (x-a_1), ... ,(x-a_n)\}$,
\be
G(0_n,x) = \frac{1}{n!} \, \log^n{x} \, , \quad G(a_1,a_2,...,a_n,x) = \int_0^x \frac{dt}{t-a_1}
G(a_2,...,a_n,t) \,,
\ee
for a certain set of $a_{i}$.

We choose to represent the GPLs of two variables, $x_q$ and $y_q$, as functions of argument $x_q$ and weights depending on $y_q$. 
The weights for the GPLs functions of $x_q$ are 
determined by the following set,
\be
\Bigl\{ x_q, 1-x_q, 1+x_q, x_y-y_q, x_y-1/y_q, x_q-R_i \Bigr\} \, ,
\ee
where the roots $R_i$ are defined by the following expressions,
\bea
R_{01} &=& \frac{y_q}{y_q^2-y_q+1} \, , \\
R_{02} &=& \frac{y_q^2-y_q+1}{y_q} \, , \\
R_{11} &=& \frac{1}{2} + \frac{1}{2y_q} - \frac{1}{2y_q} \sqrt{1 + 2 y_q - 3 y_q^2}  \, , \\
R_{12} &=& \frac{1}{2} + \frac{1}{2y_q} + \frac{1}{2y_q} \sqrt{1 + 2 y_q - 3 y_q^2}  \, , \\
R_{21} &=& \frac{1}{2} + \frac{1}{2} y_q - \frac{1}{2} \sqrt{ - 3 + 2 y_q + y_q^2} \, , \\
R_{22} &=& \frac{1}{2} + \frac{1}{2} y_q + \frac{1}{2} \sqrt{ - 3 + 2 y_q + y_q^2} \, .
\eea

We also have GPLs of the variable $y_q$. The weights of these GPLs are determined by the following set,
\be
\Bigl\{ y_q, 1-y_q, 1+y_q, y_q - c, y_q-\overline{c}, y_q-i, y_q+i \Bigr\} \, ,
\ee
where
\be
c = \frac{1-i \sqrt{3}}{2} \, , \quad \overline{c} = \frac{1+i \sqrt{3}}{2} \,,
\ee
are sixth roots of the unity.

The solution of the differential equations can be found expanding each of the canonical master integrals in $\epsilon$,
\begin{eqnarray}
f_i &= f_i^{(0)} + f_i^{(1)} \ep + f_i^{(2)} \ep^2 + \ldots
\end{eqnarray}
We can identify the terms on each side of Eq.~\eqref{sysdeq} which multiply the same power of $\epsilon$, i.e.
\begin{eqnarray}
d f_i^{(k)} = d \tilde{A}_{ij} f_j^{(k-1)}.
\label{sysdeqtrm}
\end{eqnarray}
This allows us to solve Eq.~\eqref{sysdeq} recursively order by order. As the values of $f_i^{(0)}$ are constants given entirely by the boundary conditions, we shall start by finding $f_i^{(1)}(x_q,y_q)$.
This may be done by integrating the right--hand side of Eq.~\eqref{sysdeqtrm} from the boundary point $(1,1)$ to a general point $(x_q,y_q)$, and we choose to use a stepwise path passing through $(x_q,1)$. Using the boundary conditions to fix the integration constant, gives the values for $f_i^{(1)}$. Once this is done for each of the canonical master integrals, the procedure may be repeated for $f_i^{(2)}$ and the other orders, yielding the full result for $f_i$ up to any desired order.

We checked all the expressions of the master integrals numerically against the computer program 
Fiesta \cite{Smirnov:2008py,Smirnov:2009pb,Smirnov:2013eza} finding complete agreement.

The analytic expressions of the functions $f_i$ are collected in an ancillary file of the arXiv
submission of the present paper.

A comment is due regarding analytic continuation and analytic boundary values in different regions.
In the following subsections, we explain how we obtained a formula valid in all kinematic regions.
For completeness, here we wish to comment that another possibility is to solve the differential
equations directly starting from a boundary value in the region of interest.
The boundary value we gave in Eq.~(\ref{boundary}) at $x_q = y_q =1$ is in a non-physical region,
and other regions can be reached by analytic continuation. 
In particular, when analytically continuing to negative values of $x_q$ or $y_q$, the $d\log x_q$ and $d\log y_q$
terms in $d\tilde{A}$ are responsible for imaginary parts.
As an example, let us analyze the boundary point $(x_q,y_q)=(-1,-1)$. Unlike  $(x_q,y_q)=(1,1)$, this is a singular point, so that
it has to be approached with care. Since $x_q = y_q$ is nonsingular, we can set $x_q = y_q = z$, for $z>0$ in which case the alphabet 
becomes $\{d\log z, d\log(1-z), d\log(1+z)\}$. When analytically continuing to negative values of $z$, we have to take care of the imaginary
parts due to $d\log z$, keeping in mind that $z$ has a small imaginary part $i0$. In this way, one can give an analytic boundary
value as $z\to -1$, in terms of powers of $\log(1+z)$ and a set of constants. It follows from the reduced alphabet $\{d\log z, d\log(1-z), d\log(1+z)\}$
that the only constants required are Euler sums. Up to weight four, one has $i \pi, \log 2, \zeta_3, {\rm Li}_{4}(1/2)$, and products thereof.

\subsection{An alternative choice of the functional basis}
\label{optfun}

The analytic expressions of the master integrals in terms of GPLs provide all the necessary features of
an analytic formula: possibility to expand the formulae in particular regions of the phase space,
flexibility with respect to the input physical parameters, available routines for their numerical 
evaluation \cite{Vollinga:2004sn}.
However, for the sake of a faster and more stable numerical evaluation, we describe in this section how
to represent the result in terms of a different functional basis.

Up to weight four, Goncharov polylogarithms can be rewritten in terms of the following functions (see \cite{Goncharov:2010jf} and references therein),
\begin{equation}
\label{basis}
\log(x_1) = G(0,x_1), \quad
\text{Li}_n(x_1)= - G(0_{n-1},1,x_1), \quad \text{Li}_{2,2}(x_1 ,x_2) = 
G\left(0,\frac{1}{x_2},0,\frac{1}{x_1 x_2},1\right) \, ,
\end{equation}
with $n=(2,3,4)$, and where $x_1$ and $x_2$ are rational functions of $x_q$ and $y_q$. 
$\text{Li}_n(x_1)$ has a branch cut for $x_1>1$, while $\text{Li}_{2,2}(x_1,x_2)$ has a branch cut whenever $x_1 >1$ or $x_1 x_2>1$. With this choice, the functions given in Eq.~(\ref{basis}) can be directly evaluated using the 
numerical routines of \cite{Vollinga:2004sn}. 

In order to find an expression in terms of these functions, the concept of the symbol  \cite{Goncharov:1998kja,Brown1,Goncharov:2010jf} of an iterated integral
is very useful. The symbol corresponds to the integration kernels defining the iterated integrals. 
It is completely manifest in our differential equations approach. 
It is possible to use symbol-based ideas to rewrite the master integrals in terms of a minimal function basis, see e.g. \cite{Duhr:2011zq}. Moreover, we can recover the information about the terms in the kernel of the symbol using the coproduct map \cite{Duhr:2012fh}. 

However, it is also possible to proceed in a more direct way, using the knowledge of the differential equations, 
proceeding in the following algorithmic steps.
First, one generates a list of function arguments as monomials in the letters appearing in our function alphabet (i.e. the arguments of the logarithms appearing in the one-form $d\tilde{A}$.) For the classical polylogarithms ${\rm Li}_{n}(x_1)$, one requires that $1-x_1$ factorizes over
the letters appearing in the alphabet. (A caveat is that in principle, ``spurious'' letters might be needed \cite{Duhr:2011zq}, but this was not the case here.) For ${\rm Li}_{2,2}(x_1,x_2)$, the condition is that $1-x_1,1-x_2,1-x_1 x_2$ factorize over the alphabet. Second, for each weight $k$, one chooses a maximal set of linearly independent functions for the alphabet (the linear independence can be verified using the symbol). By construction, the differential equation at oder $\eps^k$ can then be solved in terms of this set of functions.

The Li expressions for the master integrals are provided in an ancillary file and are defined for the relevant physical region $s>m_z^2$, which in terms of the new variables implies  $y<x$ or $\arg(y)< \arg(x)$, where $0 < \arg(x) \leq \pi$ is understood.

When solving the differential equations in terms of the above functions, there is a certain freedom in the choice
of the set of function arguments. The choice is usually guided by trying to make certain properties of the answer manifest,
such as simple branch cut properties. For instance, one could require that the functions are real valued in the physical region, 
so that the imaginary parts of the master integrals become explicit. 
We found that this was possible in the regions of real $x_q, y_q$. However, for $x_q$ or $y_q$ complex,
individual functions develop imaginary parts.
%
%
%

In order to obtain an expression valid over the entire physical domain, a linear combination with
rational coefficients of the Li functions is not sufficient. The reason  is that some of the functions
in Eq.~(\ref{basis}) have discontinuities in the physical region, and in order to ensure the analyticity of the master integrals it is necessary to introduce discontinuous functions (Heaviside theta functions) that
cancel branch cut discontinuities of the new basis. The same issue was discussed in Ref.~\cite{Duhr:2011zq}. In Appendix~\ref{sec:appd} we briefly introduce the issue, and we provide a method to find an expression for the master integrals in terms of Li functions valid over the entire physical region.

\section{Numerical Results}
\label{sec:numerics}

In order to assess the impact of the NLO QCD corrections, it is convenient to recall the relative contributions
of the top, bottom and electroweak loops to the leading order width.
For $m_H = 125.1$~GeV, $m_t = 173.34$~GeV, $m_b = 4.6$~GeV, $m_W = 80.398$~GeV, 
$m_Z = 91.1876$~GeV, $s_W^2 = 0.23149$, $\alpha = 1/128$, $G_F = 1.16637\cdot 10^{-5}$, 
the QCD width is 0.02~KeV at leading order. In Table~\ref{tab0}, we report the relative contributions of
the top--quark loop, the bottom--quark loop and the top--bottom interference to the leading order QCD width. 
We note that the bottom loop contributes only one per mille to the QCD width,
which is almost entirely given by the top loop, with a sizeable destructive top--bottom interference.
On the other hand, the full width is 6.67~KeV at leading order.
In Table~\ref{tab00}, we report the relative contributions of
the QCD ({\it i.e.} top and bottom) loops,
the $W$-boson loop, and the QCD--$W$-boson interference to the leading--order width. 
We see that the QCD loops contribute only three per mille to the width at leading order;
the $W$-boson loop accounts for the bulk of the width, up to a large destructive QCD--$W$-boson interference.

\begin{table}
\bc
\begin{tabular}{|c|c|c|c|}
\hline
$\Gamma^{(1l)}_{QCD}$ (KeV) & $\Gamma^{(1l)}_t/\Gamma^{(1l)}_{QCD}$ & $\Gamma^{(1l)}_b/\Gamma^{(1l)}_{QCD}$ & 
${\rm Interf}^{(1l)}_{t-b}/\Gamma^{(1l)}_{QCD}$ \cr
\hline
0.02    & 1.052    & $1 \cdot 10^{-3}$   & - 0.053 \cr
\hline
\end{tabular}
\ec
\caption{Values of the QCD leading--order width, and of the relative contributions of the top loop, the bottom loop and of the interference
between top and bottom loops, at $m_H = 125.1$~GeV.
\label{tab0}}
\end{table}

\begin{table}
\bc
\begin{tabular}{|c|c|c|c|}
\hline
$\Gamma^{(1l)}$ (KeV) & $\Gamma^{(1l)}_{QCD}/\Gamma^{(1l)}$ & $\Gamma^{(1l)}_W/\Gamma^{(1l)}$ & 
${\rm Interf}^{(1l)}_{QCD-W}/\Gamma^{(1l)}$ \cr
\hline
6.67    & $3 \cdot 10^{-3}$    &  1.112  & - 0.115 \cr
\hline
\end{tabular}
\ec
\caption{Values of the leading--order width, and of the relative contributions of the QCD loops, the $W$-boson loop and of the interference
between the QCD and the $W$-boson loops, at $m_H = 125.1$~GeV.
\label{tab00}}
\end{table}

The NLO QCD corrections to the decay width of a Higgs boson into a $Z$ boson and a photon
were computed numerically in Ref.~\cite{Spira:1991tj} and turn out to be quite mild. 
\begin{figure}
\hspace*{2.5mm}
\begingroup
  \makeatletter
  \providecommand\color[2][]{%
    \GenericError{(gnuplot) \space\space\space\@spaces}{%
      Package color not loaded in conjunction with
      terminal option `colourtext'%
    }{See the gnuplot documentation for explanation.%
    }{Either use 'blacktext' in gnuplot or load the package
      color.sty in LaTeX.}%
    \renewcommand\color[2][]{}%
  }%
  \providecommand\includegraphics[2][]{%
    \GenericError{(gnuplot) \space\space\space\@spaces}{%
      Package graphicx or graphics not loaded%
    }{See the gnuplot documentation for explanation.%
    }{The gnuplot epslatex terminal needs graphicx.sty or graphics.sty.}%
    \renewcommand\includegraphics[2][]{}%
  }%
  \providecommand\rotatebox[2]{#2}%
  \@ifundefined{ifGPcolor}{%
    \newif\ifGPcolor
    \GPcolortrue
  }{}%
  \@ifundefined{ifGPblacktext}{%
    \newif\ifGPblacktext
    \GPblacktexttrue
  }{}%
  \let\gplgaddtomacro\g@addto@macro
  \gdef\gplbacktext{}%
  \gdef\gplfronttext{}%
  \makeatother
  \ifGPblacktext
    \def\colorrgb#1{}%
    \def\colorgray#1{}%
  \else
    \ifGPcolor
      \def\colorrgb#1{\color[rgb]{#1}}%
      \def\colorgray#1{\color[gray]{#1}}%
      \expandafter\def\csname LTw\endcsname{\color{white}}%
      \expandafter\def\csname LTb\endcsname{\color{black}}%
      \expandafter\def\csname LTa\endcsname{\color{black}}%
      \expandafter\def\csname LT0\endcsname{\color[rgb]{1,0,0}}%
      \expandafter\def\csname LT1\endcsname{\color[rgb]{0,1,0}}%
      \expandafter\def\csname LT2\endcsname{\color[rgb]{0,0,1}}%
      \expandafter\def\csname LT3\endcsname{\color[rgb]{1,0,1}}%
      \expandafter\def\csname LT4\endcsname{\color[rgb]{0,1,1}}%
      \expandafter\def\csname LT5\endcsname{\color[rgb]{1,1,0}}%
      \expandafter\def\csname LT6\endcsname{\color[rgb]{0,0,0}}%
      \expandafter\def\csname LT7\endcsname{\color[rgb]{1,0.3,0}}%
      \expandafter\def\csname LT8\endcsname{\color[rgb]{0.5,0.5,0.5}}%
    \else
      \def\colorrgb#1{\color{black}}%
      \def\colorgray#1{\color[gray]{#1}}%
      \expandafter\def\csname LTw\endcsname{\color{white}}%
      \expandafter\def\csname LTb\endcsname{\color{black}}%
      \expandafter\def\csname LTa\endcsname{\color{black}}%
      \expandafter\def\csname LT0\endcsname{\color{black}}%
      \expandafter\def\csname LT1\endcsname{\color{black}}%
      \expandafter\def\csname LT2\endcsname{\color{black}}%
      \expandafter\def\csname LT3\endcsname{\color{black}}%
      \expandafter\def\csname LT4\endcsname{\color{black}}%
      \expandafter\def\csname LT5\endcsname{\color{black}}%
      \expandafter\def\csname LT6\endcsname{\color{black}}%
      \expandafter\def\csname LT7\endcsname{\color{black}}%
      \expandafter\def\csname LT8\endcsname{\color{black}}%
    \fi
  \fi
  \setlength{\unitlength}{0.0500bp}%
  \begin{picture}(7200.00,5040.00)%
    \gplgaddtomacro\gplbacktext{%
      \csname LTb\endcsname%
      \put(1210,4640){\makebox(0,0)[r]{\strut{} 50}}%
      \put(1210,4380){\makebox(0,0)[r]{\strut{} 40}}%
      \put(1210,4050){\makebox(0,0)[r]{\strut{} 30}}%
      \put(1210,3580){\makebox(0,0)[r]{\strut{} 20}}%
      \put(1210,2784){\makebox(0,0)[r]{\strut{} 10}}%
      \put(1210,2510){\makebox(0,0)[r]{\strut{} 8}}%
      \put(1210,2170){\makebox(0,0)[r]{\strut{} 6}}%
      \put(1210,1710){\makebox(0,0)[r]{\strut{} 4}}%
      \put(1210,900){\makebox(0,0)[r]{\strut{} 2}}%
      \put(1342,704){\makebox(0,0){\strut{} 115}}%
      \put(2132,704){\makebox(0,0){\strut{} 120}}%
      \put(2921,704){\makebox(0,0){\strut{} 125}}%
      \put(3711,704){\makebox(0,0){\strut{} 130}}%
      \put(4501,704){\makebox(0,0){\strut{} 135}}%
      \put(5291,704){\makebox(0,0){\strut{} 140}}%
      \put(6080,704){\makebox(0,0){\strut{} 145}}%
      \put(6870,704){\makebox(0,0){\strut{} 150}}%
      \put(600,2850){\rotatebox{90}{\makebox(0,0){\strut{} $\Gamma$ (keV) }}}%
      \put(4106,254){\makebox(0,0){\strut{} $m_{H}$ (GeV) }}%
    }%
    \gplgaddtomacro\gplfronttext{%
    }%
    \gplbacktext
    \put(0,0){\includegraphics{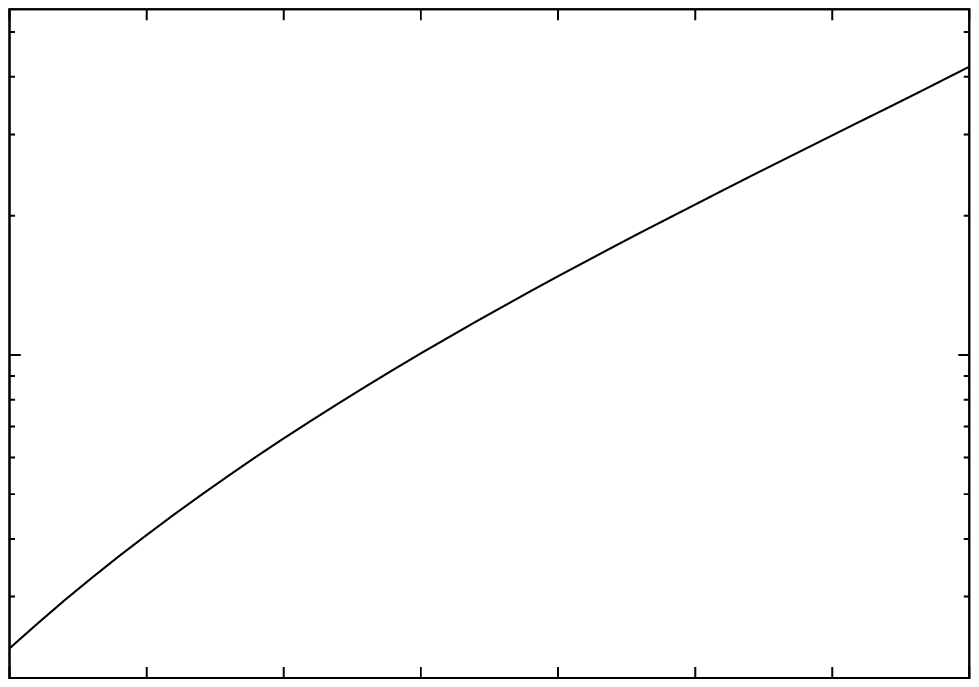}}%
    \gplfronttext
  \end{picture}%
\endgroup
\caption{\label{figGamma} $\Gamma_{H \to Z \gamma}$ including LO and NLO QCD contributions.}
\end{figure}
In Fig.~\ref{figGamma}, we plot the decay width as a function of the Higgs mass.
$\Gamma_{H \to Z \gamma}$ includes the NLO contributions of the diagrams with a top-quark loop and
a bottom-quark loop. 
In Fig.~\ref{figDelta}, we plot the two-loop QCD corrections to the decay width.
In particular, we plot the contribution of the top loop alone, the bottom loop alone, and the complete correction.
For small values of the Higgs mass, the
corrections due to the top loop and those due to the bottom have opposite sign, with the top correction being dominant.
In Table~\ref{tab1}, we quote the figures for $m_H = 125.1$~GeV and $\alpha_S(m_H^2) = 0.115$:
the two--loop QCD corrections amount to 0.22\% of the leading order width, with the top loop adding a 0.3\% and the bottom loop
subtracting a 0.08\%. The two--loop top corrections is in agreement with the corresponding plot of Ref.~\cite{Spira:1991tj} and with
the results in Ref.~\cite{GehrmannHZg}

As we recalled above, at one loop the top and bottom loops contribute 0.3\% of the leading order width. 
Since the two--loop QCD contribution is about 75\% of the one--loop QCD contribution, that implies a bad convergence of the QCD part of the width.
In principle, higher QCD orders would be necessary to stabilise the QCD perturbative series.
However, up to NLO accuracy, the QCD contribution is numerically tiny. Thus, it would be more relevant to compute the two-loop electroweak 
corrections, which we expect to be larger than the QCD ones.

\begin{figure}
\begingroup
  \makeatletter
  \providecommand\color[2][]{%
    \GenericError{(gnuplot) \space\space\space\@spaces}{%
      Package color not loaded in conjunction with
      terminal option `colourtext'%
    }{See the gnuplot documentation for explanation.%
    }{Either use 'blacktext' in gnuplot or load the package
      color.sty in LaTeX.}%
    \renewcommand\color[2][]{}%
  }%
  \providecommand\includegraphics[2][]{%
    \GenericError{(gnuplot) \space\space\space\@spaces}{%
      Package graphicx or graphics not loaded%
    }{See the gnuplot documentation for explanation.%
    }{The gnuplot epslatex terminal needs graphicx.sty or graphics.sty.}%
    \renewcommand\includegraphics[2][]{}%
  }%
  \providecommand\rotatebox[2]{#2}%
  \@ifundefined{ifGPcolor}{%
    \newif\ifGPcolor
    \GPcolortrue
  }{}%
  \@ifundefined{ifGPblacktext}{%
    \newif\ifGPblacktext
    \GPblacktexttrue
  }{}%
  \let\gplgaddtomacro\g@addto@macro
  \gdef\gplbacktext{}%
  \gdef\gplfronttext{}%
  \makeatother
  \ifGPblacktext
    \def\colorrgb#1{}%
    \def\colorgray#1{}%
  \else
    \ifGPcolor
      \def\colorrgb#1{\color[rgb]{#1}}%
      \def\colorgray#1{\color[gray]{#1}}%
      \expandafter\def\csname LTw\endcsname{\color{white}}%
      \expandafter\def\csname LTb\endcsname{\color{black}}%
      \expandafter\def\csname LTa\endcsname{\color{black}}%
      \expandafter\def\csname LT0\endcsname{\color[rgb]{1,0,0}}%
      \expandafter\def\csname LT1\endcsname{\color[rgb]{0,1,0}}%
      \expandafter\def\csname LT2\endcsname{\color[rgb]{0,0,1}}%
      \expandafter\def\csname LT3\endcsname{\color[rgb]{1,0,1}}%
      \expandafter\def\csname LT4\endcsname{\color[rgb]{0,1,1}}%
      \expandafter\def\csname LT5\endcsname{\color[rgb]{1,1,0}}%
      \expandafter\def\csname LT6\endcsname{\color[rgb]{0,0,0}}%
      \expandafter\def\csname LT7\endcsname{\color[rgb]{1,0.3,0}}%
      \expandafter\def\csname LT8\endcsname{\color[rgb]{0.5,0.5,0.5}}%
    \else
      \def\colorrgb#1{\color{black}}%
      \def\colorgray#1{\color[gray]{#1}}%
      \expandafter\def\csname LTw\endcsname{\color{white}}%
      \expandafter\def\csname LTb\endcsname{\color{black}}%
      \expandafter\def\csname LTa\endcsname{\color{black}}%
      \expandafter\def\csname LT0\endcsname{\color{black}}%
      \expandafter\def\csname LT1\endcsname{\color{black}}%
      \expandafter\def\csname LT2\endcsname{\color{black}}%
      \expandafter\def\csname LT3\endcsname{\color{black}}%
      \expandafter\def\csname LT4\endcsname{\color{black}}%
      \expandafter\def\csname LT5\endcsname{\color{black}}%
      \expandafter\def\csname LT6\endcsname{\color{black}}%
      \expandafter\def\csname LT7\endcsname{\color{black}}%
      \expandafter\def\csname LT8\endcsname{\color{black}}%
    \fi
  \fi
  \setlength{\unitlength}{0.0500bp}%
  \begin{picture}(7200.00,5040.00)%
    \gplgaddtomacro\gplbacktext{%
      \csname LTb\endcsname%
      \put(1606,1078){\makebox(0,0)[r]{\strut{}-0.001}}%
      \put(1606,1848){\makebox(0,0)[r]{\strut{} 0}}%
      \put(1606,2619){\makebox(0,0)[r]{\strut{} 0.001}}%
      \put(1606,3389){\makebox(0,0)[r]{\strut{} 0.002}}%
      \put(1606,4160){\makebox(0,0)[r]{\strut{} 0.003}}%
      \put(1738,704){\makebox(0,0){\strut{} 115}}%
      \put(2623,704){\makebox(0,0){\strut{} 120}}%
      \put(3508,704){\makebox(0,0){\strut{} 125}}%
      \put(4392,704){\makebox(0,0){\strut{} 130}}%
      \put(5277,704){\makebox(0,0){\strut{} 135}}%
      \put(6162,704){\makebox(0,0){\strut{} 140}}%
      \put(700,2850){\rotatebox{90}{\makebox(0,0){\strut{} $\delta_{QCD}$ }}}%
      \put(4304,254){\makebox(0,0){\strut{} $m_{H}$ (GeV) }}%
    }%
    \gplgaddtomacro\gplfronttext{%
      \csname LTb\endcsname%
      \put(5307,2641){\makebox(0,0)[r]{\strut{}Top+Bottom}}%
      \csname LTb\endcsname%
      \put(5307,2223){\makebox(0,0)[r]{\strut{}Only top}}%
      \csname LTb\endcsname%
      \put(5307,1805){\makebox(0,0)[r]{\strut{}Only Bottom}}%
    }%
    \gplbacktext
    \put(0,0){\includegraphics{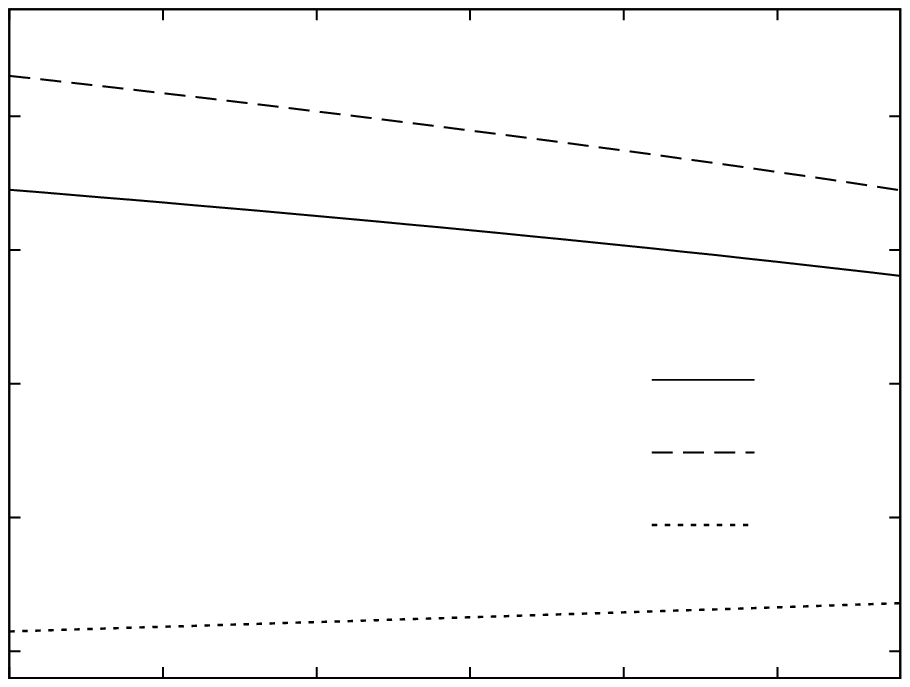}}%
    \gplfronttext
  \end{picture}%
\endgroup
\caption{\label{figDelta} The size of the NLO QCD corrections with respect to the leading order contribution,
$\delta_{QCD}$, as defined in Eq.~(\ref{deltaQCD}). In this figure we plot the 
contribution of the top loop only, the bottom loop only, and the top and bottom loops.}
\end{figure}

\begin{table}
\bc
\begin{tabular}{|c|c|c|c|}
\hline
$\Gamma$ (KeV) & $\delta_{QCD}$ (top) & $\delta_{QCD}$ (bottom) & $\delta_{QCD}$ \cr
\hline
6.68           & $3 \cdot 10^{-3}$        & $-0.8 \cdot 10^{-3}$        & $2.2 \cdot 10^{-3}$ \cr
\hline
\end{tabular}
\ec
\caption{Values of the NLO QCD width, and of the NLO QCD corrections $\delta_{QCD}$ with respect to the leading order contribution, at $m_H = 125.1$~GeV.
\label{tab1}}
\end{table}

\section{Conclusions}
\label{sec:conclusions}

In this paper we present the analytic calculation of the NLO QCD corrections to the width of a Higgs boson
decaying to a $Z$ boson and a photon. These corrections were
computed numerically in \cite{Spira:1991tj}, where the authors studied the size of the corrections that come 
from the two-loop Feynman diagrams with a top-quark loop, with respect to the leading order contribution. In our paper 
we include also the contribution due to the two-loop diagrams with a bottom-quark loop. For small values of the Higgs mass, the
corrections due to the top loop and those due to the bottom have opposite sign, with the top correction being dominant.
For $m_H = 125.1$~GeV they amount to 0.22\% of the leading order width.

The calculation was carried out using integration-by-parts identities for the reduction to master integrals of the dimensionally
regularized scalar integrals, in terms of which we expressed the amplitude. The
calculation of the master integrals was performed using the Differential Equations Method applied to a set of functions
suitably chosen to be of uniform weight.

The solution is expressed in terms of logarithms and polylogarithmic functions $\text{Li}_n$, with $n=2,3,4$, 
and $\text{Li}_{2,2}$. Their arguments are rational expressions in the two dimensionless variables $x_q$, 
$y_q$, written in terms of the three mass scales of the problem: $m_H$, $m_Z$ and $m_q$.
The numerical evaluation of the polylogarithmic functions is done using existing numerical C++ routines.

Finally, we recall that for $m_H = 125.1$~GeV, at leading order the top and the bottom loops account for about 0.3\% of the width,
the overwhelming contribution being yielded by the electroweak loop. Having found 
that the two-loop QCD corrections, which correct the heavy--quark loop, equal about 0.22\% of the leading order width,
it is reasonable to expect that the two-loop electroweak corrections, which correct the electroweak loop, will yield a larger
contribution
\footnote{The $Z$ boson may decay leptonically -- the final state of $H\to Z \gamma$ is a lepton pair, $l^- l^+$, and a photon --
or hadronically -- the final state is formed by two jets and a photon.
However, the two-jet plus photon final state is overwhelmed by the QCD background. 
Conversely, because of the smaller $pp \to l^- l^+ \gamma$ background, the decay $H \to Z \gamma \to l^- l^+ \gamma$ is much cleaner.
In order to compare to the $pp \to l^- l^+ \gamma$ background, 
it is different to consider the resonant production $H \to Z \gamma \to l^- l^+ \gamma$
or the off-resonant one $H\to l^- l^+ \gamma$, for which no on--shell $Z$--boson is produced in the intermediate state.
Estimates of the difference between the resonant and the off-resonant widths
vary~\cite{Chen:2012ju,Dicus:2013ycd,Sun:2013rqa,Passarino:2013nka},
however with realistic experimental cuts, the difference between 
the resonant and off-resonant leading--order widths for electron or muon pairs
is of the order of a few per cent~\cite{Dicus:2013lta}.
As regards the two--loop QCD corrections we computed, the off-resonant diagrams are not present.
Thus, with little effort -- just including the $Z$--boson decay into two leptons --
we could have obtained the QCD corrections to the decay $H\to l^- l^+ \gamma$.
However, as regards the two--loop electroweak corrections, there are also off-resonant diagrams.
Therefore, expecting that the still unknown two--loop electroweak corrections
may be larger than the two-loop QCD corrections, we think it is best to postpone an analysis
of the $H\to l^- l^+ \gamma$ decay to the computation of the two--loop electroweak corrections.}.


\acknowledgments{
We are greatful to D. Kara for providing the results of their independent calculation for a detailed comparison
before the publication \cite{GehrmannHZg}.
Some of the algebraic manipulations required in this work were carried out with {\tt FORM}~\cite{Kuipers:2012rf}. 
The Feynman diagrams were generated by FeynArts~\cite{Kublbeck:1990xc,Hahn:2000kx}
and drawn with {\tt Axodraw}~\cite{Vermaseren:1994je}. The work of VS was partially supported by the Alexander von Humboldt
Foundation (Humboldt Forschungspreis). The work of RB was partly supported by European Community Seventh Framework
Programme FP7/2007-2013, under grant agreement N.302997. 
The work of JMH was supported in part by the DOE grant DE-SC0009988,
and by the Marvin L. Goldberger fund.
The work of VDD, HF, JMH and VS was 
partly supported by the Research Executive Agency (REA) of the European Union, through 
the Initial Training Network LHCPhenoNet under contract PITN-GA-2010-264564.
The work of HF is supported in part by the European Commission through the HiggsTools Initial Training Network PITN-GA-2012-316704.
RB, VDD and FM would like to thank the Galileo Galilei Institute for Theoretical Physics for hospitality 
during the completion of this work.
}


\appendix

\section{Projection operators \label{sec:appa}}

The projector operators to be used in Eq.~(\ref{ffactgen}) are
\bea
P_1^{\mu \nu} & = & \frac{4}{(m_H^2-m_Z^2)^2} \, \, p_2^{\mu} p_2^{\nu}  \\
P_2^{\mu \nu} & = & \frac{4}{(m_H^2-m_Z^2)^2} \, \, p_1^{\mu} p_1^{\nu} 
              + \frac{16(d-1)m_Z^4}{(d-2)(m_H^2-m_Z^2)^4} \, \, p_2^{\mu} p_2^{\nu} \nn\\
& &
              - \frac{8(d-1)m_Z^2}{(d-2)(m_H^2-m_Z^2)^3} \, \, \left[
  	               p_1^{\mu} p_2^{\nu} 
                     + p_2^{\mu} p_1^{\nu} \right]
	      - \frac{4m_Z^2}{(d-2)(m_H^2-m_Z^2)^2} \, \, \delta^{\mu \nu} \, , \\
P_3^{\mu \nu} & = & - \frac{8(d-1)m_Z^2}{(d-2)(m_H^2-m_Z^2)^3} \, \, p_2^{\mu} p_2^{\nu}
              + \frac{4}{(d-2)(m_H^2-m_Z^2)^2} \, \, p_1^{\mu} p_2^{\nu} \nn\\
& & 
              + \frac{4(d-1)}{(d-2)(m_H^2-m_Z^2)^2} \, \,  p_2^{\mu} p_1^{\nu}
	      + \frac{2}{(d-2)(m_H^2-m_Z^2)} \, \, \delta^{\mu \nu} \, ,  \\
P_4^{\mu \nu} & = &  - \frac{8(d-1)m_Z^2}{(d-2)(m_H^2-m_Z^2)^3} \, \, p_2^{\mu} p_2^{\nu}
              + \frac{4(d-1)}{(d-2)(m_H^2-m_Z^2)^2} \, \,  p_1^{\mu} p_2^{\nu}  \nn\\
& & 
              + \frac{4}{(d-2)(m_H^2-m_Z^2)^2} \, \, p_2^{\mu} p_1^{\nu}
	      + \frac{2}{(d-2)(m_H^2-m_Z^2)} \, \, \delta^{\mu \nu} \, ,   \\
P_5^{\mu \nu} & = & - \frac{4m_Z^2}{(d-2)(m_H^2-m_Z^2)^2} \, \, p_2^{\mu} p_2^{\nu}
              + \frac{2}{(d-2)(m_H^2-m_Z^2)} \, \, \left[ 
	               p_1^{\mu} p_2^{\nu}
                     + p_2^{\mu} p_1^{\nu} \right]  \nn\\
& & 
	      + \frac{1}{(d-2)} \, \, \delta^{\mu \nu} \, ,   \\
P_6^{\mu \nu} & = & - \frac{4}{(d-2)(d-3)(m_H^2-m_Z^2)^2} \, \, 
\ep^{\mu \nu \rho \sigma} p_{1 \, \rho} p_{2 \, \sigma}  \, .
\eea

\section{Primary master integrals}
\label{sec:appb}

In this appendix we give the explicit expression of the ``primary'' master integrals shown in Fig.~\ref{misqcd}.

We define the following set of seven denominators,
\bea
D_1 & = & k_1^2+m_q^2 \, , \\
D_2 & = & k_2^2 \, , \\
D_3 & = & (k_1+k_2)^2+m_q^2 \, , \\
D_4 & = & (p_1-k_1)^2+m_q^2 \, , \\
D_5 & = & (p_2+k_1)^2+m_q^2 \, , \\
D_6 & = & (p_1-k_1-k_2)^2+m_q^2 \, , \\
D_7 & = & (p_2+k_1+k_2)^2+m_q^2 \, .
\eea

The integrals of Fig.~\ref{misqcd} are then,
\begin{align}
M_1 & = \int {\mathcal D}^D k_1 \, {\mathcal D}^D k_2 \, \frac{1}{D_1^2 \, D_3^2} \, , 
& M_2 & = \int {\mathcal D}^D k_1 \, {\mathcal D}^D k_2 \, \frac{1}{D_3^2 \, D_4^2 \, D_5} \, , \\
M_3 & = \int {\mathcal D}^D k_1 \, {\mathcal D}^D k_2 \, \frac{1}{D_1^2 \, D_3^2 \, D_4} \, , 
& M_4 & = \int {\mathcal D}^D k_1 \, {\mathcal D}^D k_2 \, \frac{1}{D_2 \, D_5^2 \, D_6^2} \, , \\
M_5 & = \int {\mathcal D}^D k_1 \, {\mathcal D}^D k_2 \, \frac{1}{D_2 \, D_3^2 \, D_4^2} \, , 
& M_6 & = \int {\mathcal D}^D k_1 \, {\mathcal D}^D k_2 \, \frac{1}{D_2^2 \, D_5^2 \, D_6} \, , \\
M_7 & = \int {\mathcal D}^D k_1 \, {\mathcal D}^D k_2 \, \frac{1}{D_2^2 \, D_3^2 \, D_4} \, , 
& M_8 & = \int {\mathcal D}^D k_1 \, {\mathcal D}^D k_2 \, \frac{1}{D_1 \, D_3^2 \, D_4 \, D_5} \, , \\
M_9 & = \int {\mathcal D}^D k_1 \, {\mathcal D}^D k_2 \, \frac{1}{D_4^2 \, D_5 \, D_6^2 \, D_7} \, , 
& M_{10} & = \int {\mathcal D}^D k_1 \, {\mathcal D}^D k_2 \, \frac{1}{D_1^2 \, D_3^2 \, D_4 \, D_6} \, , \\
M_{11} & = \int {\mathcal D}^D k_1 \, {\mathcal D}^D k_2 \, \frac{1}{D_3^2 \, D_4^2 \, D_5 \, D_6} \, , 
& M_{12} & = \int {\mathcal D}^D k_1 \, {\mathcal D}^D k_2 \, \frac{1}{D_1 \, D_2 \, D_5 \, D_6^2} \, , \\
M_{13} & = \int {\mathcal D}^D k_1 \, {\mathcal D}^D k_2 \, \frac{1}{D_2 \, D_3^2 \, D_4 \, D_5} \, , 
& M_{14} & = \int {\mathcal D}^D k_1 \, {\mathcal D}^D k_2 \, \frac{1}{D_2 \, D_3^3 \, D_4 \, D_5} \, , \\
M_{15} & = \int {\mathcal D}^D k_1 \, {\mathcal D}^D k_2 \, \frac{k_1^2+m_q^2}{D_2 \, D_3^2 \, D_4^2 \, D_5} \, , 
& M_{16} & = \int {\mathcal D}^D k_1 \, {\mathcal D}^D k_2 \, \frac{1}{D_2 \, D_3 \, D_5^2 \, D_6} \, , \\
M_{17} & = \int {\mathcal D}^D k_1 \, {\mathcal D}^D k_2 \, \frac{1}{D_2 \, D_3 \, D_5^3 \, D_6} \, , 
& M_{18} & = \int {\mathcal D}^D k_1 \, {\mathcal D}^D k_2 \, \frac{(p_2+k_1+k_2)^2}{D_2 \, D_3^2 \, D_5^2 \, D_6} \, , \\
M_{19} & = \int {\mathcal D}^D k_1 \, {\mathcal D}^D k_2 \, \frac{1}{D_1 \, D_4 \, D_5 \, D_6^2 \, D_7} \, , 
& M_{20} & = \int {\mathcal D}^D k_1 \, {\mathcal D}^D k_2 \, \frac{1}{D_1 \, D_3^2 \, D_4 \, D_5 \, D_6} \, , \\
M_{21} & = \int {\mathcal D}^D k_1 \, {\mathcal D}^D k_2 \, \frac{1}{D_1 \, D_2 \, D_3 \, D_5 \, D_6} \, , 
& M_{22} & = \int {\mathcal D}^D k_1 \, {\mathcal D}^D k_2 \, \frac{1}{D_1 \, D_2 \, D_3 \, D_5 \, D_6^2} \, , \\
M_{23} & = \int {\mathcal D}^D k_1 \, {\mathcal D}^D k_2 \, \frac{1}{D_2 \, D_3 \, D_4 \, D_5 \, D_7} \, , 
& M_{24} & = \int {\mathcal D}^D k_1 \, {\mathcal D}^D k_2 \, \frac{1}{D_2 \, D_3 \, D_4^2 \, D_5 \, D_7} \, , \\
M_{25} & = \int {\mathcal D}^D k_1 \, {\mathcal D}^D k_2 \, \frac{1}{D_2 \, D_3 \, D_4 \, D_5 \, D_6} \, , 
& M_{26} & = \int {\mathcal D}^D k_1 \, {\mathcal D}^D k_2 \, \frac{1}{D_2 \, D_3^2 \, D_4 \, D_5 \, D_6} \, , \\
M_{27} & = \int {\mathcal D}^D k_1 \, {\mathcal D}^D k_2 \, \frac{1}{D_2 \, D_3 \, D_4 \, D_5^2 \, D_6} \, , 
& M_{28} & = \int {\mathcal D}^D k_1 \, {\mathcal D}^D k_2 \, \frac{1}{D_2 \, D_3^2 \, D_4 \, D_5^2 \, D_6} \, .
\end{align}

\section{Matrices for the System of Differential Equations}
\label{sec:appc}

In this appendix we collect the matrices $\mathbb{S}_1$--$\mathbb{S}_{12}$ defined in Eq.~(\ref{differential}).

\bea
\mathbb{S}_1 &=& 
\left(
\scriptsize{
\begin{array}{cccccccccccccccccccccccccccc}
 0 & 0 & 0 & 0 & 0 & 0 & 0 & 0 & 0 & 0 & 0 & 0 & 0 & 0 & 0 & 0 & 0 & 0 & 0 & 0 & 0 & 0 & 0 & 0 & 0 & 0 & 0 & 0 \\
 0 & -1 & -2 & 0 & 0 & 0 & 0 & 0 & 0 & 0 & 0 & 0 & 0 & 0 & 0 & 0 & 0 & 0 & 0 & 0 & 0 & 0 & 0 & 0 & 0 & 0 & 0 & 0 \\
 1 & 3 & 4 & 0 & 0 & 0 & 0 & 0 & 0 & 0 & 0 & 0 & 0 & 0 & 0 & 0 & 0 & 0 & 0 & 0 & 0 & 0 & 0 & 0 & 0 & 0 & 0 & 0 \\
 0 & 0 & 0 & 0 & 0 & 0 & 0 & 0 & 0 & 0 & 0 & 0 & 0 & 0 & 0 & 0 & 0 & 0 & 0 & 0 & 0 & 0 & 0 & 0 & 0 & 0 & 0 & 0 \\
 0 & 0 & 0 & 0 & 0 & 0 & 0 & 0 & 0 & 0 & 0 & 0 & 0 & 0 & 0 & 0 & 0 & 0 & 0 & 0 & 0 & 0 & 0 & 0 & 0 & 0 & 0 & 0 \\
 0 & 0 & 0 & 0 & 0 & 0 & 0 & 0 & 0 & 0 & 0 & 0 & 0 & 0 & 0 & 0 & 0 & 0 & 0 & 0 & 0 & 0 & 0 & 0 & 0 & 0 & 0 & 0 \\
 -1 & 0 & 0 & 0 & 0 & 0 & 1 & 0 & 0 & 0 & 0 & 0 & 0 & 0 & 0 & 0 & 0 & 0 & 0 & 0 & 0 & 0 & 0 & 0 & 0 & 0 & 0 & 0 \\
 0 & -\frac{1}{2} & 0 & 0 & 0 & 0 & 0 & 1 & -2 & 0 & 0 & 0 & 0 & 0 & 0 & 0 & 0 & 0 & 0 & 0 & 0 & 0 & 0 & 0 & 0 & 0 & 0 & 0 \\
 0 & -\frac{3}{4} & -\frac{1}{2} & 0 & 0 & 0 & 0 & 0 & 0 & 0 & 0 & 0 & 0 & 0 & 0 & 0 & 0 & 0 & 0 & 0 & 0 & 0 & 0 & 0 & 0 & 0 & 0 & 0 \\
 0 & 0 & 0 & 0 & 0 & -1 & 0 & 0 & 0 & 1 & 0 & 0 & 0 & 0 & 0 & 0 & 0 & 0 & 0 & 0 & 0 & 0 & 0 & 0 & 0 & 0 & 0 & 0 \\
 0 & 0 & 0 & 1 & 0 & 0 & 0 & 0 & 0 & 0 & -1 & 0 & 2 & 0 & 0 & 0 & 0 & 0 & 0 & 0 & 0 & 0 & 0 & 0 & 0 & 0 & 0 & 0 \\
 0 & 0 & 0 & \frac{3}{4} & 0 & 0 & \frac{1}{2} & 0 & 0 & 0 & -\frac{3}{2} & 1 & 1 & 0 & 0 & 0 & 0 & 0 & 0 & 0 & 0 & 0 & 0 & 0 & 0 & 0 & 0 & 0 \\
 0 & 0 & 0 & \frac{3}{4} & 0 & 0 & -\frac{1}{2} & 0 & 0 & 0 & 0 & -1 & 2 & 0 & 0 & 0 & 0 & 0 & 0 & 0 & 0 & 0 & 0 & 0 & 0 & 0 & 0 & 0 \\
 0 & -1 & 0 & 0 & 0 & 0 & 0 & 0 & 0 & 0 & 0 & 0 & 0 & 0 & 0 & 0 & 0 & 0 & 0 & 0 & 0 & 0 & 0 & 0 & 0 & 0 & 0 & 0 \\
 0 & 0 & 0 & 0 & 0 & 0 & 0 & 0 & 0 & 0 & 0 & 0 & 0 & 0 & 0 & 0 & 0 & 0 & 0 & 0 & 0 & 0 & 0 & 0 & 0 & 0 & 0 & 0 \\
 0 & 0 & 0 & 0 & 0 & -1 & 0 & 0 & 0 & 0 & 0 & 0 & 0 & 0 & 0 & 1 & 0 & 0 & 0 & 0 & 0 & 0 & 0 & 0 & 0 & 0 & 0 & 0 \\
 0 & 0 & 0 & 0 & 0 & 0 & 0 & 0 & 0 & 0 & 0 & 0 & 0 & 0 & 0 & 1 & 0 & 0 & 0 & 0 & 0 & 0 & 0 & 0 & 0 & 0 & 0 & 0 \\
 0 & 0 & 0 & 0 & 0 & 0 & 0 & -1 & 0 & 0 & 0 & 0 & 0 & 0 & 0 & 0 & 0 & 0 & 0 & 0 & 0 & 0 & 0 & 0 & 0 & 0 & 0 & 0 \\
 0 & 0 & 0 & 0 & 0 & 0 & 0 & 0 & 0 & 2 & 0 & 0 & 0 & 0 & 0 & 0 & 0 & 0 & 0 & 0 & 0 & 0 & 0 & 0 & 0 & 0 & 0 & 0 \\
 0 & \frac{1}{2} & 0 & \frac{1}{2} & 0 & 0 & 0 & -\frac{1}{2} & 0 & 0 & 0 & 0 & 0 & 0 & 0 & 0 & 0 & 0 & 0 & -1 & 0 & \frac{1}{2} & \frac{1}{2} & 0 & 0 & 0 & 0 & 0 \\
 0 & 0 & 0 & 0 & 0 & 0 & 0 & 0 & 0 & -2 & 0 & 0 & 0 & 0 & 0 & 2 & 0 & 0 & 0 & 0 & 0 & 0 & 0 & 0 & 0 & 0 & 0 & 0 \\
 0 & 0 & 0 & \frac{3}{2} & 0 & 0 & 0 & 1 & -4 & 0 & -2 & 2 & 0 & 0 & 0 & 0 & 0 & 0 & 0 & -2 & 0 & 1 & 1 & 0 & 0 & 0 & 0 & 0 \\
 0 & \frac{3}{2} & 2 & 0 & 0 & 0 & 0 & -2 & 2 & 0 & 1 & 0 & -2 & 0 & 0 & 0 & 0 & 0 & 0 & -2 & 0 & 1 & 1 & 0 & 0 & 0 & 0 & 0 \\
 0 & 0 & 0 & 0 & 0 & 0 & 0 & 0 & 0 & 0 & -\frac{1}{2} & 0 & 0 & -\frac{1}{2} & 0 & 0 & 0 & 0 & 0 & 0 & 0 & 0 & 0 & 0 & \frac{1}{2} & 0 & 0 & 0 \\
 0 & -2 & 0 & \frac{1}{2} & 0 & 0 & 0 & 0 & 0 & 0 & -3 & 2 & 0 & -1 & 0 & 0 & 0 & 0 & 0 & 0 & 0 & 0 & 0 & 0 & 1 & 0 & 0 & 0 \\
 0 & 0 & 0 & 0 & 0 & 0 & 1 & 0 & 0 & 0 & 0 & 0 & 0 & 0 & 0 & 0 & 0 & 0 & 0 & 0 & 0 & 0 & 0 & 0 & 0 & 0 & 0 & 0 \\
 0 & 0 & 0 & 0 & 0 & 0 & 2 & 0 & 0 & 0 & 0 & 0 & 0 & 0 & 0 & 0 & 0 & 0 & 0 & 0 & 0 & 0 & 0 & 0 & 0 & 0 & 2 & 0 \\
 0 & 0 & 0 & 0 & 0 & 0 & 0 & 0 & 0 & 0 & 0 & 0 & 0 & 0 & 0 & 0 & 0 & 0 & 0 & 0 & 0 & 0 & 0 & 0 & 0 & -1 & -1 & 1 \\
\end{array}}
\right) \, , \nonumber \\
& & \nonumber \\
& & \nonumber \\
\mathbb{S}_2 &=& 
\left(
\scriptsize{
\begin{array}{cccccccccccccccccccccccccccc}
 0 & 0 & 0 & 0 & 0 & 0 & 0 & 0 & 0 & 0 & 0 & 0 & 0 & 0 & 0 & 0 & 0 & 0 & 0 & 0 & 0 & 0 & 0 & 0 & 0 & 0 & 0 & 0 \\
 0 & 2 & 0 & 0 & 0 & 0 & 0 & 0 & 0 & 0 & 0 & 0 & 0 & 0 & 0 & 0 & 0 & 0 & 0 & 0 & 0 & 0 & 0 & 0 & 0 & 0 & 0 & 0 \\
 0 & 0 & -2 & 0 & 0 & 0 & 0 & 0 & 0 & 0 & 0 & 0 & 0 & 0 & 0 & 0 & 0 & 0 & 0 & 0 & 0 & 0 & 0 & 0 & 0 & 0 & 0 & 0 \\
 0 & 0 & 0 & 0 & 0 & 0 & 0 & 0 & 0 & 0 & 0 & 0 & 0 & 0 & 0 & 0 & 0 & 0 & 0 & 0 & 0 & 0 & 0 & 0 & 0 & 0 & 0 & 0 \\
 0 & 0 & 0 & 0 & 0 & 0 & 0 & 0 & 0 & 0 & 0 & 0 & 0 & 0 & 0 & 0 & 0 & 0 & 0 & 0 & 0 & 0 & 0 & 0 & 0 & 0 & 0 & 0 \\
 0 & 0 & 0 & 0 & 0 & 0 & 0 & 0 & 0 & 0 & 0 & 0 & 0 & 0 & 0 & 0 & 0 & 0 & 0 & 0 & 0 & 0 & 0 & 0 & 0 & 0 & 0 & 0 \\
 0 & 0 & 0 & 0 & 0 & 0 & 0 & 0 & 0 & 0 & 0 & 0 & 0 & 0 & 0 & 0 & 0 & 0 & 0 & 0 & 0 & 0 & 0 & 0 & 0 & 0 & 0 & 0 \\
 0 & -2 & 0 & 0 & 0 & 0 & 0 & 0 & 0 & 0 & 0 & 0 & 0 & 0 & 0 & 0 & 0 & 0 & 0 & 0 & 0 & 0 & 0 & 0 & 0 & 0 & 0 & 0 \\
 0 & -\frac{3}{2} & 0 & 0 & 0 & 0 & 0 & 0 & 0 & 0 & 0 & 0 & 0 & 0 & 0 & 0 & 0 & 0 & 0 & 0 & 0 & 0 & 0 & 0 & 0 & 0 & 0 & 0 \\
 0 & 0 & 0 & 0 & 0 & 0 & 0 & 0 & 0 & 0 & 0 & 0 & 0 & 0 & 0 & 0 & 0 & 0 & 0 & 0 & 0 & 0 & 0 & 0 & 0 & 0 & 0 & 0 \\
 0 & 0 & 0 & -\frac{1}{2} & 0 & 0 & 0 & 0 & 0 & 0 & 1 & -2 & 0 & 0 & 0 & 0 & 0 & 0 & 0 & 0 & 0 & 0 & 0 & 0 & 0 & 0 & 0 & 0 \\
 0 & 0 & 0 & 0 & 0 & 0 & 0 & 0 & 0 & 0 & 0 & 0 & 0 & 0 & 0 & 0 & 0 & 0 & 0 & 0 & 0 & 0 & 0 & 0 & 0 & 0 & 0 & 0 \\
 0 & 0 & 0 & 0 & 0 & 0 & -1 & 0 & 0 & 0 & 0 & 0 & 1 & 0 & 0 & 0 & 0 & 0 & 0 & 0 & 0 & 0 & 0 & 0 & 0 & 0 & 0 & 0 \\
 0 & 2 & 0 & 0 & 0 & 0 & 0 & 0 & 0 & 0 & 0 & 0 & 0 & 0 & 0 & 0 & 0 & 0 & 0 & 0 & 0 & 0 & 0 & 0 & 0 & 0 & 0 & 0 \\
 0 & 0 & 0 & 0 & 0 & 0 & 0 & 0 & 0 & 0 & 0 & 0 & 0 & 0 & 0 & 0 & 0 & 0 & 0 & 0 & 0 & 0 & 0 & 0 & 0 & 0 & 0 & 0 \\
 0 & 0 & 0 & 0 & 0 & 0 & 0 & 0 & 0 & 0 & 0 & 0 & 0 & 0 & 0 & 0 & 0 & 0 & 0 & 0 & 0 & 0 & 0 & 0 & 0 & 0 & 0 & 0 \\
 0 & 0 & 0 & 0 & 0 & 0 & 0 & 0 & 0 & 0 & 0 & 0 & 0 & 0 & 0 & 0 & 0 & 0 & 0 & 0 & 0 & 0 & 0 & 0 & 0 & 0 & 0 & 0 \\
 0 & 0 & 0 & 0 & 0 & 0 & 0 & 0 & 0 & 0 & 0 & 0 & 0 & 0 & 0 & 0 & 0 & 0 & 0 & 0 & 0 & 0 & 0 & 0 & 0 & 0 & 0 & 0 \\
 0 & 0 & 0 & 0 & 0 & 0 & 0 & 0 & 0 & 0 & 0 & 0 & 0 & 0 & 0 & 0 & 0 & 0 & 0 & 0 & 0 & 0 & 0 & 0 & 0 & 0 & 0 & 0 \\
 0 & -1 & 0 & -1 & 0 & 0 & 0 & 1 & 0 & 0 & 0 & 0 & 0 & 0 & 0 & 0 & 0 & 0 & 0 & -2 & 0 & 0 & -1 & 0 & 0 & 0 & 0 & 0 \\
 0 & 0 & 0 & 0 & 0 & 0 & 0 & 0 & 0 & 0 & 0 & 0 & 0 & 0 & 0 & 0 & 0 & 0 & 0 & 0 & 0 & 0 & 0 & 0 & 0 & 0 & 0 & 0 \\
 0 & 0 & 0 & 0 & 0 & 0 & 0 & 0 & 0 & 0 & 0 & 0 & 2 & 0 & 0 & 0 & 0 & 0 & 0 & 0 & 0 & 0 & 0 & 0 & 0 & 0 & 0 & 0 \\
 0 & 0 & 0 & \frac{3}{2} & 0 & 0 & 0 & -2 & 0 & 0 & 1 & -2 & 0 & 0 & 0 & 0 & 0 & 0 & 0 & 4 & 0 & 0 & 2 & 0 & 0 & 0 & 0 & 0 \\
 0 & 0 & 0 & 0 & 0 & 0 & 0 & 0 & 0 & 0 & -1 & 0 & 0 & 1 & 0 & 0 & 0 & 0 & 0 & 0 & 0 & 0 & 0 & 0 & 0 & 0 & 0 & 0 \\
 0 & 0 & 0 & 0 & 0 & 0 & 0 & 0 & 0 & 0 & 0 & 0 & -2 & 0 & 0 & 0 & 0 & 0 & 0 & 0 & 0 & 0 & 0 & 0 & 0 & 0 & 0 & 0 \\
 0 & 0 & 0 & 0 & 0 & 0 & 0 & 0 & 0 & 0 & 0 & 0 & 0 & 0 & 0 & 0 & 0 & 0 & 0 & 0 & 0 & 0 & 0 & 0 & 0 & 0 & 0 & 0 \\
 0 & 0 & 0 & 0 & 0 & 0 & 0 & 0 & 0 & 0 & 0 & 0 & 0 & 0 & 0 & 0 & 0 & 0 & 0 & 0 & 0 & 0 & 0 & 0 & 0 & 0 & 0 & 0 \\
 0 & 0 & 0 & 0 & 0 & 0 & 0 & 0 & 0 & 0 & 0 & 0 & 0 & 0 & 0 & 0 & 0 & 0 & 0 & 0 & 0 & 0 & 0 & 0 & 0 & 0 & 0 & 0 \\
\end{array}
}
\right) \, , \nonumber \\
~ \nonumber \\
\mathbb{S}_3 &=& 
\left(
\scriptsize{
\begin{array}{cccccccccccccccccccccccccccc}
 0 & 0 & 0 & 0 & 0 & 0 & 0 & 0 & 0 & 0 & 0 & 0 & 0 & 0 & 0 & 0 & 0 & 0 & 0 & 0 & 0 & 0 & 0 & 0 & 0 & 0 & 0 & 0 \\
 0 & 0 & 0 & 0 & 0 & 0 & 0 & 0 & 0 & 0 & 0 & 0 & 0 & 0 & 0 & 0 & 0 & 0 & 0 & 0 & 0 & 0 & 0 & 0 & 0 & 0 & 0 & 0 \\
 0 & 0 & -6 & 0 & 0 & 0 & 0 & 0 & 0 & 0 & 0 & 0 & 0 & 0 & 0 & 0 & 0 & 0 & 0 & 0 & 0 & 0 & 0 & 0 & 0 & 0 & 0 & 0 \\
 0 & 0 & 0 & 0 & 0 & 0 & 0 & 0 & 0 & 0 & 0 & 0 & 0 & 0 & 0 & 0 & 0 & 0 & 0 & 0 & 0 & 0 & 0 & 0 & 0 & 0 & 0 & 0 \\
 0 & 0 & 0 & 0 & 0 & 0 & 0 & 0 & 0 & 0 & 0 & 0 & 0 & 0 & 0 & 0 & 0 & 0 & 0 & 0 & 0 & 0 & 0 & 0 & 0 & 0 & 0 & 0 \\
 0 & 0 & 0 & 0 & 0 & 0 & 0 & 0 & 0 & 0 & 0 & 0 & 0 & 0 & 0 & 0 & 0 & 0 & 0 & 0 & 0 & 0 & 0 & 0 & 0 & 0 & 0 & 0 \\
 0 & 0 & 0 & 0 & 0 & 0 & -2 & 0 & 0 & 0 & 0 & 0 & 0 & 0 & 0 & 0 & 0 & 0 & 0 & 0 & 0 & 0 & 0 & 0 & 0 & 0 & 0 & 0 \\
 0 & 0 & 0 & 0 & 0 & 0 & 0 & 0 & 0 & 0 & 0 & 0 & 0 & 0 & 0 & 0 & 0 & 0 & 0 & 0 & 0 & 0 & 0 & 0 & 0 & 0 & 0 & 0 \\
 0 & 0 & 0 & 0 & 0 & 0 & 0 & 0 & 0 & 0 & 0 & 0 & 0 & 0 & 0 & 0 & 0 & 0 & 0 & 0 & 0 & 0 & 0 & 0 & 0 & 0 & 0 & 0 \\
 0 & 0 & 0 & 0 & 0 & 0 & 0 & 0 & 0 & 0 & 0 & 0 & 0 & 0 & 0 & 0 & 0 & 0 & 0 & 0 & 0 & 0 & 0 & 0 & 0 & 0 & 0 & 0 \\
 0 & 0 & 0 & 0 & 0 & 0 & 0 & 0 & 0 & 0 & 0 & 0 & 0 & 0 & 0 & 0 & 0 & 0 & 0 & 0 & 0 & 0 & 0 & 0 & 0 & 0 & 0 & 0 \\
 0 & 0 & 0 & 0 & 0 & 0 & 0 & 0 & 0 & 0 & 0 & 0 & 0 & 0 & 0 & 0 & 0 & 0 & 0 & 0 & 0 & 0 & 0 & 0 & 0 & 0 & 0 & 0 \\
 0 & 0 & 0 & 0 & 0 & 0 & 0 & 0 & 0 & 0 & 0 & 0 & -2 & 0 & 0 & 0 & 0 & 0 & 0 & 0 & 0 & 0 & 0 & 0 & 0 & 0 & 0 & 0 \\
 0 & 0 & 0 & 0 & 0 & 0 & 0 & 0 & 0 & 0 & 0 & 0 & 0 & 0 & 0 & 0 & 0 & 0 & 0 & 0 & 0 & 0 & 0 & 0 & 0 & 0 & 0 & 0 \\
 0 & 0 & 0 & 0 & 0 & 0 & 0 & 0 & 0 & 0 & 0 & 0 & 0 & 0 & 0 & 0 & 0 & 0 & 0 & 0 & 0 & 0 & 0 & 0 & 0 & 0 & 0 & 0 \\
 0 & 0 & 0 & 0 & 0 & 0 & 0 & 0 & 0 & 0 & 0 & 0 & 0 & 0 & 0 & -2 & 0 & 0 & 0 & 0 & 0 & 0 & 0 & 0 & 0 & 0 & 0 & 0 \\
 0 & 0 & 0 & 0 & 0 & 0 & 0 & 0 & 0 & 0 & 0 & 0 & 0 & 0 & 0 & 0 & 0 & 0 & 0 & 0 & 0 & 0 & 0 & 0 & 0 & 0 & 0 & 0 \\
 0 & 0 & 0 & 0 & 0 & 0 & 0 & 0 & 0 & 0 & 0 & 0 & 0 & 0 & 0 & 0 & 0 & 0 & 0 & 0 & 0 & 0 & 0 & 0 & 0 & 0 & 0 & 0 \\
 0 & 0 & 0 & 0 & 0 & 0 & 0 & 0 & 0 & 0 & 0 & 0 & 0 & 0 & 0 & 0 & 0 & 0 & 0 & 0 & 0 & 0 & 0 & 0 & 0 & 0 & 0 & 0 \\
 0 & 0 & 0 & 0 & 0 & 0 & 0 & 0 & 0 & 0 & 0 & 0 & 0 & 0 & 0 & 0 & 0 & 0 & 0 & 0 & 0 & 0 & 0 & 0 & 0 & 0 & 0 & 0 \\
 0 & 0 & 0 & 0 & 0 & 0 & 0 & 0 & 0 & 0 & 0 & 0 & 0 & 0 & 0 & 0 & 0 & 0 & 0 & 0 & 0 & 0 & 0 & 0 & 0 & 0 & 0 & 0 \\
 0 & 0 & 0 & 0 & 0 & 0 & 0 & 0 & 0 & 0 & 0 & 0 & 0 & 0 & 0 & 0 & 0 & 0 & 0 & 0 & 0 & -2 & 0 & 0 & 0 & 0 & 0 & 0 \\
 0 & 0 & 0 & 0 & 0 & 0 & 0 & 0 & 0 & 0 & 0 & 0 & 0 & 0 & 0 & 0 & 0 & 0 & 0 & 0 & 0 & 0 & 0 & 0 & 0 & 0 & 0 & 0 \\
 0 & 0 & 0 & 0 & 0 & 0 & 0 & 0 & 0 & 0 & 0 & 0 & 0 & 0 & 0 & 0 & 0 & 0 & 0 & 0 & 0 & 0 & 0 & 0 & 0 & 0 & 0 & 0 \\
 0 & 0 & 0 & 0 & 0 & 0 & 0 & 0 & 0 & 0 & 0 & 0 & 0 & 0 & 0 & 0 & 0 & 0 & 0 & 0 & 0 & 0 & 0 & 0 & -2 & 0 & 0 & 0 \\
 0 & 0 & 0 & 0 & 0 & 0 & 0 & 0 & 0 & 0 & 0 & 0 & 0 & 0 & 0 & 0 & 0 & 0 & 0 & 0 & 0 & 0 & 0 & 0 & 0 & 0 & 0 & 0 \\
 0 & 0 & 0 & 0 & 0 & 0 & 0 & 0 & 0 & 0 & 0 & 0 & 0 & 0 & 0 & 0 & 0 & 0 & 0 & 0 & 0 & 0 & 0 & 0 & 0 & 0 & -4 & 0 \\
 0 & 0 & 0 & 0 & 0 & 0 & 0 & 0 & 0 & 0 & 0 & 0 & 0 & 0 & 0 & 0 & 0 & 0 & 0 & 0 & 0 & 0 & 0 & 0 & 0 & 0 & 0 & -2 \\
\end{array}
}
\right) \, , \nonumber \\
& & \nonumber \\
& & \nonumber \\
\mathbb{S}_4 &=& 
\left(
\scriptsize{
\begin{array}{cccccccccccccccccccccccccccc}
 0 & 0 & 0 & 0 & 0 & 0 & 0 & 0 & 0 & 0 & 0 & 0 & 0 & 0 & 0 & 0 & 0 & 0 & 0 & 0 & 0 & 0 & 0 & 0 & 0 & 0 & 0 & 0 \\
 0 & 0 & 0 & 0 & 0 & 0 & 0 & 0 & 0 & 0 & 0 & 0 & 0 & 0 & 0 & 0 & 0 & 0 & 0 & 0 & 0 & 0 & 0 & 0 & 0 & 0 & 0 & 0 \\
 0 & 0 & 0 & 0 & 0 & 0 & 0 & 0 & 0 & 0 & 0 & 0 & 0 & 0 & 0 & 0 & 0 & 0 & 0 & 0 & 0 & 0 & 0 & 0 & 0 & 0 & 0 & 0 \\
 0 & 0 & 0 & -1 & -2 & 0 & 0 & 0 & 0 & 0 & 0 & 0 & 0 & 0 & 0 & 0 & 0 & 0 & 0 & 0 & 0 & 0 & 0 & 0 & 0 & 0 & 0 & 0 \\
 1 & 0 & 0 & 3 & 4 & 0 & 0 & 0 & 0 & 0 & 0 & 0 & 0 & 0 & 0 & 0 & 0 & 0 & 0 & 0 & 0 & 0 & 0 & 0 & 0 & 0 & 0 & 0 \\
 -1 & 0 & 0 & 0 & 0 & 1 & 0 & 0 & 0 & 0 & 0 & 0 & 0 & 0 & 0 & 0 & 0 & 0 & 0 & 0 & 0 & 0 & 0 & 0 & 0 & 0 & 0 & 0 \\
 0 & 0 & 0 & 0 & 0 & 0 & 0 & 0 & 0 & 0 & 0 & 0 & 0 & 0 & 0 & 0 & 0 & 0 & 0 & 0 & 0 & 0 & 0 & 0 & 0 & 0 & 0 & 0 \\
 0 & -1 & 0 & 0 & 0 & 0 & 0 & -1 & 0 & -2 & 0 & 0 & 0 & 0 & 0 & 0 & 0 & 0 & 0 & 0 & 0 & 0 & 0 & 0 & 0 & 0 & 0 & 0 \\
 0 & -\frac{3}{4} & 0 & 0 & 0 & -\frac{1}{2} & 0 & -\frac{3}{2} & 1 & -1 & 0 & 0 & 0 & 0 & 0 & 0 & 0 & 0 & 0 & 0 & 0 & 0 & 0 & 0 & 0 & 0 & 0 & 0 \\
 0 & \frac{3}{4} & 0 & 0 & 0 & -\frac{1}{2} & 0 & 0 & 1 & 2 & 0 & 0 & 0 & 0 & 0 & 0 & 0 & 0 & 0 & 0 & 0 & 0 & 0 & 0 & 0 & 0 & 0 & 0 \\
 0 & 0 & 0 & \frac{1}{2} & 0 & 0 & 0 & 0 & 0 & 0 & 1 & -2 & 0 & 0 & 0 & 0 & 0 & 0 & 0 & 0 & 0 & 0 & 0 & 0 & 0 & 0 & 0 & 0 \\
 0 & 0 & 0 & \frac{3}{4} & \frac{1}{2} & 0 & 0 & 0 & 0 & 0 & 0 & 0 & 0 & 0 & 0 & 0 & 0 & 0 & 0 & 0 & 0 & 0 & 0 & 0 & 0 & 0 & 0 & 0 \\
 0 & 0 & 0 & 0 & 0 & 0 & -1 & 0 & 0 & 0 & 0 & 0 & 1 & 0 & 0 & 0 & 0 & 0 & 0 & 0 & 0 & 0 & 0 & 0 & 0 & 0 & 0 & 0 \\
 0 & 0 & 0 & 1 & 0 & 0 & 0 & 0 & 0 & 0 & 0 & 0 & 0 & 0 & 0 & 0 & 0 & 0 & 0 & 0 & 0 & 0 & 0 & 0 & 0 & 0 & 0 & 0 \\
 0 & 0 & 0 & 0 & 0 & 2 & 0 & 0 & 0 & 0 & 0 & 0 & 0 & 0 & 2 & 0 & 0 & 0 & 0 & 0 & 0 & 0 & 0 & 0 & 0 & 0 & 0 & 0 \\
 0 & 0 & 0 & 0 & 0 & 0 & -1 & 0 & 0 & 0 & 0 & 0 & 0 & 0 & 0 & 1 & 0 & 0 & 0 & 0 & 0 & 0 & 0 & 0 & 0 & 0 & 0 & 0 \\
 0 & 0 & 0 & 0 & 0 & 0 & 0 & 0 & 0 & 0 & 0 & 0 & 0 & 0 & 1 & 0 & 1 & 0 & 0 & 0 & 0 & 0 & 0 & 0 & 0 & -1 & 0 & 0 \\
 0 & 0 & 0 & 0 & 0 & 0 & 0 & -\frac{1}{2} & 0 & 0 & 0 & 0 & 0 & -\frac{1}{2} & 0 & 0 & 0 & 0 & \frac{1}{2} & 0 & 0 & 0 & 0 & 0 & 0 & 0 & 0 & 0 \\
 0 & -\frac{1}{2} & 0 & 2 & 0 & 0 & 0 & -3 & 2 & 0 & 0 & 0 & 0 & -1 & 0 & 0 & 0 & 0 & 1 & 0 & 0 & 0 & 0 & 0 & 0 & 0 & 0 & 0 \\
 0 & -\frac{1}{2} & 0 & -\frac{1}{2} & 0 & 0 & 0 & 0 & 0 & 0 & -\frac{1}{2} & 0 & 0 & 0 & 0 & 0 & 0 & 0 & 0 & -1 & \frac{1}{2} & 0 & -\frac{1}{2} & 0 & 0 & 0 & 0 & 0 \\
 0 & -\frac{3}{2} & 0 & 0 & 0 & 0 & 0 & -2 & 2 & 0 & 1 & -4 & 0 & 0 & 0 & 0 & 0 & 0 & 0 & -2 & 1 & 0 & -1 & 0 & 0 & 0 & 0 & 0 \\
 0 & 0 & 0 & 0 & 0 & 0 & 0 & 0 & 0 & 0 & 0 & 0 & 2 & 0 & 0 & -2 & 0 & 0 & 0 & 0 & 0 & 0 & 0 & 0 & 0 & 0 & 0 & 0 \\
 0 & 0 & 0 & \frac{3}{2} & 2 & 0 & 0 & -1 & 0 & -2 & 2 & -2 & 0 & 0 & 0 & 0 & 0 & 0 & 0 & 2 & -1 & 0 & 1 & 0 & 0 & 0 & 0 & 0 \\
 0 & 0 & 0 & 0 & 0 & 0 & 0 & 0 & 0 & 0 & -1 & 0 & 0 & 0 & 0 & 0 & 0 & 0 & 0 & 0 & 0 & 0 & 0 & 0 & 0 & 0 & 0 & 0 \\
 0 & 0 & 0 & 0 & 0 & 0 & 0 & 0 & 0 & 0 & 0 & 0 & -2 & 0 & 0 & 0 & 0 & 0 & 0 & 0 & 0 & 0 & 0 & 0 & 0 & 0 & 0 & 0 \\
 0 & 0 & 0 & 0 & 0 & -1 & 0 & 0 & 0 & 0 & 0 & 0 & 0 & 0 & 0 & 0 & 0 & 0 & 0 & 0 & 0 & 0 & 0 & 0 & 0 & 0 & 0 & 0 \\
 0 & 0 & 0 & 0 & 0 & 0 & 0 & 0 & 0 & 0 & 0 & 0 & 0 & 0 & 0 & 0 & 0 & 0 & 0 & 0 & 0 & 0 & 0 & 0 & 0 & 0 & 0 & 0 \\
 0 & 0 & 0 & 0 & 0 & 0 & 0 & 0 & 0 & 0 & 0 & 0 & 0 & 0 & 0 & -1 & 0 & 0 & 0 & 0 & 0 & 0 & 0 & 0 & 0 & 0 & 0 & 0 \\
\end{array}
}
\right) \, , \nonumber \\
~ \nonumber \\
\mathbb{S}_5 &=& 
\left(
\scriptsize{
\begin{array}{cccccccccccccccccccccccccccc}
 0 & 0 & 0 & 0 & 0 & 0 & 0 & 0 & 0 & 0 & 0 & 0 & 0 & 0 & 0 & 0 & 0 & 0 & 0 & 0 & 0 & 0 & 0 & 0 & 0 & 0 & 0 & 0 \\
 0 & 0 & 0 & 0 & 0 & 0 & 0 & 0 & 0 & 0 & 0 & 0 & 0 & 0 & 0 & 0 & 0 & 0 & 0 & 0 & 0 & 0 & 0 & 0 & 0 & 0 & 0 & 0 \\
 0 & 0 & 0 & 0 & 0 & 0 & 0 & 0 & 0 & 0 & 0 & 0 & 0 & 0 & 0 & 0 & 0 & 0 & 0 & 0 & 0 & 0 & 0 & 0 & 0 & 0 & 0 & 0 \\
 0 & 0 & 0 & 2 & 0 & 0 & 0 & 0 & 0 & 0 & 0 & 0 & 0 & 0 & 0 & 0 & 0 & 0 & 0 & 0 & 0 & 0 & 0 & 0 & 0 & 0 & 0 & 0 \\
 0 & 0 & 0 & 0 & -2 & 0 & 0 & 0 & 0 & 0 & 0 & 0 & 0 & 0 & 0 & 0 & 0 & 0 & 0 & 0 & 0 & 0 & 0 & 0 & 0 & 0 & 0 & 0 \\
 0 & 0 & 0 & 0 & 0 & 0 & 0 & 0 & 0 & 0 & 0 & 0 & 0 & 0 & 0 & 0 & 0 & 0 & 0 & 0 & 0 & 0 & 0 & 0 & 0 & 0 & 0 & 0 \\
 0 & 0 & 0 & 0 & 0 & 0 & 0 & 0 & 0 & 0 & 0 & 0 & 0 & 0 & 0 & 0 & 0 & 0 & 0 & 0 & 0 & 0 & 0 & 0 & 0 & 0 & 0 & 0 \\
 0 & \frac{1}{2} & 0 & 0 & 0 & 0 & 0 & 1 & -2 & 0 & 0 & 0 & 0 & 0 & 0 & 0 & 0 & 0 & 0 & 0 & 0 & 0 & 0 & 0 & 0 & 0 & 0 & 0 \\
 0 & 0 & 0 & 0 & 0 & 0 & 0 & 0 & 0 & 0 & 0 & 0 & 0 & 0 & 0 & 0 & 0 & 0 & 0 & 0 & 0 & 0 & 0 & 0 & 0 & 0 & 0 & 0 \\
 0 & 0 & 0 & 0 & 0 & -1 & 0 & 0 & 0 & 1 & 0 & 0 & 0 & 0 & 0 & 0 & 0 & 0 & 0 & 0 & 0 & 0 & 0 & 0 & 0 & 0 & 0 & 0 \\
 0 & 0 & 0 & 2 & 0 & 0 & 0 & 0 & 0 & 0 & 0 & 0 & 0 & 0 & 0 & 0 & 0 & 0 & 0 & 0 & 0 & 0 & 0 & 0 & 0 & 0 & 0 & 0 \\
 0 & 0 & 0 & \frac{3}{2} & 0 & 0 & 0 & 0 & 0 & 0 & 0 & 0 & 0 & 0 & 0 & 0 & 0 & 0 & 0 & 0 & 0 & 0 & 0 & 0 & 0 & 0 & 0 & 0 \\
 0 & 0 & 0 & 0 & 0 & 0 & 0 & 0 & 0 & 0 & 0 & 0 & 0 & 0 & 0 & 0 & 0 & 0 & 0 & 0 & 0 & 0 & 0 & 0 & 0 & 0 & 0 & 0 \\
 0 & 0 & 0 & -2 & 0 & 0 & 0 & 0 & 0 & 0 & 0 & 0 & 0 & 0 & 0 & 0 & 0 & 0 & 0 & 0 & 0 & 0 & 0 & 0 & 0 & 0 & 0 & 0 \\
 0 & 0 & 0 & 0 & 0 & 0 & 0 & 0 & 0 & 0 & 0 & 0 & 0 & 0 & 0 & 0 & 0 & 0 & 0 & 0 & 0 & 0 & 0 & 0 & 0 & 0 & 0 & 0 \\
 0 & 0 & 0 & 0 & 0 & 0 & 0 & 0 & 0 & 0 & 0 & 0 & 0 & 0 & 0 & 0 & 0 & 0 & 0 & 0 & 0 & 0 & 0 & 0 & 0 & 0 & 0 & 0 \\
 0 & 0 & 0 & 0 & 0 & 0 & 0 & 0 & 0 & 0 & 0 & 0 & 0 & 0 & 0 & 0 & 0 & 0 & 0 & 0 & 0 & 0 & 0 & 0 & 0 & 0 & 0 & 0 \\
 0 & 0 & 0 & 0 & 0 & 0 & 0 & -1 & 0 & 0 & 0 & 0 & 0 & 1 & 0 & 0 & 0 & 0 & 0 & 0 & 0 & 0 & 0 & 0 & 0 & 0 & 0 & 0 \\
 0 & 0 & 0 & 0 & 0 & 0 & 0 & 0 & 0 & 2 & 0 & 0 & 0 & 0 & 0 & 0 & 0 & 0 & 0 & 0 & 0 & 0 & 0 & 0 & 0 & 0 & 0 & 0 \\
 0 & 1 & 0 & 1 & 0 & 0 & 0 & 0 & 0 & 0 & 1 & 0 & 0 & 0 & 0 & 0 & 0 & 0 & 0 & -2 & 0 & 0 & 1 & 0 & 0 & 0 & 0 & 0 \\
 0 & 0 & 0 & 0 & 0 & 0 & 0 & 0 & 0 & -2 & 0 & 0 & 0 & 0 & 0 & 0 & 0 & 0 & 0 & 0 & 0 & 0 & 0 & 0 & 0 & 0 & 0 & 0 \\
 0 & 0 & 0 & 0 & 0 & 0 & 0 & 0 & 0 & 0 & 0 & 0 & 0 & 0 & 0 & 0 & 0 & 0 & 0 & 0 & 0 & 0 & 0 & 0 & 0 & 0 & 0 & 0 \\
 0 & \frac{3}{2} & 0 & 0 & 0 & 0 & 0 & -1 & 2 & 0 & 2 & 0 & 0 & 0 & 0 & 0 & 0 & 0 & 0 & -4 & 0 & 0 & 2 & 0 & 0 & 0 & 0 & 0 \\
 0 & 0 & 0 & 0 & 0 & 0 & 0 & 0 & 0 & 0 & 0 & 0 & 0 & 0 & 0 & 0 & 0 & 0 & 0 & 0 & 0 & 0 & 0 & 0 & 0 & 0 & 0 & 0 \\
 0 & 0 & 0 & 0 & 0 & 0 & 0 & 0 & 0 & 0 & 0 & 0 & 0 & 0 & 0 & 0 & 0 & 0 & 0 & 0 & 0 & 0 & 0 & 0 & 0 & 0 & 0 & 0 \\
 0 & 0 & 0 & 0 & 0 & 0 & 0 & 0 & 0 & 0 & 0 & 0 & 0 & 0 & 0 & 0 & 0 & 0 & 0 & 0 & 0 & 0 & 0 & 0 & 0 & 0 & 0 & 0 \\
 0 & 0 & 0 & 0 & 0 & 0 & 0 & 0 & 0 & 0 & 0 & 0 & 0 & 0 & 0 & 0 & 0 & 0 & 0 & 0 & 0 & 0 & 0 & 0 & 0 & 0 & 0 & 0 \\
 0 & 0 & 0 & 0 & 0 & 0 & 0 & 0 & 0 & 0 & 0 & 0 & 0 & 0 & 0 & 0 & 0 & 0 & 0 & 0 & 0 & 0 & 0 & 0 & 0 & 0 & 0 & 0 \\
\end{array}
}
\right) \, , \nonumber \\
& & \nonumber \\
& & \nonumber \\
\mathbb{S}_6 &=& 
\left(
\scriptsize{
\begin{array}{cccccccccccccccccccccccccccc}
 0 & 0 & 0 & 0 & 0 & 0 & 0 & 0 & 0 & 0 & 0 & 0 & 0 & 0 & 0 & 0 & 0 & 0 & 0 & 0 & 0 & 0 & 0 & 0 & 0 & 0 & 0 & 0 \\
 0 & 0 & 0 & 0 & 0 & 0 & 0 & 0 & 0 & 0 & 0 & 0 & 0 & 0 & 0 & 0 & 0 & 0 & 0 & 0 & 0 & 0 & 0 & 0 & 0 & 0 & 0 & 0 \\
 0 & 0 & 0 & 0 & 0 & 0 & 0 & 0 & 0 & 0 & 0 & 0 & 0 & 0 & 0 & 0 & 0 & 0 & 0 & 0 & 0 & 0 & 0 & 0 & 0 & 0 & 0 & 0 \\
 0 & 0 & 0 & 0 & 0 & 0 & 0 & 0 & 0 & 0 & 0 & 0 & 0 & 0 & 0 & 0 & 0 & 0 & 0 & 0 & 0 & 0 & 0 & 0 & 0 & 0 & 0 & 0 \\
 0 & 0 & 0 & 0 & -6 & 0 & 0 & 0 & 0 & 0 & 0 & 0 & 0 & 0 & 0 & 0 & 0 & 0 & 0 & 0 & 0 & 0 & 0 & 0 & 0 & 0 & 0 & 0 \\
 0 & 0 & 0 & 0 & 0 & -2 & 0 & 0 & 0 & 0 & 0 & 0 & 0 & 0 & 0 & 0 & 0 & 0 & 0 & 0 & 0 & 0 & 0 & 0 & 0 & 0 & 0 & 0 \\
 0 & 0 & 0 & 0 & 0 & 0 & 0 & 0 & 0 & 0 & 0 & 0 & 0 & 0 & 0 & 0 & 0 & 0 & 0 & 0 & 0 & 0 & 0 & 0 & 0 & 0 & 0 & 0 \\
 0 & 0 & 0 & 0 & 0 & 0 & 0 & 0 & 0 & 0 & 0 & 0 & 0 & 0 & 0 & 0 & 0 & 0 & 0 & 0 & 0 & 0 & 0 & 0 & 0 & 0 & 0 & 0 \\
 0 & 0 & 0 & 0 & 0 & 0 & 0 & 0 & 0 & 0 & 0 & 0 & 0 & 0 & 0 & 0 & 0 & 0 & 0 & 0 & 0 & 0 & 0 & 0 & 0 & 0 & 0 & 0 \\
 0 & 0 & 0 & 0 & 0 & 0 & 0 & 0 & 0 & -2 & 0 & 0 & 0 & 0 & 0 & 0 & 0 & 0 & 0 & 0 & 0 & 0 & 0 & 0 & 0 & 0 & 0 & 0 \\
 0 & 0 & 0 & 0 & 0 & 0 & 0 & 0 & 0 & 0 & 0 & 0 & 0 & 0 & 0 & 0 & 0 & 0 & 0 & 0 & 0 & 0 & 0 & 0 & 0 & 0 & 0 & 0 \\
 0 & 0 & 0 & 0 & 0 & 0 & 0 & 0 & 0 & 0 & 0 & 0 & 0 & 0 & 0 & 0 & 0 & 0 & 0 & 0 & 0 & 0 & 0 & 0 & 0 & 0 & 0 & 0 \\
 0 & 0 & 0 & 0 & 0 & 0 & 0 & 0 & 0 & 0 & 0 & 0 & 0 & 0 & 0 & 0 & 0 & 0 & 0 & 0 & 0 & 0 & 0 & 0 & 0 & 0 & 0 & 0 \\
 0 & 0 & 0 & 0 & 0 & 0 & 0 & 0 & 0 & 0 & 0 & 0 & 0 & 0 & 0 & 0 & 0 & 0 & 0 & 0 & 0 & 0 & 0 & 0 & 0 & 0 & 0 & 0 \\
 0 & 0 & 0 & 0 & 0 & 0 & 0 & 0 & 0 & 0 & 0 & 0 & 0 & 0 & -4 & 0 & 0 & 0 & 0 & 0 & 0 & 0 & 0 & 0 & 0 & 0 & 0 & 0 \\
 0 & 0 & 0 & 0 & 0 & 0 & 0 & 0 & 0 & 0 & 0 & 0 & 0 & 0 & 0 & -2 & 0 & 0 & 0 & 0 & 0 & 0 & 0 & 0 & 0 & 0 & 0 & 0 \\
 0 & 0 & 0 & 0 & 0 & 0 & 0 & 0 & 0 & 0 & 0 & 0 & 0 & 0 & 0 & 0 & -2 & 0 & 0 & 0 & 0 & 0 & 0 & 0 & 0 & 0 & 0 & 0 \\
 0 & 0 & 0 & 0 & 0 & 0 & 0 & 0 & 0 & 0 & 0 & 0 & 0 & 0 & 0 & 0 & 0 & 0 & 0 & 0 & 0 & 0 & 0 & 0 & 0 & 0 & 0 & 0 \\
 0 & 0 & 0 & 0 & 0 & 0 & 0 & 0 & 0 & 0 & 0 & 0 & 0 & 0 & 0 & 0 & 0 & 0 & -2 & 0 & 0 & 0 & 0 & 0 & 0 & 0 & 0 & 0 \\
 0 & 0 & 0 & 0 & 0 & 0 & 0 & 0 & 0 & 0 & 0 & 0 & 0 & 0 & 0 & 0 & 0 & 0 & 0 & 0 & 0 & 0 & 0 & 0 & 0 & 0 & 0 & 0 \\
 0 & 0 & 0 & 0 & 0 & 0 & 0 & 0 & 0 & 0 & 0 & 0 & 0 & 0 & 0 & 0 & 0 & 0 & 0 & 0 & -2 & 0 & 0 & 0 & 0 & 0 & 0 & 0 \\
 0 & 0 & 0 & 0 & 0 & 0 & 0 & 0 & 0 & 0 & 0 & 0 & 0 & 0 & 0 & 0 & 0 & 0 & 0 & 0 & 0 & 0 & 0 & 0 & 0 & 0 & 0 & 0 \\
 0 & 0 & 0 & 0 & 0 & 0 & 0 & 0 & 0 & 0 & 0 & 0 & 0 & 0 & 0 & 0 & 0 & 0 & 0 & 0 & 0 & 0 & 0 & 0 & 0 & 0 & 0 & 0 \\
 0 & 0 & 0 & 0 & 0 & 0 & 0 & 0 & 0 & 0 & 0 & 0 & 0 & 0 & 0 & 0 & 0 & 0 & 0 & 0 & 0 & 0 & 0 & 0 & 0 & 0 & 0 & 0 \\
 0 & 0 & 0 & 0 & 0 & 0 & 0 & 0 & 0 & 0 & 0 & 0 & 0 & 0 & 0 & 0 & 0 & 0 & 0 & 0 & 0 & 0 & 0 & 0 & 0 & 0 & 0 & 0 \\
 0 & 0 & 0 & 0 & 0 & 0 & 0 & 0 & 0 & 0 & 0 & 0 & 0 & 0 & 0 & 0 & 0 & 0 & 0 & 0 & 0 & 0 & 0 & 0 & 0 & 0 & 0 & 0 \\
 0 & 0 & 0 & 0 & 0 & 0 & 0 & 0 & 0 & 0 & 0 & 0 & 0 & 0 & 0 & 0 & 0 & 0 & 0 & 0 & 0 & 0 & 0 & 0 & 0 & 0 & 0 & 0 \\
 0 & 0 & 0 & 0 & 0 & 0 & 0 & 0 & 0 & 0 & 0 & 0 & 0 & 0 & 0 & 0 & 0 & 0 & 0 & 0 & 0 & 0 & 0 & 0 & 0 & 0 & 0 & 0 \\
\end{array}
}
\right) \, , \nonumber \\
~ \nonumber \\
\mathbb{S}_7 &=& 
\left(
\scriptsize{
\begin{array}{cccccccccccccccccccccccccccc}
 0 & 0 & 0 & 0 & 0 & 0 & 0 & 0 & 0 & 0 & 0 & 0 & 0 & 0 & 0 & 0 & 0 & 0 & 0 & 0 & 0 & 0 & 0 & 0 & 0 & 0 & 0 & 0 \\
 0 & 0 & 0 & 0 & 0 & 0 & 0 & 0 & 0 & 0 & 0 & 0 & 0 & 0 & 0 & 0 & 0 & 0 & 0 & 0 & 0 & 0 & 0 & 0 & 0 & 0 & 0 & 0 \\
 0 & 0 & 0 & 0 & 0 & 0 & 0 & 0 & 0 & 0 & 0 & 0 & 0 & 0 & 0 & 0 & 0 & 0 & 0 & 0 & 0 & 0 & 0 & 0 & 0 & 0 & 0 & 0 \\
 0 & 0 & 0 & 0 & 0 & 0 & 0 & 0 & 0 & 0 & 0 & 0 & 0 & 0 & 0 & 0 & 0 & 0 & 0 & 0 & 0 & 0 & 0 & 0 & 0 & 0 & 0 & 0 \\
 0 & 0 & 0 & 0 & 0 & 0 & 0 & 0 & 0 & 0 & 0 & 0 & 0 & 0 & 0 & 0 & 0 & 0 & 0 & 0 & 0 & 0 & 0 & 0 & 0 & 0 & 0 & 0 \\
 0 & 0 & 0 & 0 & 0 & 0 & 0 & 0 & 0 & 0 & 0 & 0 & 0 & 0 & 0 & 0 & 0 & 0 & 0 & 0 & 0 & 0 & 0 & 0 & 0 & 0 & 0 & 0 \\
 0 & 0 & 0 & 0 & 0 & 0 & 0 & 0 & 0 & 0 & 0 & 0 & 0 & 0 & 0 & 0 & 0 & 0 & 0 & 0 & 0 & 0 & 0 & 0 & 0 & 0 & 0 & 0 \\
 0 & 0 & 0 & 0 & 0 & 0 & 0 & 2 & 0 & 0 & 0 & 0 & 0 & 0 & 0 & 0 & 0 & 0 & 0 & 0 & 0 & 0 & 0 & 0 & 0 & 0 & 0 & 0 \\
 0 & 0 & 0 & 0 & 0 & 0 & 0 & 3 & -2 & 0 & 0 & 0 & 0 & 0 & 0 & 0 & 0 & 0 & 0 & 0 & 0 & 0 & 0 & 0 & 0 & 0 & 0 & 0 \\
 0 & 0 & 1 & 0 & 0 & 1 & 0 & 0 & 0 & -2 & 0 & 0 & 0 & 0 & 0 & 0 & 0 & 0 & 0 & 0 & 0 & 0 & 0 & 0 & 0 & 0 & 0 & 0 \\
 0 & 0 & 0 & 0 & 0 & 0 & 0 & 0 & 0 & 0 & 2 & 0 & 0 & 0 & 0 & 0 & 0 & 0 & 0 & 0 & 0 & 0 & 0 & 0 & 0 & 0 & 0 & 0 \\
 0 & 0 & 0 & 0 & 0 & 0 & 0 & 0 & 0 & 0 & 3 & -2 & 0 & 0 & 0 & 0 & 0 & 0 & 0 & 0 & 0 & 0 & 0 & 0 & 0 & 0 & 0 & 0 \\
 0 & 0 & 0 & 0 & 1 & 0 & 1 & 0 & 0 & 0 & 0 & 0 & -2 & 0 & 0 & 0 & 0 & 0 & 0 & 0 & 0 & 0 & 0 & 0 & 0 & 0 & 0 & 0 \\
 0 & 0 & 0 & 0 & 0 & 0 & 0 & 0 & 0 & 0 & 0 & 0 & 0 & 0 & 0 & 0 & 0 & 0 & 0 & 0 & 0 & 0 & 0 & 0 & 0 & 0 & 0 & 0 \\
 0 & 0 & 0 & 0 & 0 & 0 & 0 & 0 & 0 & 0 & 0 & 0 & 0 & 0 & 0 & 0 & 0 & 0 & 0 & 0 & 0 & 0 & 0 & 0 & 0 & 0 & 0 & 0 \\
 0 & 0 & 0 & 0 & 0 & 0 & 0 & 0 & 0 & 0 & 0 & 0 & 0 & 0 & 0 & 0 & 0 & 0 & 0 & 0 & 0 & 0 & 0 & 0 & 0 & 0 & 0 & 0 \\
 0 & 0 & 0 & 0 & 0 & 0 & 0 & 0 & 0 & 0 & 0 & 0 & 0 & 0 & 0 & 0 & 0 & 0 & 0 & 0 & 0 & 0 & 0 & 0 & 0 & 0 & 0 & 0 \\
 0 & 0 & 0 & 0 & 0 & 0 & 0 & 1 & 0 & 0 & 0 & 0 & 0 & 0 & 0 & 0 & 0 & 0 & 0 & 0 & 0 & 0 & 0 & 0 & 0 & 0 & 0 & 0 \\
 0 & 0 & 0 & 0 & 0 & 0 & 0 & 0 & 0 & 0 & 0 & 0 & 0 & 0 & 0 & 0 & 0 & 0 & 0 & 0 & 0 & 0 & 0 & 0 & 0 & 0 & 0 & 0 \\
 0 & 0 & 0 & 0 & 0 & 0 & 0 & 0 & 0 & 0 & 0 & 0 & 0 & 0 & 0 & 0 & 0 & 0 & 0 & 2 & 0 & 0 & 0 & 0 & 0 & 0 & 0 & 0 \\
 0 & 0 & 0 & 0 & 0 & 0 & 0 & 0 & 0 & 0 & 0 & 0 & 0 & 0 & 0 & 0 & 0 & 0 & 0 & 0 & 0 & 2 & 0 & 0 & 0 & 0 & 0 & 0 \\
 0 & 0 & 0 & 0 & 0 & 0 & 0 & 0 & 0 & 0 & 0 & 0 & 0 & 0 & 0 & 0 & 0 & 0 & 0 & 0 & 2 & 0 & 0 & 0 & 0 & 0 & 0 & 0 \\
 0 & 0 & 0 & 0 & 0 & 0 & 0 & 0 & 0 & 0 & 0 & 0 & 0 & 0 & 0 & 4 & 0 & 0 & 0 & 0 & 0 & 0 & -2 & 0 & 0 & 0 & 0 & 0 \\
 0 & 0 & 0 & 0 & 0 & 0 & 0 & 0 & 0 & 0 & 1 & 0 & 0 & 0 & 0 & 0 & 0 & 0 & 0 & 0 & 0 & 0 & 0 & 0 & 0 & 0 & 0 & 0 \\
 0 & 0 & 0 & 0 & 0 & 0 & 0 & 0 & 0 & 0 & 0 & 0 & 0 & 0 & 0 & 0 & 0 & 0 & 0 & 0 & 0 & 0 & 0 & 0 & 0 & 0 & 0 & 0 \\
 0 & 0 & 0 & 0 & 0 & 0 & 0 & 0 & 0 & 0 & 0 & 0 & 0 & 0 & 0 & 0 & 0 & 0 & 0 & 0 & 0 & 0 & 0 & 0 & 0 & 0 & 0 & 0 \\
 0 & 0 & 0 & 0 & 0 & 0 & 0 & 0 & 0 & 0 & 0 & 0 & 0 & 0 & 0 & 0 & 0 & 0 & 0 & 0 & 0 & 0 & 0 & 0 & 0 & 0 & 0 & 0 \\
 0 & 0 & 0 & 0 & 0 & 0 & 0 & 0 & 0 & 0 & 0 & 0 & 0 & 0 & 0 & 0 & 0 & 0 & 0 & 0 & 0 & 0 & 0 & 0 & 0 & 0 & 0 & 0 \\
\end{array}
}
\right) \, , \nonumber \\
& & \nonumber \\
& & \nonumber \\
\mathbb{S}_8 &=& 
\left(
\scriptsize{
\begin{array}{cccccccccccccccccccccccccccc}
 0 & 0 & 0 & 0 & 0 & 0 & 0 & 0 & 0 & 0 & 0 & 0 & 0 & 0 & 0 & 0 & 0 & 0 & 0 & 0 & 0 & 0 & 0 & 0 & 0 & 0 & 0 & 0 \\
 0 & 0 & 0 & 0 & 0 & 0 & 0 & 0 & 0 & 0 & 0 & 0 & 0 & 0 & 0 & 0 & 0 & 0 & 0 & 0 & 0 & 0 & 0 & 0 & 0 & 0 & 0 & 0 \\
 0 & 0 & 0 & 0 & 0 & 0 & 0 & 0 & 0 & 0 & 0 & 0 & 0 & 0 & 0 & 0 & 0 & 0 & 0 & 0 & 0 & 0 & 0 & 0 & 0 & 0 & 0 & 0 \\
 0 & 0 & 0 & 0 & 0 & 0 & 0 & 0 & 0 & 0 & 0 & 0 & 0 & 0 & 0 & 0 & 0 & 0 & 0 & 0 & 0 & 0 & 0 & 0 & 0 & 0 & 0 & 0 \\
 0 & 0 & 0 & 0 & 0 & 0 & 0 & 0 & 0 & 0 & 0 & 0 & 0 & 0 & 0 & 0 & 0 & 0 & 0 & 0 & 0 & 0 & 0 & 0 & 0 & 0 & 0 & 0 \\
 0 & 0 & 0 & 0 & 0 & 0 & 0 & 0 & 0 & 0 & 0 & 0 & 0 & 0 & 0 & 0 & 0 & 0 & 0 & 0 & 0 & 0 & 0 & 0 & 0 & 0 & 0 & 0 \\
 0 & 0 & 0 & 0 & 0 & 0 & 0 & 0 & 0 & 0 & 0 & 0 & 0 & 0 & 0 & 0 & 0 & 0 & 0 & 0 & 0 & 0 & 0 & 0 & 0 & 0 & 0 & 0 \\
 0 & 0 & 0 & 0 & 0 & 0 & 0 & 2 & 0 & 0 & 0 & 0 & 0 & 0 & 0 & 0 & 0 & 0 & 0 & 0 & 0 & 0 & 0 & 0 & 0 & 0 & 0 & 0 \\
 0 & 0 & 0 & 0 & 0 & 0 & 0 & 3 & -2 & 0 & 0 & 0 & 0 & 0 & 0 & 0 & 0 & 0 & 0 & 0 & 0 & 0 & 0 & 0 & 0 & 0 & 0 & 0 \\
 0 & 0 & -1 & 0 & 0 & 1 & 0 & 0 & 0 & -2 & 0 & 0 & 0 & 0 & 0 & 0 & 0 & 0 & 0 & 0 & 0 & 0 & 0 & 0 & 0 & 0 & 0 & 0 \\
 0 & 0 & 0 & 0 & 0 & 0 & 0 & 0 & 0 & 0 & 2 & 0 & 0 & 0 & 0 & 0 & 0 & 0 & 0 & 0 & 0 & 0 & 0 & 0 & 0 & 0 & 0 & 0 \\
 0 & 0 & 0 & 0 & 0 & 0 & 0 & 0 & 0 & 0 & 3 & -2 & 0 & 0 & 0 & 0 & 0 & 0 & 0 & 0 & 0 & 0 & 0 & 0 & 0 & 0 & 0 & 0 \\
 0 & 0 & 0 & 0 & -1 & 0 & 1 & 0 & 0 & 0 & 0 & 0 & -2 & 0 & 0 & 0 & 0 & 0 & 0 & 0 & 0 & 0 & 0 & 0 & 0 & 0 & 0 & 0 \\
 0 & 0 & 0 & 0 & 0 & 0 & 0 & 0 & 0 & 0 & 0 & 0 & 0 & 0 & 0 & 0 & 0 & 0 & 0 & 0 & 0 & 0 & 0 & 0 & 0 & 0 & 0 & 0 \\
 0 & 0 & 0 & 0 & 0 & 0 & 0 & 0 & 0 & 0 & 0 & 0 & 0 & 0 & 0 & 0 & 0 & 0 & 0 & 0 & 0 & 0 & 0 & 0 & 0 & 0 & 0 & 0 \\
 0 & 0 & 0 & 0 & 0 & 0 & 0 & 0 & 0 & 0 & 0 & 0 & 0 & 0 & 0 & 0 & 0 & 0 & 0 & 0 & 0 & 0 & 0 & 0 & 0 & 0 & 0 & 0 \\
 0 & 0 & 0 & 0 & 0 & 0 & 0 & 0 & 0 & 0 & 0 & 0 & 0 & 0 & 0 & 0 & 0 & 0 & 0 & 0 & 0 & 0 & 0 & 0 & 0 & 0 & 0 & 0 \\
 0 & 0 & 0 & 0 & 0 & 0 & 0 & 1 & 0 & 0 & 0 & 0 & 0 & 0 & 0 & 0 & 0 & 0 & 0 & 0 & 0 & 0 & 0 & 0 & 0 & 0 & 0 & 0 \\
 0 & 0 & 0 & 0 & 0 & 0 & 0 & 0 & 0 & 0 & 0 & 0 & 0 & 0 & 0 & 0 & 0 & 0 & 0 & 0 & 0 & 0 & 0 & 0 & 0 & 0 & 0 & 0 \\
 0 & 0 & 0 & 0 & 0 & 0 & 0 & 0 & 0 & 0 & 0 & 0 & 0 & 0 & 0 & 0 & 0 & 0 & 0 & 2 & 0 & 0 & 0 & 0 & 0 & 0 & 0 & 0 \\
 0 & 0 & 0 & 0 & 0 & 0 & 0 & 0 & 0 & 0 & 0 & 0 & 0 & 0 & 0 & 0 & 0 & 0 & 0 & 0 & 0 & -2 & 0 & 0 & 0 & 0 & 0 & 0 \\
 0 & 0 & 0 & 0 & 0 & 0 & 0 & 0 & 0 & 0 & 0 & 0 & 0 & 0 & 0 & 0 & 0 & 0 & 0 & 0 & -2 & 0 & 0 & 0 & 0 & 0 & 0 & 0 \\
 0 & 0 & 0 & 0 & 0 & 0 & 0 & 0 & 0 & 0 & 0 & 0 & 0 & 0 & 0 & -4 & 0 & 0 & 0 & 0 & 0 & 0 & -2 & 0 & 0 & 0 & 0 & 0 \\
 0 & 0 & 0 & 0 & 0 & 0 & 0 & 0 & 0 & 0 & 1 & 0 & 0 & 0 & 0 & 0 & 0 & 0 & 0 & 0 & 0 & 0 & 0 & 0 & 0 & 0 & 0 & 0 \\
 0 & 0 & 0 & 0 & 0 & 0 & 0 & 0 & 0 & 0 & 0 & 0 & 0 & 0 & 0 & 0 & 0 & 0 & 0 & 0 & 0 & 0 & 0 & 0 & 0 & 0 & 0 & 0 \\
 0 & 0 & 0 & 0 & 0 & 0 & 0 & 0 & 0 & 0 & 0 & 0 & 0 & 0 & 0 & 0 & 0 & 0 & 0 & 0 & 0 & 0 & 0 & 0 & 0 & 0 & 0 & 0 \\
 0 & 0 & 0 & 0 & 0 & 0 & 0 & 0 & 0 & 0 & 0 & 0 & 0 & 0 & 0 & 0 & 0 & 0 & 0 & 0 & 0 & 0 & 0 & 0 & 0 & 0 & 0 & 0 \\
 0 & 0 & 0 & 0 & 0 & 0 & 0 & 0 & 0 & 0 & 0 & 0 & 0 & 0 & 0 & 0 & 0 & 0 & 0 & 0 & 0 & 0 & 0 & 0 & 0 & 0 & 0 & 0 \\
\end{array}
}
\right) \, , \nonumber \\
~ \nonumber \\
\mathbb{S}_9 &=& 
\left(
\scriptsize{
\begin{array}{cccccccccccccccccccccccccccc}
 0 & 0 & 0 & 0 & 0 & 0 & 0 & 0 & 0 & 0 & 0 & 0 & 0 & 0 & 0 & 0 & 0 & 0 & 0 & 0 & 0 & 0 & 0 & 0 & 0 & 0 & 0 & 0 \\
 0 & 0 & 0 & 0 & 0 & 0 & 0 & 0 & 0 & 0 & 0 & 0 & 0 & 0 & 0 & 0 & 0 & 0 & 0 & 0 & 0 & 0 & 0 & 0 & 0 & 0 & 0 & 0 \\
 0 & 0 & 0 & 0 & 0 & 0 & 0 & 0 & 0 & 0 & 0 & 0 & 0 & 0 & 0 & 0 & 0 & 0 & 0 & 0 & 0 & 0 & 0 & 0 & 0 & 0 & 0 & 0 \\
 0 & 0 & 0 & 0 & 0 & 0 & 0 & 0 & 0 & 0 & 0 & 0 & 0 & 0 & 0 & 0 & 0 & 0 & 0 & 0 & 0 & 0 & 0 & 0 & 0 & 0 & 0 & 0 \\
 0 & 0 & 0 & 0 & 0 & 0 & 0 & 0 & 0 & 0 & 0 & 0 & 0 & 0 & 0 & 0 & 0 & 0 & 0 & 0 & 0 & 0 & 0 & 0 & 0 & 0 & 0 & 0 \\
 0 & 0 & 0 & 0 & 0 & 0 & 0 & 0 & 0 & 0 & 0 & 0 & 0 & 0 & 0 & 0 & 0 & 0 & 0 & 0 & 0 & 0 & 0 & 0 & 0 & 0 & 0 & 0 \\
 0 & 0 & 0 & 0 & 0 & 0 & 0 & 0 & 0 & 0 & 0 & 0 & 0 & 0 & 0 & 0 & 0 & 0 & 0 & 0 & 0 & 0 & 0 & 0 & 0 & 0 & 0 & 0 \\
 0 & \frac{3}{4} & 0 & 0 & 0 & 0 & 0 & -\frac{3}{2} & 1 & -1 & 0 & 0 & 0 & 0 & 0 & 0 & 0 & 0 & 0 & 0 & 0 & 0 & 0 & 0 & 0 & 0 & 0 & 0 \\
 0 & \frac{3}{4} & 0 & 0 & 0 & 0 & 0 & -\frac{3}{2} & 1 & -1 & 0 & 0 & 0 & 0 & 0 & 0 & 0 & 0 & 0 & 0 & 0 & 0 & 0 & 0 & 0 & 0 & 0 & 0 \\
 0 & -\frac{3}{8} & 0 & 0 & 0 & 0 & 0 & \frac{3}{4} & -\frac{1}{2} & \frac{1}{2} & 0 & 0 & 0 & 0 & 0 & 0 & 0 & 0 & 0 & 0 & 0 & 0 & 0 & 0 & 0 & 0 & 0 & 0 \\
 0 & 0 & 0 & 0 & 0 & 0 & 0 & 0 & 0 & 0 & 0 & 0 & 0 & 0 & 0 & 0 & 0 & 0 & 0 & 0 & 0 & 0 & 0 & 0 & 0 & 0 & 0 & 0 \\
 0 & 0 & 0 & 0 & 0 & 0 & 0 & 0 & 0 & 0 & 0 & 0 & 0 & 0 & 0 & 0 & 0 & 0 & 0 & 0 & 0 & 0 & 0 & 0 & 0 & 0 & 0 & 0 \\
 0 & 0 & 0 & 0 & 0 & 0 & 0 & 0 & 0 & 0 & 0 & 0 & 0 & 0 & 0 & 0 & 0 & 0 & 0 & 0 & 0 & 0 & 0 & 0 & 0 & 0 & 0 & 0 \\
 0 & 0 & 0 & 0 & 0 & 0 & 0 & 0 & 0 & 0 & 0 & 0 & 0 & 0 & 0 & 0 & 0 & 0 & 0 & 0 & 0 & 0 & 0 & 0 & 0 & 0 & 0 & 0 \\
 0 & 0 & 0 & 0 & 0 & 0 & 0 & 0 & 0 & 0 & 0 & 0 & 0 & 0 & 0 & 0 & 0 & 0 & 0 & 0 & 0 & 0 & 0 & 0 & 0 & 0 & 0 & 0 \\
 0 & 0 & 0 & 0 & 0 & 0 & 0 & 0 & 0 & 0 & 0 & 0 & 0 & 0 & 0 & 0 & 0 & 0 & 0 & 0 & 0 & 0 & 0 & 0 & 0 & 0 & 0 & 0 \\
 0 & 0 & 0 & 0 & 0 & 0 & 0 & 0 & 0 & 0 & 0 & 0 & 0 & 0 & 0 & 0 & 0 & 0 & 0 & 0 & 0 & 0 & 0 & 0 & 0 & 0 & 0 & 0 \\
 0 & 0 & 0 & 0 & 0 & 0 & 0 & 0 & 0 & 0 & 0 & 0 & 0 & 0 & 0 & 0 & 0 & 0 & 0 & 0 & 0 & 0 & 0 & 0 & 0 & 0 & 0 & 0 \\
 0 & \frac{3}{4} & 0 & 0 & 0 & 0 & 0 & -\frac{3}{2} & 1 & -1 & 0 & 0 & 0 & 0 & 0 & 0 & 0 & 0 & 0 & 0 & 0 & 0 & 0 & 0 & 0 & 0 & 0 & 0 \\
 0 & 0 & 0 & 0 & 0 & 0 & 0 & 0 & 0 & 0 & 0 & 0 & 0 & 0 & 0 & 0 & 0 & 0 & 0 & 0 & 0 & 0 & 0 & 0 & 0 & 0 & 0 & 0 \\
 0 & -\frac{3}{4} & 0 & 0 & 0 & 0 & 0 & \frac{3}{2} & -1 & 1 & 0 & 0 & 0 & 0 & 0 & 0 & 0 & 0 & 0 & 0 & 0 & 0 & 0 & 0 & 0 & 0 & 0 & 0 \\
 0 & 0 & 0 & 0 & 0 & 0 & 0 & 0 & 0 & 0 & 0 & 0 & 0 & 0 & 0 & 0 & 0 & 0 & 0 & 0 & 0 & 0 & 0 & 0 & 0 & 0 & 0 & 0 \\
 0 & -\frac{3}{4} & 0 & 0 & 0 & 0 & 0 & \frac{3}{2} & -1 & 1 & 0 & 0 & 0 & 0 & 0 & 0 & 0 & 0 & 0 & 0 & 0 & 0 & 0 & 0 & 0 & 0 & 0 & 0 \\
 0 & 0 & 0 & 0 & 0 & 0 & 0 & 0 & 0 & 0 & 0 & 0 & 0 & 0 & 0 & 0 & 0 & 0 & 0 & 0 & 0 & 0 & 0 & 0 & 0 & 0 & 0 & 0 \\
 0 & 0 & 0 & 0 & 0 & 0 & 0 & 0 & 0 & 0 & 0 & 0 & 0 & 0 & 0 & 0 & 0 & 0 & 0 & 0 & 0 & 0 & 0 & 0 & 0 & 0 & 0 & 0 \\
 0 & 0 & 0 & 0 & 0 & 0 & 0 & 0 & 0 & 0 & 0 & 0 & 0 & 0 & 0 & 0 & 0 & 0 & 0 & 0 & 0 & 0 & 0 & 0 & 0 & 0 & 0 & 0 \\
 0 & 0 & 0 & 0 & 0 & 0 & 0 & 0 & 0 & 0 & 0 & 0 & 0 & 0 & 0 & 0 & 0 & 0 & 0 & 0 & 0 & 0 & 0 & 0 & 0 & 0 & 0 & 0 \\
 0 & 0 & 0 & 0 & 0 & 0 & 0 & 0 & 0 & 0 & 0 & 0 & 0 & 0 & 0 & 0 & 0 & 0 & 0 & 0 & 0 & 0 & 0 & 0 & 0 & 0 & 0 & 0 \\
\end{array}
}
\right) \, , \nonumber \\
& & \nonumber \\
& & \nonumber \\
\mathbb{S}_{10} &=& 
\left(
\scriptsize{
\begin{array}{cccccccccccccccccccccccccccc}
 0 & 0 & 0 & 0 & 0 & 0 & 0 & 0 & 0 & 0 & 0 & 0 & 0 & 0 & 0 & 0 & 0 & 0 & 0 & 0 & 0 & 0 & 0 & 0 & 0 & 0 & 0 & 0 \\
 0 & 0 & 0 & 0 & 0 & 0 & 0 & 0 & 0 & 0 & 0 & 0 & 0 & 0 & 0 & 0 & 0 & 0 & 0 & 0 & 0 & 0 & 0 & 0 & 0 & 0 & 0 & 0 \\
 0 & 0 & 0 & 0 & 0 & 0 & 0 & 0 & 0 & 0 & 0 & 0 & 0 & 0 & 0 & 0 & 0 & 0 & 0 & 0 & 0 & 0 & 0 & 0 & 0 & 0 & 0 & 0 \\
 0 & 0 & 0 & 0 & 0 & 0 & 0 & 0 & 0 & 0 & 0 & 0 & 0 & 0 & 0 & 0 & 0 & 0 & 0 & 0 & 0 & 0 & 0 & 0 & 0 & 0 & 0 & 0 \\
 0 & 0 & 0 & 0 & 0 & 0 & 0 & 0 & 0 & 0 & 0 & 0 & 0 & 0 & 0 & 0 & 0 & 0 & 0 & 0 & 0 & 0 & 0 & 0 & 0 & 0 & 0 & 0 \\
 0 & 0 & 0 & 0 & 0 & 0 & 0 & 0 & 0 & 0 & 0 & 0 & 0 & 0 & 0 & 0 & 0 & 0 & 0 & 0 & 0 & 0 & 0 & 0 & 0 & 0 & 0 & 0 \\
 0 & 0 & 0 & 0 & 0 & 0 & 0 & 0 & 0 & 0 & 0 & 0 & 0 & 0 & 0 & 0 & 0 & 0 & 0 & 0 & 0 & 0 & 0 & 0 & 0 & 0 & 0 & 0 \\
 0 & 0 & 0 & 0 & 0 & 0 & 0 & 0 & 0 & 0 & 0 & 0 & 0 & 0 & 0 & 0 & 0 & 0 & 0 & 0 & 0 & 0 & 0 & 0 & 0 & 0 & 0 & 0 \\
 0 & 0 & 0 & 0 & 0 & 0 & 0 & 0 & 0 & 0 & 0 & 0 & 0 & 0 & 0 & 0 & 0 & 0 & 0 & 0 & 0 & 0 & 0 & 0 & 0 & 0 & 0 & 0 \\
 0 & 0 & 0 & 0 & 0 & 0 & 0 & 0 & 0 & 0 & 0 & 0 & 0 & 0 & 0 & 0 & 0 & 0 & 0 & 0 & 0 & 0 & 0 & 0 & 0 & 0 & 0 & 0 \\
 0 & 0 & 0 & -\frac{3}{4} & 0 & 0 & 0 & 0 & 0 & 0 & -\frac{3}{2} & 1 & 1 & 0 & 0 & 0 & 0 & 0 & 0 & 0 & 0 & 0 & 0 & 0 & 0 & 0 & 0 & 0 \\
 0 & 0 & 0 & -\frac{3}{4} & 0 & 0 & 0 & 0 & 0 & 0 & -\frac{3}{2} & 1 & 1 & 0 & 0 & 0 & 0 & 0 & 0 & 0 & 0 & 0 & 0 & 0 & 0 & 0 & 0 & 0 \\
 0 & 0 & 0 & -\frac{3}{8} & 0 & 0 & 0 & 0 & 0 & 0 & -\frac{3}{4} & \frac{1}{2} & \frac{1}{2} & 0 & 0 & 0 & 0 & 0 & 0 & 0 & 0 & 0 & 0 & 0 & 0 & 0 & 0 & 0 \\
 0 & 0 & 0 & 0 & 0 & 0 & 0 & 0 & 0 & 0 & 0 & 0 & 0 & 0 & 0 & 0 & 0 & 0 & 0 & 0 & 0 & 0 & 0 & 0 & 0 & 0 & 0 & 0 \\
 0 & 0 & 0 & 0 & 0 & 0 & 0 & 0 & 0 & 0 & 0 & 0 & 0 & 0 & 0 & 0 & 0 & 0 & 0 & 0 & 0 & 0 & 0 & 0 & 0 & 0 & 0 & 0 \\
 0 & 0 & 0 & 0 & 0 & 0 & 0 & 0 & 0 & 0 & 0 & 0 & 0 & 0 & 0 & 0 & 0 & 0 & 0 & 0 & 0 & 0 & 0 & 0 & 0 & 0 & 0 & 0 \\
 0 & 0 & 0 & 0 & 0 & 0 & 0 & 0 & 0 & 0 & 0 & 0 & 0 & 0 & 0 & 0 & 0 & 0 & 0 & 0 & 0 & 0 & 0 & 0 & 0 & 0 & 0 & 0 \\
 0 & 0 & 0 & 0 & 0 & 0 & 0 & 0 & 0 & 0 & 0 & 0 & 0 & 0 & 0 & 0 & 0 & 0 & 0 & 0 & 0 & 0 & 0 & 0 & 0 & 0 & 0 & 0 \\
 0 & 0 & 0 & 0 & 0 & 0 & 0 & 0 & 0 & 0 & 0 & 0 & 0 & 0 & 0 & 0 & 0 & 0 & 0 & 0 & 0 & 0 & 0 & 0 & 0 & 0 & 0 & 0 \\
 0 & 0 & 0 & 0 & 0 & 0 & 0 & 0 & 0 & 0 & 0 & 0 & 0 & 0 & 0 & 0 & 0 & 0 & 0 & 0 & 0 & 0 & 0 & 0 & 0 & 0 & 0 & 0 \\
 0 & 0 & 0 & 0 & 0 & 0 & 0 & 0 & 0 & 0 & 0 & 0 & 0 & 0 & 0 & 0 & 0 & 0 & 0 & 0 & 0 & 0 & 0 & 0 & 0 & 0 & 0 & 0 \\
 0 & 0 & 0 & \frac{3}{4} & 0 & 0 & 0 & 0 & 0 & 0 & \frac{3}{2} & -1 & -1 & 0 & 0 & 0 & 0 & 0 & 0 & 0 & 0 & 0 & 0 & 0 & 0 & 0 & 0 & 0 \\
 0 & 0 & 0 & -\frac{3}{4} & 0 & 0 & 0 & 0 & 0 & 0 & -\frac{3}{2} & 1 & 1 & 0 & 0 & 0 & 0 & 0 & 0 & 0 & 0 & 0 & 0 & 0 & 0 & 0 & 0 & 0 \\
 0 & 0 & 0 & 0 & 0 & 0 & 0 & 0 & 0 & 0 & 0 & 0 & 0 & 0 & 0 & 0 & 0 & 0 & 0 & 0 & 0 & 0 & 0 & 0 & 0 & 0 & 0 & 0 \\
 0 & 0 & 0 & -\frac{3}{4} & 0 & 0 & 0 & 0 & 0 & 0 & -\frac{3}{2} & 1 & 1 & 0 & 0 & 0 & 0 & 0 & 0 & 0 & 0 & 0 & 0 & 0 & 0 & 0 & 0 & 0 \\
 0 & 0 & 0 & 0 & 0 & 0 & 0 & 0 & 0 & 0 & 0 & 0 & 0 & 0 & 0 & 0 & 0 & 0 & 0 & 0 & 0 & 0 & 0 & 0 & 0 & 0 & 0 & 0 \\
 0 & 0 & 0 & 0 & 0 & 0 & 0 & 0 & 0 & 0 & 0 & 0 & 0 & 0 & 0 & 0 & 0 & 0 & 0 & 0 & 0 & 0 & 0 & 0 & 0 & 0 & 0 & 0 \\
 0 & 0 & 0 & 0 & 0 & 0 & 0 & 0 & 0 & 0 & 0 & 0 & 0 & 0 & 0 & 0 & 0 & 0 & 0 & 0 & 0 & 0 & 0 & 0 & 0 & 0 & 0 & 0 \\
\end{array}
}
\right) \, , \nonumber \\
~ \nonumber \\
\mathbb{S}_{11} &=& 
\left(
\scriptsize{
\begin{array}{cccccccccccccccccccccccccccc}
 0 & 0 & 0 & 0 & 0 & 0 & 0 & 0 & 0 & 0 & 0 & 0 & 0 & 0 & 0 & 0 & 0 & 0 & 0 & 0 & 0 & 0 & 0 & 0 & 0 & 0 & 0 & 0 \\
 0 & 0 & 0 & 0 & 0 & 0 & 0 & 0 & 0 & 0 & 0 & 0 & 0 & 0 & 0 & 0 & 0 & 0 & 0 & 0 & 0 & 0 & 0 & 0 & 0 & 0 & 0 & 0 \\
 0 & 0 & 0 & 0 & 0 & 0 & 0 & 0 & 0 & 0 & 0 & 0 & 0 & 0 & 0 & 0 & 0 & 0 & 0 & 0 & 0 & 0 & 0 & 0 & 0 & 0 & 0 & 0 \\
 0 & 0 & 0 & 0 & 0 & 0 & 0 & 0 & 0 & 0 & 0 & 0 & 0 & 0 & 0 & 0 & 0 & 0 & 0 & 0 & 0 & 0 & 0 & 0 & 0 & 0 & 0 & 0 \\
 0 & 0 & 0 & 0 & 0 & 0 & 0 & 0 & 0 & 0 & 0 & 0 & 0 & 0 & 0 & 0 & 0 & 0 & 0 & 0 & 0 & 0 & 0 & 0 & 0 & 0 & 0 & 0 \\
 0 & 0 & 0 & 0 & 0 & 0 & 0 & 0 & 0 & 0 & 0 & 0 & 0 & 0 & 0 & 0 & 0 & 0 & 0 & 0 & 0 & 0 & 0 & 0 & 0 & 0 & 0 & 0 \\
 0 & 0 & 0 & 0 & 0 & 0 & 0 & 0 & 0 & 0 & 0 & 0 & 0 & 0 & 0 & 0 & 0 & 0 & 0 & 0 & 0 & 0 & 0 & 0 & 0 & 0 & 0 & 0 \\
 0 & \frac{3}{4} & 0 & 0 & 0 & 0 & 0 & -\frac{3}{2} & 1 & 1 & 0 & 0 & 0 & 0 & 0 & 0 & 0 & 0 & 0 & 0 & 0 & 0 & 0 & 0 & 0 & 0 & 0 & 0 \\
 0 & \frac{3}{4} & 0 & 0 & 0 & 0 & 0 & -\frac{3}{2} & 1 & 1 & 0 & 0 & 0 & 0 & 0 & 0 & 0 & 0 & 0 & 0 & 0 & 0 & 0 & 0 & 0 & 0 & 0 & 0 \\
 0 & \frac{3}{8} & 0 & 0 & 0 & 0 & 0 & -\frac{3}{4} & \frac{1}{2} & \frac{1}{2} & 0 & 0 & 0 & 0 & 0 & 0 & 0 & 0 & 0 & 0 & 0 & 0 & 0 & 0 & 0 & 0 & 0 & 0 \\
 0 & 0 & 0 & 0 & 0 & 0 & 0 & 0 & 0 & 0 & 0 & 0 & 0 & 0 & 0 & 0 & 0 & 0 & 0 & 0 & 0 & 0 & 0 & 0 & 0 & 0 & 0 & 0 \\
 0 & 0 & 0 & 0 & 0 & 0 & 0 & 0 & 0 & 0 & 0 & 0 & 0 & 0 & 0 & 0 & 0 & 0 & 0 & 0 & 0 & 0 & 0 & 0 & 0 & 0 & 0 & 0 \\
 0 & 0 & 0 & 0 & 0 & 0 & 0 & 0 & 0 & 0 & 0 & 0 & 0 & 0 & 0 & 0 & 0 & 0 & 0 & 0 & 0 & 0 & 0 & 0 & 0 & 0 & 0 & 0 \\
 0 & 0 & 0 & 0 & 0 & 0 & 0 & 0 & 0 & 0 & 0 & 0 & 0 & 0 & 0 & 0 & 0 & 0 & 0 & 0 & 0 & 0 & 0 & 0 & 0 & 0 & 0 & 0 \\
 0 & 0 & 0 & 0 & 0 & 0 & 0 & 0 & 0 & 0 & 0 & 0 & 0 & 0 & 0 & 0 & 0 & 0 & 0 & 0 & 0 & 0 & 0 & 0 & 0 & 0 & 0 & 0 \\
 0 & 0 & 0 & 0 & 0 & 0 & 0 & 0 & 0 & 0 & 0 & 0 & 0 & 0 & 0 & 0 & 0 & 0 & 0 & 0 & 0 & 0 & 0 & 0 & 0 & 0 & 0 & 0 \\
 0 & 0 & 0 & 0 & 0 & 0 & 0 & 0 & 0 & 0 & 0 & 0 & 0 & 0 & 0 & 0 & 0 & 0 & 0 & 0 & 0 & 0 & 0 & 0 & 0 & 0 & 0 & 0 \\
 0 & 0 & 0 & 0 & 0 & 0 & 0 & 0 & 0 & 0 & 0 & 0 & 0 & 0 & 0 & 0 & 0 & 0 & 0 & 0 & 0 & 0 & 0 & 0 & 0 & 0 & 0 & 0 \\
 0 & -\frac{3}{4} & 0 & 0 & 0 & 0 & 0 & \frac{3}{2} & -1 & -1 & 0 & 0 & 0 & 0 & 0 & 0 & 0 & 0 & 0 & 0 & 0 & 0 & 0 & 0 & 0 & 0 & 0 & 0 \\
 0 & 0 & 0 & 0 & 0 & 0 & 0 & 0 & 0 & 0 & 0 & 0 & 0 & 0 & 0 & 0 & 0 & 0 & 0 & 0 & 0 & 0 & 0 & 0 & 0 & 0 & 0 & 0 \\
 0 & \frac{3}{4} & 0 & 0 & 0 & 0 & 0 & -\frac{3}{2} & 1 & 1 & 0 & 0 & 0 & 0 & 0 & 0 & 0 & 0 & 0 & 0 & 0 & 0 & 0 & 0 & 0 & 0 & 0 & 0 \\
 0 & 0 & 0 & 0 & 0 & 0 & 0 & 0 & 0 & 0 & 0 & 0 & 0 & 0 & 0 & 0 & 0 & 0 & 0 & 0 & 0 & 0 & 0 & 0 & 0 & 0 & 0 & 0 \\
 0 & -\frac{3}{4} & 0 & 0 & 0 & 0 & 0 & \frac{3}{2} & -1 & -1 & 0 & 0 & 0 & 0 & 0 & 0 & 0 & 0 & 0 & 0 & 0 & 0 & 0 & 0 & 0 & 0 & 0 & 0 \\
 0 & 0 & 0 & 0 & 0 & 0 & 0 & 0 & 0 & 0 & 0 & 0 & 0 & 0 & 0 & 0 & 0 & 0 & 0 & 0 & 0 & 0 & 0 & 0 & 0 & 0 & 0 & 0 \\
 0 & 0 & 0 & 0 & 0 & 0 & 0 & 0 & 0 & 0 & 0 & 0 & 0 & 0 & 0 & 0 & 0 & 0 & 0 & 0 & 0 & 0 & 0 & 0 & 0 & 0 & 0 & 0 \\
 0 & 0 & 0 & 0 & 0 & 0 & 0 & 0 & 0 & 0 & 0 & 0 & 0 & 0 & 0 & 0 & 0 & 0 & 0 & 0 & 0 & 0 & 0 & 0 & 0 & 0 & 0 & 0 \\
 0 & 0 & 0 & 0 & 0 & 0 & 0 & 0 & 0 & 0 & 0 & 0 & 0 & 0 & 0 & 0 & 0 & 0 & 0 & 0 & 0 & 0 & 0 & 0 & 0 & 0 & 0 & 0 \\
 0 & 0 & 0 & 0 & 0 & 0 & 0 & 0 & 0 & 0 & 0 & 0 & 0 & 0 & 0 & 0 & 0 & 0 & 0 & 0 & 0 & 0 & 0 & 0 & 0 & 0 & 0 & 0 \\
\end{array}
}
\right) \, , \nonumber \\
& & \nonumber \\
& & \nonumber \\
\mathbb{S}_{12} &=& 
\left(
\scriptsize{
\begin{array}{cccccccccccccccccccccccccccc}
 0 & 0 & 0 & 0 & 0 & 0 & 0 & 0 & 0 & 0 & 0 & 0 & 0 & 0 & 0 & 0 & 0 & 0 & 0 & 0 & 0 & 0 & 0 & 0 & 0 & 0 & 0 & 0 \\
 0 & 0 & 0 & 0 & 0 & 0 & 0 & 0 & 0 & 0 & 0 & 0 & 0 & 0 & 0 & 0 & 0 & 0 & 0 & 0 & 0 & 0 & 0 & 0 & 0 & 0 & 0 & 0 \\
 0 & 0 & 0 & 0 & 0 & 0 & 0 & 0 & 0 & 0 & 0 & 0 & 0 & 0 & 0 & 0 & 0 & 0 & 0 & 0 & 0 & 0 & 0 & 0 & 0 & 0 & 0 & 0 \\
 0 & 0 & 0 & 0 & 0 & 0 & 0 & 0 & 0 & 0 & 0 & 0 & 0 & 0 & 0 & 0 & 0 & 0 & 0 & 0 & 0 & 0 & 0 & 0 & 0 & 0 & 0 & 0 \\
 0 & 0 & 0 & 0 & 0 & 0 & 0 & 0 & 0 & 0 & 0 & 0 & 0 & 0 & 0 & 0 & 0 & 0 & 0 & 0 & 0 & 0 & 0 & 0 & 0 & 0 & 0 & 0 \\
 0 & 0 & 0 & 0 & 0 & 0 & 0 & 0 & 0 & 0 & 0 & 0 & 0 & 0 & 0 & 0 & 0 & 0 & 0 & 0 & 0 & 0 & 0 & 0 & 0 & 0 & 0 & 0 \\
 0 & 0 & 0 & 0 & 0 & 0 & 0 & 0 & 0 & 0 & 0 & 0 & 0 & 0 & 0 & 0 & 0 & 0 & 0 & 0 & 0 & 0 & 0 & 0 & 0 & 0 & 0 & 0 \\
 0 & 0 & 0 & 0 & 0 & 0 & 0 & 0 & 0 & 0 & 0 & 0 & 0 & 0 & 0 & 0 & 0 & 0 & 0 & 0 & 0 & 0 & 0 & 0 & 0 & 0 & 0 & 0 \\
 0 & 0 & 0 & 0 & 0 & 0 & 0 & 0 & 0 & 0 & 0 & 0 & 0 & 0 & 0 & 0 & 0 & 0 & 0 & 0 & 0 & 0 & 0 & 0 & 0 & 0 & 0 & 0 \\
 0 & 0 & 0 & 0 & 0 & 0 & 0 & 0 & 0 & 0 & 0 & 0 & 0 & 0 & 0 & 0 & 0 & 0 & 0 & 0 & 0 & 0 & 0 & 0 & 0 & 0 & 0 & 0 \\
 0 & 0 & 0 & -\frac{3}{4} & 0 & 0 & 0 & 0 & 0 & 0 & -\frac{3}{2} & 1 & -1 & 0 & 0 & 0 & 0 & 0 & 0 & 0 & 0 & 0 & 0 & 0 & 0 & 0 & 0 & 0 \\
 0 & 0 & 0 & -\frac{3}{4} & 0 & 0 & 0 & 0 & 0 & 0 & -\frac{3}{2} & 1 & -1 & 0 & 0 & 0 & 0 & 0 & 0 & 0 & 0 & 0 & 0 & 0 & 0 & 0 & 0 & 0 \\
 0 & 0 & 0 & \frac{3}{8} & 0 & 0 & 0 & 0 & 0 & 0 & \frac{3}{4} & -\frac{1}{2} & \frac{1}{2} & 0 & 0 & 0 & 0 & 0 & 0 & 0 & 0 & 0 & 0 & 0 & 0 & 0 & 0 & 0 \\
 0 & 0 & 0 & 0 & 0 & 0 & 0 & 0 & 0 & 0 & 0 & 0 & 0 & 0 & 0 & 0 & 0 & 0 & 0 & 0 & 0 & 0 & 0 & 0 & 0 & 0 & 0 & 0 \\
 0 & 0 & 0 & 0 & 0 & 0 & 0 & 0 & 0 & 0 & 0 & 0 & 0 & 0 & 0 & 0 & 0 & 0 & 0 & 0 & 0 & 0 & 0 & 0 & 0 & 0 & 0 & 0 \\
 0 & 0 & 0 & 0 & 0 & 0 & 0 & 0 & 0 & 0 & 0 & 0 & 0 & 0 & 0 & 0 & 0 & 0 & 0 & 0 & 0 & 0 & 0 & 0 & 0 & 0 & 0 & 0 \\
 0 & 0 & 0 & 0 & 0 & 0 & 0 & 0 & 0 & 0 & 0 & 0 & 0 & 0 & 0 & 0 & 0 & 0 & 0 & 0 & 0 & 0 & 0 & 0 & 0 & 0 & 0 & 0 \\
 0 & 0 & 0 & 0 & 0 & 0 & 0 & 0 & 0 & 0 & 0 & 0 & 0 & 0 & 0 & 0 & 0 & 0 & 0 & 0 & 0 & 0 & 0 & 0 & 0 & 0 & 0 & 0 \\
 0 & 0 & 0 & 0 & 0 & 0 & 0 & 0 & 0 & 0 & 0 & 0 & 0 & 0 & 0 & 0 & 0 & 0 & 0 & 0 & 0 & 0 & 0 & 0 & 0 & 0 & 0 & 0 \\
 0 & 0 & 0 & 0 & 0 & 0 & 0 & 0 & 0 & 0 & 0 & 0 & 0 & 0 & 0 & 0 & 0 & 0 & 0 & 0 & 0 & 0 & 0 & 0 & 0 & 0 & 0 & 0 \\
 0 & 0 & 0 & 0 & 0 & 0 & 0 & 0 & 0 & 0 & 0 & 0 & 0 & 0 & 0 & 0 & 0 & 0 & 0 & 0 & 0 & 0 & 0 & 0 & 0 & 0 & 0 & 0 \\
 0 & 0 & 0 & -\frac{3}{4} & 0 & 0 & 0 & 0 & 0 & 0 & -\frac{3}{2} & 1 & -1 & 0 & 0 & 0 & 0 & 0 & 0 & 0 & 0 & 0 & 0 & 0 & 0 & 0 & 0 & 0 \\
 0 & 0 & 0 & -\frac{3}{4} & 0 & 0 & 0 & 0 & 0 & 0 & -\frac{3}{2} & 1 & -1 & 0 & 0 & 0 & 0 & 0 & 0 & 0 & 0 & 0 & 0 & 0 & 0 & 0 & 0 & 0 \\
 0 & 0 & 0 & 0 & 0 & 0 & 0 & 0 & 0 & 0 & 0 & 0 & 0 & 0 & 0 & 0 & 0 & 0 & 0 & 0 & 0 & 0 & 0 & 0 & 0 & 0 & 0 & 0 \\
 0 & 0 & 0 & \frac{3}{4} & 0 & 0 & 0 & 0 & 0 & 0 & \frac{3}{2} & -1 & 1 & 0 & 0 & 0 & 0 & 0 & 0 & 0 & 0 & 0 & 0 & 0 & 0 & 0 & 0 & 0 \\
 0 & 0 & 0 & 0 & 0 & 0 & 0 & 0 & 0 & 0 & 0 & 0 & 0 & 0 & 0 & 0 & 0 & 0 & 0 & 0 & 0 & 0 & 0 & 0 & 0 & 0 & 0 & 0 \\
 0 & 0 & 0 & 0 & 0 & 0 & 0 & 0 & 0 & 0 & 0 & 0 & 0 & 0 & 0 & 0 & 0 & 0 & 0 & 0 & 0 & 0 & 0 & 0 & 0 & 0 & 0 & 0 \\
 0 & 0 & 0 & 0 & 0 & 0 & 0 & 0 & 0 & 0 & 0 & 0 & 0 & 0 & 0 & 0 & 0 & 0 & 0 & 0 & 0 & 0 & 0 & 0 & 0 & 0 & 0 & 0 \\
\end{array}
}
\right) \, . \nonumber
\eea

\section{Solution of the differential equations in terms of Li functions}
\label{sec:appd}
By definition the solution of the differential equations can be valid only in a region were the differential equations are non-singular. When a singular point of the differential equations is reached the solution might develop a branch cut, depending on the functional basis used to represent the solution. However in a physical process the master integrals are analytic in the physical region and those branch cuts must be spurious. In order to ensure the right analytic properties of the master integrals, we can replace the discontinuous functions with different, analytic branches.

Solving the differential equations in terms of Li functions, we can in principle find a  basis without branch cuts in the physical region, so that the solution has the right analytic structure and no spurious branch cuts are present. However in our case the set of analytic functions in the physical region is too small to represent the master integrals, and we are forced to employ also discontinuous functions. For simplicity they are chosen to be the following two logarithms and five $\text{Li}_3 $,
\begin{equation}
\log (x_q-y_q),\; \log(-y_q x_q^2+y_q x_q-y_q+x_q) \, , \; \text{Li}_3 (a_i) \; i=1,...,5 \, .
\label{setadd}
\end{equation}
where the $a_i$'s are
\begin{align}
\vec{a}=&  \left\{ \frac{x_q y_q^2-x_q y_q+x_q-y_q}{(x_q+1) (1-y_q)^2},\frac{-x_q y_q^2+x_q y_q-x_q+y_q}{(1-x_q) (x_q+1) y_q},
\frac{-x_q y_q^2+x_q y_q-x_q+y_q}{(x_q+1) (y_q-x_q) (1-x_q y_q)}\right. ,\nonumber \\
 &\left.  \frac{x_q y_q^2-x_q y_q+x_q-y_q}{x_q y_q-y_q^2+y_q-1},\frac{x_q y_q^2-x_q y_q+x_q-y_q}{x_q \left(-x_q y_q+y_q^2-y_q+1\right) }\right\}\,.
\end{align}
Let us show how to find the analytic branch of $\log(x_q-y_q)$ in the physical region. 
The values of the argument cover a region in the second, third and fourth quadrant of the complex plane, positive real axis included, intersecting the cut of the logarithm. An analytic branch can be found, for every polylogarithm, subtracting the discontinuity of the function across the branch cut. In the present case $\text{Disc}(\log(x))=2 \pi i$ so that the analytic branch, $\log ^*(x)$, is
\begin{equation}
\label{logcont}
\log^{*}(x_q-y_q)= \log(x_q-y_q) - 2 i \pi \,\theta(\Im (x_q-y_q))) \,,
\end{equation}
where $\theta(x)$ is the Heaviside $\theta$ function. 
Note that employing $\theta$ functions we introduce a (removable) discontinuity on the real axis. In the next section we show how to remove this discontinuity.

For $\text{Li}_3 (a_i)$ the $a_i$'s lie on the branch cut whenever $\Re(a_i) > 1$. We remove the cut ambiguity recalling that the physical region is defined via the Feynman prescription,
and in order to get an analytic function we use the analytic continuation of $\text{Li}_3$ whenever
$\Re(a_i)>1$. This can be achieved, again, by means of $\theta$ functions,
\begin{equation}
\label{li3cont}
\text{Li}^*_3 (a_i) = \theta(1-\Re(a_i))\text{Li}_3 (a_i) + \theta(\Re(a_i)-1) \text{C}_3(a_i,\sigma_i) \,,
\end{equation}
where $\sigma_i$ is the sign of the imaginary part of $a_i$ due to Feynman prescription,
\begin{equation}
\vec{\sigma }= \{-1, -1, -1, 1, 1\}\,,
\end{equation}
and $C_3$ is the analytic continuation of $\text{Li}_3$,
\begin{equation}
\text{C}_3(a_i,\sigma_i)= \text{Li}_3\left(\frac{1}{a_i}\right)-\frac{1}{6} \log ^3(a_i)+\sigma_i \frac{1}{2} i \pi   \log ^2(a_i)+\frac{1}{3} \pi ^2 \log (a_i) \,.
\end{equation}

\subsection{Analytic branches in the physical region}

The representations (\ref{logcont}) and (\ref{li3cont}) are analytic everywhere in the physical region, 
but if the argument of the $\theta$ function is zero the expression is not defined. Nevertheless, by construction, the new
basis has no branch cuts and the limits to the discontinuous points are well defined. In order to remove these (spurious) discontinuities we impose the limiting values as long as the arguments of the $\theta$ functions vanish.

The limiting values to the real axis of Eq.~(\ref{logcont}) are such that $\log^*(-1)= -i \pi$ and $\log^*(x>0)$ is positive. This uniquely defines Eq.~(\ref{logcont}) over the entire physical region. With a similar reasoning we can find the analytic branch, $\log^\dagger (x)$,     of the second logarithm of Eq.~(\ref{setadd}). Defining $a=x_q-y_q$ and $b = -y_q x_q^2+y_q x_q-y_q+x_q$ we arrive at
\begin{align}
\label{fullcont}
\log^*(a)=& \; \left[1 - \delta\left(\Im(a)\right)\right] c(a) + \delta\left(\Im(a)\right) [\Re(\log(a))-i \pi],\nonumber \\
\log^\dagger (b)= &\; \left[1 - \delta\left(\Im(b)\right)\right] d(b) + \delta\left(\Im(b)\right) [\Re(\log(b))-i \pi(1+\theta_{1/2}(x_q,y_q)) ]\,,
\end{align}
where 
\begin{equation}
\delta(x) = \left\{ 
\begin{array}{ll} 
1 & \mbox{for}\, x=0 \\
0 & \mbox{otherwise}
	\end{array}
\right. \,,
\end{equation}
and 
\begin{align}
c(a) = &\; \log(a)-2 i \pi\, \theta (\Im(a)),\nonumber\\
d(b)= & \;\log(b)-2 i \pi\, \theta (\Im(b)) -2 i \pi \theta_{1/2}(x_q,y_q) \theta (-\Im(b)),\nonumber \\
\theta_{1/2}(x_q,y_q)=& \;\theta_0(\pi/2-\arg(x_q)) \theta_0(\pi/2-\arg(y_q)) \,,
\end{align}
where $0 < \arg(x) \leq \pi$.

In Eq.~(\ref{li3cont}), $\text{Li}_3(1)=\text{C}_3(1,\sigma_i)$, then
\begin{equation}
\text{Li}^*_3 (a_i) = \theta_0(1-\Re(a_i))\text{Li}_3 (a_i) + \theta_1(\Re(a_i)-1) \text{C}_3(a_i,\sigma_i),
\end{equation}
where $\theta_0(x)$ and $\theta_1(x)$ are modified Heaviside $\theta$ functions, 
such that $\theta_0(0) = 0$ and $\theta_1(0) = 1$.

Note that Eqs.~(\ref{fullcont}) do not hold if $x_q,y_q$ are both real, where the standard log has to be used. To avoid clutter we omit here the extra terms extending the validity of Eq.~(\ref{fullcont}) to that region, while they are present in the ancillary files of the arXiv submission. Although the above representations might look cumbersome, they are nothing but simple combinations of elementary functions like the Heaviside $\theta$, and implementing them for numeric evaluations is straightforward computer algebra. The expressions in the ancillary files make only use of built-in {\tt Mathematica} functions.

\bibliographystyle{JHEP}
\bibliography{refs}

\providecommand{\href}[2]{#2}\begingroup\raggedright\begin{thebibliography}{10}

\bibitem{Aad:2012tfa}
{\bf ATLAS} Collaboration, G.~Aad et~al., {\it {Observation of a new particle
  in the search for the Standard Model Higgs boson with the ATLAS detector at
  the LHC}},  {\em Phys.Lett.} {\bf B716} (2012) 1--29,
  [\href{http://arxiv.org/abs/1207.7214}{{\tt arXiv:1207.7214}}].

\bibitem{Chatrchyan:2012ufa}
{\bf CMS} Collaboration, S.~Chatrchyan et~al., {\it {Observation of a new boson
  at a mass of 125 GeV with the CMS experiment at the LHC}},  {\em Phys.Lett.}
  {\bf B716} (2012) 30--61, [\href{http://arxiv.org/abs/1207.7235}{{\tt
  arXiv:1207.7235}}].

\bibitem{Heinemeyer:2013tqa}
{\bf LHC Higgs Cross Section Working Group} Collaboration, S.~Heinemeyer
  et~al., {\it {Handbook of LHC Higgs Cross Sections: 3. Higgs Properties}},
  \href{http://arxiv.org/abs/1307.1347}{{\tt arXiv:1307.1347}}.

\bibitem{Chatrchyan:2013vaa}
{\bf CMS} Collaboration, S.~Chatrchyan et~al., {\it {Search for a Higgs boson
  decaying into a Z and a photon in pp collisions at $\sqrt{s}$ = 7 and 8
  TeV}},  {\em Phys.Lett.} {\bf B726} (2013) 587--609,
  [\href{http://arxiv.org/abs/1307.5515}{{\tt arXiv:1307.5515}}].

\bibitem{Aad:2014fia}
{\bf ATLAS} Collaboration, G.~Aad et~al., {\it {Search for Higgs boson decays
  to a photon and a Z boson in pp collisions at $\sqrt{s}$=7 and 8 TeV with the
  ATLAS detector}},  {\em Phys.Lett.} {\bf B732} (2014) 8--27,
  [\href{http://arxiv.org/abs/1402.3051}{{\tt arXiv:1402.3051}}].

\bibitem{Cahn:1978nz}
R.~Cahn, M.~S. Chanowitz, and N.~Fleishon, {\it {Higgs Particle Production by Z
  to H Gamma}},  {\em Phys.Lett.} {\bf B82} (1979) 113.

\bibitem{Bergstrom:1985hp}
L.~Bergstrom and G.~Hulth, {\it {Induced Higgs Couplings to Neutral Bosons in
  $e^+ e^-$ Collisions}},  {\em Nucl.Phys.} {\bf B259} (1985) 137.

\bibitem{Low:2011gn}
I.~Low, J.~Lykken, and G.~Shaughnessy, {\it {Singlet scalars as Higgs imposters
  at the Large Hadron Collider}},  {\em Phys.Rev.} {\bf D84} (2011) 035027,
  [\href{http://arxiv.org/abs/1105.4587}{{\tt arXiv:1105.4587}}].

\bibitem{Low:2012rj}
I.~Low, J.~Lykken, and G.~Shaughnessy, {\it {Have We Observed the Higgs
  (Imposter)?}},  {\em Phys.Rev.} {\bf D86} (2012) 093012,
  [\href{http://arxiv.org/abs/1207.1093}{{\tt arXiv:1207.1093}}].

\bibitem{Azatov:2013ura}
A.~Azatov, R.~Contino, A.~Di~Iura, and J.~Galloway, {\it {New Prospects for
  Higgs Compositeness in $h \to Z\gamma$}},  {\em Phys.Rev.} {\bf D88} (2013),
  no.~7 075019, [\href{http://arxiv.org/abs/1308.2676}{{\tt arXiv:1308.2676}}].

\bibitem{Carena:2012xa}
M.~Carena, I.~Low, and C.~E. Wagner, {\it {Implications of a Modified Higgs to
  Diphoton Decay Width}},  {\em JHEP} {\bf 1208} (2012) 060,
  [\href{http://arxiv.org/abs/1206.1082}{{\tt arXiv:1206.1082}}].

\bibitem{Chiang:2012qz}
C.-W. Chiang and K.~Yagyu, {\it {Higgs boson decays to $\gamma\gamma$ and $Z
  \gamma$ in models with Higgs extensions}},  {\em Phys.Rev.} {\bf D87} (2013),
  no.~3 033003, [\href{http://arxiv.org/abs/1207.1065}{{\tt arXiv:1207.1065}}].

\bibitem{Chen:2013vi}
C.-S. Chen, C.-Q. Geng, D.~Huang, and L.-H. Tsai, {\it {New Scalar
  Contributions to $h\to Z\gamma$}},  {\em Phys.Rev.} {\bf D87} (2013) 075019,
  [\href{http://arxiv.org/abs/1301.4694}{{\tt arXiv:1301.4694}}].

\bibitem{Spira:1991tj}
M.~Spira, A.~Djouadi, and P.~Zerwas, {\it {QCD corrections to the H Z gamma
  coupling}},  {\em Phys.Lett.} {\bf B276} (1992) 350--353.

\bibitem{Bardeen:1978yd}
W.~A. Bardeen, A.~Buras, D.~Duke, and T.~Muta, {\it {Deep Inelastic Scattering
  Beyond the Leading Order in Asymptotically Free Gauge Theories}},  {\em
  Phys.Rev.} {\bf D18} (1978) 3998.

\bibitem{'tHooft:1972fi}
G.~'t~Hooft and M.~Veltman, {\it {Regularization and Renormalization of Gauge
  Fields}},  {\em Nucl.Phys.} {\bf B44} (1972) 189--213.

\bibitem{Bollini:1972bi}
C.~Bollini and J.~Giambiagi, {\it {Lowest order divergent graphs in
  nu-dimensional space}},  {\em Phys.Lett.} {\bf B40} (1972) 566--568.

\bibitem{Bollini:1972ui}
C.~Bollini and J.~Giambiagi, {\it {Dimensional Renormalization: The Number of
  Dimensions as a Regularizing Parameter}},  {\em Nuovo Cim.} {\bf 12} (1972),
  no.~1 20--26.

\bibitem{Ashmore:1972uj}
J.~Ashmore, {\it {A Method of Gauge Invariant Regularization}},  {\em
  Lett.Nuovo Cim.} {\bf 4} (1972) 289--290.

\bibitem{Cicuta:1972jf}
G.~Cicuta and E.~Montaldi, {\it {Analytic renormalization via continuous space
  dimension}},  {\em Lett.Nuovo Cim.} {\bf 4} (1972) 329--332.

\bibitem{Gastmans:1973uv}
R.~Gastmans and R.~Meuldermans, {\it {Dimensional regularization of the
  infrared problem}},  {\em Nucl.Phys.} {\bf B63} (1973) 277--284.

\bibitem{Smirnov:2008iw}
A.~Smirnov, {\it {Algorithm FIRE -- Feynman Integral REduction}},  {\em JHEP}
  {\bf 0810} (2008) 107, [\href{http://arxiv.org/abs/0807.3243}{{\tt
  arXiv:0807.3243}}].

\bibitem{Smirnov:2013dia}
A.~Smirnov and V.~Smirnov, {\it {FIRE4, LiteRed and accompanying tools to solve
  integration by parts relations}},  {\em Comput.Phys.Commun.} {\bf 184} (2013)
  2820--2827, [\href{http://arxiv.org/abs/1302.5885}{{\tt arXiv:1302.5885}}].

\bibitem{Smirnov:2014hma}
A.~V. Smirnov, {\it {FIRE5: a C++ implementation of Feynman Integral
  REduction}},  {\em Comput.Phys.Commun.} {\bf 189} (2014) 182--191,
  [\href{http://arxiv.org/abs/1408.2372}{{\tt arXiv:1408.2372}}].

\bibitem{Studerus:2009ye}
C.~Studerus, {\it {Reduze-Feynman Integral Reduction in C++}},  {\em
  Comput.Phys.Commun.} {\bf 181} (2010) 1293--1300,
  [\href{http://arxiv.org/abs/0912.2546}{{\tt arXiv:0912.2546}}].

\bibitem{vonManteuffel:2012np}
A.~von Manteuffel and C.~Studerus, {\it {Reduze 2 - Distributed Feynman
  Integral Reduction}},  \href{http://arxiv.org/abs/1201.4330}{{\tt
  arXiv:1201.4330}}.

\bibitem{Tkachov:1981wb}
F.~Tkachov, {\it {A Theorem on Analytical Calculability of Four Loop
  Renormalization Group Functions}},  {\em Phys.Lett.} {\bf B100} (1981)
  65--68.

\bibitem{Chetyrkin:1981qh}
K.~Chetyrkin and F.~Tkachov, {\it {Integration by Parts: The Algorithm to
  Calculate beta Functions in 4 Loops}},  {\em Nucl.Phys.} {\bf B192} (1981)
  159--204.

\bibitem{Kotikov:1990kg}
A.~Kotikov, {\it {Differential equations method: New technique for massive
  Feynman diagrams calculation}},  {\em Phys.Lett.} {\bf B254} (1991) 158--164.

\bibitem{Kotikov:1991hm}
A.~Kotikov, {\it {Differential equations method: The Calculation of vertex type
  Feynman diagrams}},  {\em Phys.Lett.} {\bf B259} (1991) 314--322.

\bibitem{Kotikov:1991pm}
A.~Kotikov, {\it {Differential equation method: The Calculation of N point
  Feynman diagrams}},  {\em Phys.Lett.} {\bf B267} (1991) 123--127.

\bibitem{Remiddi:1997ny}
E.~Remiddi, {\it {Differential equations for Feynman graph amplitudes}},  {\em
  Nuovo Cim.} {\bf A110} (1997) 1435--1452,
  [\href{http://arxiv.org/abs/hep-th/9711188}{{\tt hep-th/9711188}}].

\bibitem{Caffo:1998du}
M.~Caffo, H.~Czyz, S.~Laporta, and E.~Remiddi, {\it {The Master differential
  equations for the two loop sunrise selfmass amplitudes}},  {\em Nuovo Cim.}
  {\bf A111} (1998) 365--389, [\href{http://arxiv.org/abs/hep-th/9805118}{{\tt
  hep-th/9805118}}].

\bibitem{Gehrmann:1999as}
T.~Gehrmann and E.~Remiddi, {\it {Differential equations for two loop four
  point functions}},  {\em Nucl.Phys.} {\bf B580} (2000) 485--518,
  [\href{http://arxiv.org/abs/hep-ph/9912329}{{\tt hep-ph/9912329}}].

\bibitem{Argeri:2007up}
M.~Argeri and P.~Mastrolia, {\it {Feynman Diagrams and Differential
  Equations}},  {\em Int.J.Mod.Phys.} {\bf A22} (2007) 4375--4436,
  [\href{http://arxiv.org/abs/0707.4037}{{\tt arXiv:0707.4037}}].

\bibitem{Henn:2013pwa}
J.~M. Henn, {\it {Multiloop integrals in dimensional regularization made
  simple}},  {\em Phys.Rev.Lett.} {\bf 110} (2013), no.~25 251601,
  [\href{http://arxiv.org/abs/1304.1806}{{\tt arXiv:1304.1806}}].

\bibitem{Henn:2013woa}
J.~M. Henn and V.~A. Smirnov, {\it {Analytic results for two-loop master
  integrals for Bhabha scattering I}},  {\em JHEP} {\bf 1311} (2013) 041,
  [\href{http://arxiv.org/abs/1307.4083}{{\tt arXiv:1307.4083}}].

\bibitem{Henn:2013nsa}
J.~M. Henn, A.~V. Smirnov, and V.~A. Smirnov, {\it {Evaluating single-scale
  and/or non-planar diagrams by differential equations}},  {\em JHEP} {\bf
  1403} (2014) 088, [\href{http://arxiv.org/abs/1312.2588}{{\tt
  arXiv:1312.2588}}].

\bibitem{Henn:2014lfa}
J.~M. Henn, K.~Melnikov, and V.~A. Smirnov, {\it {Two-loop planar master
  integrals for the production of off-shell vector bosons in hadron
  collisions}},  {\em JHEP} {\bf 1405} (2014) 090,
  [\href{http://arxiv.org/abs/1402.7078}{{\tt arXiv:1402.7078}}].

\bibitem{Caola:2014lpa}
F.~Caola, J.~M. Henn, K.~Melnikov, and V.~A. Smirnov, {\it {Non-planar master
  integrals for the production of two off-shell vector bosons in collisions of
  massless partons}},  {\em JHEP} {\bf 1409} (2014) 043,
  [\href{http://arxiv.org/abs/1404.5590}{{\tt arXiv:1404.5590}}].

\bibitem{Caron-Huot:2014lda}
S.~Caron-Huot and J.~M. Henn, {\it {Iterative structure of finite loop
  integrals}},  {\em JHEP} {\bf 1406} (2014) 114,
  [\href{http://arxiv.org/abs/1404.2922}{{\tt arXiv:1404.2922}}].

\bibitem{Gehrmann:2014bfa}
T.~Gehrmann, A.~von Manteuffel, L.~Tancredi, and E.~Weihs, {\it {The two-loop
  master integrals for $q\overline{q} \to VV$}},  {\em JHEP} {\bf 1406} (2014)
  032, [\href{http://arxiv.org/abs/1404.4853}{{\tt arXiv:1404.4853}}].

\bibitem{Argeri:2014qva}
M.~Argeri, S.~Di~Vita, P.~Mastrolia, E.~Mirabella, J.~Schlenk, et~al., {\it
  {Magnus and Dyson Series for Master Integrals}},  {\em JHEP} {\bf 1403}
  (2014) 082, [\href{http://arxiv.org/abs/1401.2979}{{\tt arXiv:1401.2979}}].

\bibitem{Hoschele:2014qsa}
M.~Hoeschele, J.~Hoff, and T.~Ueda, {\it {Adequate bases of phase space master
  integrals for gg $\to$ h at NNLO and beyond}},  {\em JHEP} {\bf 1409} (2014)
  116, [\href{http://arxiv.org/abs/1407.4049}{{\tt arXiv:1407.4049}}].

\bibitem{Dulat:2014mda}
F.~Dulat and B.~Mistlberger, {\it {Real-Virtual-Virtual contributions to the
  inclusive Higgs cross section at N3LO}},
  \href{http://arxiv.org/abs/1411.3586}{{\tt arXiv:1411.3586}}.

\bibitem{Bell:2014zya}
G.~Bell and T.~Huber, {\it {Master integrals for the two-loop penguin
  contribution in non-leptonic $B$-decays}},  {\em JHEP} {\bf 1412} (2014) 129,
  [\href{http://arxiv.org/abs/1410.2804}{{\tt arXiv:1410.2804}}].

\bibitem{Henn:2014qga}
J.~M. Henn, {\it {Lectures on differential equations for Feynman integrals}},
  {\em J.Phys.} {\bf A48} (2015), no.~15 153001,
  [\href{http://arxiv.org/abs/1412.2296}{{\tt arXiv:1412.2296}}].

\bibitem{Lee:2014ioa}
R.~N. Lee, {\it {Reducing differential equations for multiloop master
  integrals}},  \href{http://arxiv.org/abs/1411.0911}{{\tt arXiv:1411.0911}}.

\bibitem{Christian}
C.~Kurz, {\it {Two-Loop QCD Corrections for the Production and Decays of Higgs
  Bosons}},  {\em {\rm Diploma Thesis, University of Freiburg, Germany.}}
  (December 2005).

\bibitem{Chen}
K.-T. Chen, {\it {Iterated path integrals}},  {\em Bull. Amer. Math. Soc.} {\bf
  83} (1977) 831--879.

\bibitem{Goncharov:1998kja}
A.~B. Goncharov, {\it {Multiple polylogarithms, cyclotomy and modular
  complexes}},  {\em Math.Res.Lett.} {\bf 5} (1998) 497--516,
  [\href{http://arxiv.org/abs/1105.2076}{{\tt arXiv:1105.2076}}].

\bibitem{Broadhurst:1998rz}
D.~J. Broadhurst, {\it {Massive three - loop Feynman diagrams reducible to SC*
  primitives of algebras of the sixth root of unity}},  {\em Eur.Phys.J.} {\bf
  C8} (1999) 311--333, [\href{http://arxiv.org/abs/hep-th/9803091}{{\tt
  hep-th/9803091}}].

\bibitem{Remiddi:1999ew}
E.~Remiddi and J.~Vermaseren, {\it {Harmonic polylogarithms}},  {\em
  Int.J.Mod.Phys.} {\bf A15} (2000) 725--754,
  [\href{http://arxiv.org/abs/hep-ph/9905237}{{\tt hep-ph/9905237}}].

\bibitem{Smirnov:2008py}
A.~Smirnov and M.~Tentyukov, {\it {Feynman Integral Evaluation by a Sector
  decomposiTion Approach (FIESTA)}},  {\em Comput.Phys.Commun.} {\bf 180}
  (2009) 735--746, [\href{http://arxiv.org/abs/0807.4129}{{\tt
  arXiv:0807.4129}}].

\bibitem{Smirnov:2009pb}
A.~Smirnov, V.~Smirnov, and M.~Tentyukov, {\it {FIESTA 2: Parallelizeable
  multiloop numerical calculations}},  {\em Comput.Phys.Commun.} {\bf 182}
  (2011) 790--803, [\href{http://arxiv.org/abs/0912.0158}{{\tt
  arXiv:0912.0158}}].

\bibitem{Smirnov:2013eza}
A.~V. Smirnov, {\it {FIESTA 3: cluster-parallelizable multiloop numerical
  calculations in physical regions}},  {\em Comput.Phys.Commun.} {\bf 185}
  (2014) 2090--2100, [\href{http://arxiv.org/abs/1312.3186}{{\tt
  arXiv:1312.3186}}].

\bibitem{Vollinga:2004sn}
J.~Vollinga and S.~Weinzierl, {\it {Numerical evaluation of multiple
  polylogarithms}},  {\em Comput.Phys.Commun.} {\bf 167} (2005) 177,
  [\href{http://arxiv.org/abs/hep-ph/0410259}{{\tt hep-ph/0410259}}].

\bibitem{Goncharov:2010jf}
A.~B. Goncharov, M.~Spradlin, C.~Vergu, and A.~Volovich, {\it {Classical
  Polylogarithms for Amplitudes and Wilson Loops}},  {\em Phys.Rev.Lett.} {\bf
  105} (2010) 151605, [\href{http://arxiv.org/abs/1006.5703}{{\tt
  arXiv:1006.5703}}].

\bibitem{Brown1}
F.~C.~S. {Brown}, {\it {Multiple zeta values and periods of moduli spaces
  ${\mathfrak M}_{0,n}$}},  {\em ArXiv Mathematics e-prints} (June, 2006)
  [\href{http://arxiv.org/abs/math/0606419}{{\tt math/0606419}}].

\bibitem{Duhr:2011zq}
C.~Duhr, H.~Gangl, and J.~R. Rhodes, {\it {From polygons and symbols to
  polylogarithmic functions}},  {\em JHEP} {\bf 1210} (2012) 075,
  [\href{http://arxiv.org/abs/1110.0458}{{\tt arXiv:1110.0458}}].

\bibitem{Duhr:2012fh}
C.~Duhr, {\it {Hopf algebras, coproducts and symbols: an application to Higgs
  boson amplitudes}},  {\em JHEP} {\bf 1208} (2012) 043,
  [\href{http://arxiv.org/abs/1203.0454}{{\tt arXiv:1203.0454}}].

\bibitem{GehrmannHZg}
T.~Gehrmann, S.~Guns, and D.~Kara, {\it {The rare decay $H->Z\gamma$ in
  perturbative QCD}},  {\em {\rm in preparation}}.

\bibitem{Chen:2012ju}
L.-B. Chen, C.-F. Qiao, and R.-L. Zhu, {\it {Reconstructing the 125 GeV SM
  Higgs Boson Through $\ell\bar{\ell}\gamma$}},  {\em Phys.Lett.} {\bf B726}
  (2013) 306--311, [\href{http://arxiv.org/abs/1211.6058}{{\tt
  arXiv:1211.6058}}].

\bibitem{Dicus:2013ycd}
D.~A. Dicus and W.~W. Repko, {\it {Calculation of the decay ${H\to e{\bar
  e}\gamma}$}},  {\em Phys.Rev.} {\bf D87} (2013), no.~7 077301,
  [\href{http://arxiv.org/abs/1302.2159}{{\tt arXiv:1302.2159}}].

\bibitem{Sun:2013rqa}
Y.~Sun, H.-R. Chang, and D.-N. Gao, {\it {Higgs decays to gamma l+ l- in the
  standard model}},  {\em JHEP} {\bf 1305} (2013) 061,
  [\href{http://arxiv.org/abs/1303.2230}{{\tt arXiv:1303.2230}}].

\bibitem{Passarino:2013nka}
G.~Passarino, {\it {Higgs Boson Production and Decay: Dalitz Sector}},  {\em
  Phys.Lett.} {\bf B727} (2013) 424--431,
  [\href{http://arxiv.org/abs/1308.0422}{{\tt arXiv:1308.0422}}].

\bibitem{Dicus:2013lta}
D.~A. Dicus, C.~Kao, and W.~W. Repko, {\it {Comparison of
  $H\to\ell\bar{\ell}\gamma$ and $H\to\gamma\,Z, Z\to\ell\bar{\ell}$ including
  the ATLAS cuts}},  {\em Phys.Rev.} {\bf D89} (2014) 033013,
  [\href{http://arxiv.org/abs/1310.4380}{{\tt arXiv:1310.4380}}].

\bibitem{Kuipers:2012rf}
J.~Kuipers, T.~Ueda, J.~Vermaseren, and J.~Vollinga, {\it {FORM version 4.0}},
  {\em Comput.Phys.Commun.} {\bf 184} (2013) 1453--1467,
  [\href{http://arxiv.org/abs/1203.6543}{{\tt arXiv:1203.6543}}].

\bibitem{Kublbeck:1990xc}
J.~Kublbeck, M.~Bohm, and A.~Denner, {\it {Feyn Arts: Computer Algebraic
  Generation of Feynman Graphs and Amplitudes}},  {\em Comput.Phys.Commun.}
  {\bf 60} (1990) 165--180.

\bibitem{Hahn:2000kx}
T.~Hahn, {\it {Generating Feynman diagrams and amplitudes with FeynArts 3}},
  {\em Comput.Phys.Commun.} {\bf 140} (2001) 418--431,
  [\href{http://arxiv.org/abs/hep-ph/0012260}{{\tt hep-ph/0012260}}].

\bibitem{Vermaseren:1994je}
J.~Vermaseren, {\it {Axodraw}},  {\em Comput.Phys.Commun.} {\bf 83} (1994)
  45--58.

\end{thebibliography}\endgroup

\end{document}